\documentclass[pdftex,twocolumn,epjc3]{svjour3}          
\RequirePackage[T1]{fontenc}

\smartqed  

\RequirePackage{graphicx}
\RequirePackage{mathptmx}      
\RequirePackage{flushend}
\usepackage{dcolumn}
\usepackage{bbold}
\usepackage{tabularx}
\RequirePackage[numbers,sort&compress]{natbib}
\RequirePackage[colorlinks,citecolor=blue,urlcolor=blue,linkcolor=blue,breaklinks=true]{hyperref}

\usepackage{epsfig,bbm,cancel,ulem}
\usepackage{epstopdf}
\usepackage{bm}
\usepackage{color}
\usepackage{slashed}
\usepackage{physics}
\usepackage{amssymb}
\usepackage{multirow}
\usepackage{float}
\usepackage{wasysym}
\usepackage{derivative}
\usepackage{rotating}
\usepackage{booktabs}
\usepackage{multirow}
\usepackage{amsmath}
\usepackage{makecell}
\usepackage{adjustbox}
\usepackage[caption=false]{subfig}
\usepackage[utf8]{inputenc}
\usepackage[T1]{fontenc}
\usepackage{textcomp}
\usepackage{array}
\usepackage{makecell}
\newcolumntype{x}[1]{>{\centering\arraybackslash}p{#1}}
\usepackage{tikz}

\newcommand {\bea}{\begin{eqnarray}}
	\newcommand {\eea}{\end{eqnarray}}
\newcommand {\be}{\begin{equation}}
	\newcommand {\ee}{\end{equation}}

\journalname{-}

\begin{document}
	\renewcommand{\hbar}{\mathchar'26\mkern-9mu h}
	\title{The Milky Way and M31 rotation curves in Yukawa gravity: phenomenology and Bayesian analysis}

	\author{Davendra S. Hassan\thanksref{e1,addr1}
		\and		
		M Dio Danarianto\thanksref{e2,addr2}
		\and
		Anto Sulaksono\thanksref{e3,addr3} 
	}
	
	\thankstext{e1}{e-mail:davendra.hassan@u.nus.edu }	
	\thankstext{e2}{e-mail: m.dio.danarianto@brin.go.id}
	\thankstext{e3}{e-mail: anto.sulaksono@sci.ui.ac.id}	
	
	\institute{Department of Physics, National University of Singapore, 117551, Singapore\label{addr1}
		\and	
		Research Center for Computing, National Research and Innovation Agency (BRIN), Bandung 40173, Indonesia\label{addr2}
        \and
        Departemen Fisika, FMIPA, Universitas Indonesia, Depok 16424, Indonesia\label{addr3}}

\maketitle

\begin{abstract}
Yukawa gravity provides a generalized framework for modeling gravity modification. We investigate the rotation curve profiles of spiral galaxies under Yukawa-like theories governed by the coupling strength $\beta$ and the interaction range $\lambda$. We develop a unified analytical and numerical framework to calculate rotational velocities under Yukawa gravity, which includes contributions from all major galactic components: stellar bulge, disk, dark matter (DM) halo, and central supermassive black hole. The calculations show that $\beta$ and $\lambda$ strongly influence velocity distributions, shifting peaks, creating double-peak structures, or enhancing dark matter dominance in the bulge or disk. To assess observational implications, we perform Bayesian analyses using data from the Milky Way (MW) and Andromeda (M31), which offer complementary characteristics: MW provides precise velocity profiles across multiple scales, while M31 includes broader morphological constraints. We examine four scenarios: Yukawa gravity without dark matter, dark matter with non-trivial coupling, fully modified gravity, and standard Newtonian gravity. Results show that MW models with $\lambda < 1$ kpc yield high Bayes factors but risk overfitting, as dark matter mimics baryonic kinematics, while M31's photometric priors from conjugate observations mitigate this, yielding robust parameter estimates. However, in M31, Bayes factors favor Newtonian gravity, suggesting that current data lack the precision to resolve more complex models. This finding highlights two key needs: (i) realistic, physically or empirically informed priors to avoid biased constraints, and (ii) high-precision data with independent photometry to guard against overfitting. Our framework offers a scalable approach for testing gravity with large galactic rotation curve datasets.
\end{abstract}

\section{Introduction}
\label{sec:level1}
Rotation curves (RCs) of a galaxy represent the average circular velocity of matter as a function of its distance from the galactic center and, under certain assumptions, can serve as a fundamental measurement of the approximate mass within a galaxy \cite{Bertone2018}. Assuming the mass is concentrated in the cores of galaxies, based on Newtonian gravity and general relativity (GR), these curves are expected to follow Kepler's laws, where the velocity declines as $v \propto 1/\sqrt{r}$. However, empirical data in general, such as that gathered through optical (H$\alpha$) observations and HI line spectroscopy \cite{Honma1997}, reveal flat RCs in the outer regions of the galaxies, i.e., $v \propto$ constant, which highlights the mass discrepancies and possible modification of GR theory in galaxies. The explicit conclusions regarding the need for invisible matter, known as dark matter (DM) and or modified gravity (MG) theories, began to emerge. See Ref. \cite{Bertone2018} for a historical review of DM, Ref. \cite{Cirelli2024} for a complete review of the theoretical, observational, and experimental aspects of DM, and \cite{Will2018} for an overview of theory and experiments regarding gravitational physics.

Despite the compelling nature of the DM hypothesis, it faces significant challenges, including the absence of direct observational evidence \cite{Bertone2018,Bertone2005}. DM is predominant in the galaxy's outer regions, with the contribution of mass significantly surpassing that of luminous matter to account for the flatness of RCs \cite{Mannheim2006}. Thus, the nature of DM remains unknown. It is usually assumed that an unknown particle is not contained in the Standard Model of particle physics, where the possible mass of proposed DM particles depends strongly on the specific model and ranges span dozens of orders of magnitude. The most widely studied is cold DM (CDM), which has successfully predicted properties of the large-scale structure and some CMB anisotropies. An alternative model is Fuzzy DM (FDM), which has attracted attention recently because it predicts the scaling relation between DM particles and density parameters. Please see Ref. \cite{Khelashvili2023} for a recent study on using FDM to understand SPARC galaxies.

However, while GR has passed numerous tests, some argue that the DM problem comes from the nature of gravity on galactic and cosmological scales, which indicates a deviation from GR. Exploring alternatives to GR offers a better way to probe its validity, ensure its universal applicability, and establish more precise constraints. Within the context of the gravity scale, galaxies and galaxy clusters offer an extensive range of gravitational potentials and curvatures ideal for testing gravity. However, high-resolution observations of the central regions are currently limited to the Milky Way and a few neighboring massive galaxies \cite{Baker2015}, such as Andromeda  M31 and the Milky Way, which are both massive spiral galaxies, and hence useful for comparisons. The Milky Way and M31, Andromeda, are prime examples of spiral galaxies characterized by their distinctive spiral patterns. These galaxies feature relatively flat disk structures and are often called disk galaxies. The precise mechanisms underlying the formation of these disk and spiral structures remain a subject of ongoing problems within the field \cite{Binney2008}. This fact makes the Milky Way and M31 galaxies interesting subjects for testing theories of gravity, so that we can test gravity throughout all regions of the galaxy, compared to the more simplistic dataset in other spiral galaxies, such as that derived in the SPARC dataset used in previous studies. Using these galaxies, we can study the impact of somewhat precise rotational velocity $v(r)$ observations/derivations, such as the MW rotation curve, and extragalactic observations, such as that of the M31, enabling independent galactic morphology measurements.

Various MG theories, such as $f(R)$, scalar-tensor, and massive gravity theories, result in an additional exponential decay function to the gravitational potential, similar to the Yukawa potential in particle physics. Please see the examples in Refs. \cite{Almeida2018,Clifton2008,Stabile2011,Moffat2013,Rahvar2014,Henrichs2021} and the references therein. The Yukawa-like correction is often used to constrain gravity in experimental tests and literature, as it can represent a simple deviation from the Newtonian inverse square law. It is a generic representation of the weak-field limit of the aforementioned MG theories. The physical motivation behind introducing this Yukawa-correction can stem from a modified gravitational interaction between DM and ordinary matter (baryons) \cite{Piazza2003,Jusufi2023,Berezhiani2009}. Some of these studies have tried to relieve the core-cusp problem in galaxies through the Yukawa correction, hence it has become a central focus in recent cosmological research. Recent studies have investigated the Yukawa-like correction in late-type galaxies to establish a model that fits better with observed data. For instance, Ref. \cite{Henrichs2021} examined the Milky Way, while Ref. \cite{Almeida2018} examined the SPARC catalog of spiral and irregular galaxies. However, these studies are conducted assuming that baryons are weakly coupled to gravity to meet local gravity constraints, which impacts the constraints on the Yukawa parameters at the small-scale range of the galaxies, resulting in the analysis of the Yukawa correction only in the DM halo of the galaxies, with no studies specifically analyzing the baryonic sector. Moreover, they overlook the presence of a central supermassive black hole (SMBH) and the established experimental constraints on the Yukawa parameters, such as in Refs. \cite{Murata2015,Jovanovic2023}. Hence, this motivates the consideration of the Yukawa correction in all components of the galaxies, not just the DM distribution, but also in baryonic matter components.

Building upon our previous preliminary analysis of dark matter-gravity coupling in the Milky Way \cite{Hassan2024}, the objectives of this research are to study the dynamical behaviors of the Milky Way and M31 Andromeda galaxy's components (including the central SMBH) when affected by the Yukawa-like correction to the gravitational potential in both baryonic components and the DM halo; to determine the probability distribution of model parameters, including the strength and range parameters in the Yukawa term; and to determine the optimal parameter value based on observational data of both galaxies, from which we investigate the compatibility of the model via the Bayes factor.

This study significantly expands the scope and methodology of previous research. Firstly, we do not assume that baryons are weakly coupled to gravity. Unlike earlier works such as Refs. \cite{Almeida2018, Henrichs2021}, which only consider dark matter (DM) as effectively coupled to gravity and limit Yukawa analysis to DM components, we propose that both baryonic and dark matter components are coupled to gravity. We systematically derive the explicit Yukawa-corrected rotation curves for each galactic component, including the supermassive black hole (SMBH). Secondly, we investigate various scenarios, starting with standard Newtonian gravity as a baseline. We consider Yukawa corrections without DM distribution (Y-noDM), Yukawa corrections applied only to the DM distribution (non-trivial dark matter coupling, NTDMC), and a full modified gravity (MG) case where all components experience Yukawa corrections. This classification allows for systematic comparisons with the scenarios examined in previous studies. Thirdly, we improve the statistical methodology by employing Bayesian analysis. While prior studies, including Ref. \cite{Almeida2018}, used the standard MCMC method for parameter estimation, our approach utilizes nested sampling to provide more accurate Bayesian evidence values for model comparison. Fourthly, we analyze two contrasting galactic datasets: the exact Milky Way (MW) dataset and the morphologically well-constrained M31 dataset. This result differs from previous studies, such as Refs. \cite{Almeida2018, Khelashvili2023}, which relied solely on SPARC datasets that lack independent morphological constraints. In contrast to the earlier Yukawa analysis of the MW performed by Ref. \cite{Henrichs2021}, we utilize a more recent dataset by Sofue \cite{Sofue2020} and incorporate the latest Gaia and GRAVITY measurements for improved accuracy. Lastly, we employed broad priors for the Yukawa parameters that were not confined by local solar system constraints.

The structure of this paper is as follows: Section \ref{sect:summary} provides a summary of this study to highlight the main points of this work. Section \ref{yukawa} discusses an overview of the Yukawa correction model considered in this paper. Section \ref{galacticstruc} presents the outline of galactic structure modeling, which details the matter distribution used for each component. The derivation of the circular velocity of the components under Newtonian and Yukawa gravity is explained in Section \ref{rotvel}. Section \ref{rotcurdecompose} describes the statistical framework for decomposing the Yukawa-corrected velocity contributions of the components from galactic rotation curve data and calculating model preference. We finalize with a discussion of our results in Section \ref{discuss} and conclude in Section \ref{conclu}. More detailed mathematical derivations can be found in the Appendix.

\section{Executive summary}
\label{sect:summary}
The circular velocity of galactic components, such as stars and gases, is related to the overall gravitational potential distribution, denoted as $\Phi(\mathbf{r})$, through the equation $V^2_{circ}(r) = r (d\Phi(r)/dr) $. In Yukawa gravity, an additional force is introduced alongside standard Newtonian gravity as an exponential term in the gravitational potential. This potential is influenced by the scale length of the additional Yukawa force, represented by $\lambda$, and its magnitude, denoted as $\beta$.

In this work, we aim to support the investigation of Yukawa-like modified gravity by addressing two interconnected components: (i) the comprehensive mathematical framework to calculate the squared velocity ($V^2$) contributions of each galactic component (bulge, disk, dark matter halo, and supermassive blackhole) under Yukawa-like gravity and (ii) the systematic exploration of their consequences for galactic modified gravity models.

We present a comprehensive derivation of analytical expressions as well as their numerical framework to solve the squared rotational velocity, $V^2(r)$, of all major galactic components -- bulge, disk, dark matter (DM) halo, and central supermassive black hole (SMBH) -- within the context of Yukawa-like modified gravity. While previous studies have primarily focused on applying Yukawa corrections to dark matter alone~\cite{Almeida2018, Henrichs2021}, our approach aims to provide a more unified treatment by explicitly incorporating Yukawa corrections to:
\begin{itemize}
    \item Baryonic components, modeled via S\'ersic bulge and exponential disk gravitational potentials,
    \item Central supermassive black holes, treated as Yukawa-corrected point mass,
    \item Dark matter halos, represented through generalized NFW profiles.
\end{itemize}
This framework allows for a component-wise decomposition of the total velocity profile
$V^2_{\text{tot}}(r) = \sum_i V^2_{\text{N},i}(r) + V^2_{\text{Y},i}(r),$
highlighting how the Yukawa parameters $\beta$ and $\lambda$ influence each component's contribution to the rotation curve (see Fig. \ref{fig:mw_ppc}). Complete expressions for the rotational velocities of these components are provided in Section \ref{rotvel}. We examine four scenarios based on the underlying theoretical framework, which are summarized in Table \ref{tab:cases}.

Our mathematical solutions enabled us to interrogate galactic structure against modified gravity. One notable characteristic is that the combination of the Yukawa parameters, $\beta$ and $\lambda$, can modify the velocity distribution in various ways, depending on the component's distribution (see Fig. \ref{fig:varyuk_all}). These solutions include shifting the peak velocity and creating a double-peaked velocity distribution for a single constituent. Conversely, a small value of $\lambda \sim 1$ kpc combined with a significant value of $\beta \sim 10$ allows the dark matter component to mimic or dominate the distribution of the bulge and disk. Additionally, the repulsive behavior associated with a negative $\beta$ can suppress the velocity to a small value, constrained by the stability condition $V_{tot}^2 > 0$.  

The proposed framework is compared with observational Milky Way (MW) and the Andromeda galaxy (M31) data. While both are broadly classified as spiral galaxies, they offer complementary observational perspectives: MW data provide relatively precise, multi-scale measurements across a wide range of distances, whereas M31 offers morphological information representing global features. 

Employing Bayesian analysis through the \textit{nested sampling} method, we discovered that, in the Milky Way, a small $\lambda$ leads to a high Bayes factor, indicating a risk of overfitting due to the dark matter component "taking over" the disk and bulge regions. However, this behavior does not occur with the M31 data, which is restricted by additional morphological information that strictly constrains the galaxy's shape.

This study highlights two key aspects of constraining gravity through galactic rotation curves. First, using overly restrictive priors can lead to inaccurate constraints; thus, applying realistic priors, i.e., either based on physical principles or driven by data, rather than relying on \textit{ad-hoc} assumptions, is essential. Second, the precision and completeness of data are crucial for effectively constraining gravity through these curves. Additionally, independent morphological measurements obtained from photometric observations enhance the robustness of our findings and help mitigate the risk of overfitting. By adhering to these principles, this study offers a framework for testing gravity using the more extensive galactic rotation curve database.

\begin{table}[]
	\caption{Cases considered in this study. We calculate the Newtonian case of the four-component galactic model as a base comparison. The case where Yukawa gravity is used to explain the flat rotation curve is denoted by Yukawa gravity without dark matter (Y-noDM). In the non-trivial dark matter coupling (NTDMC), the Yukawa correction is assumed only to affect the dark matter component. In contrast, in full modified gravity (MG), the Yukawa correction affects all components with equal coupling strength.}
	\label{tab:cases}
	\begin{tabular}{lcccc}
		\hline \hline
		Component ($i$)  & Newtonian & Y-noDM& NTDMC & MG    \\ \hline
		Bulge      & $\fullmoon$     & $\newmoon$ & $\fullmoon$ & $\newmoon$ \\ 
		Disk       & $\fullmoon$     & $\newmoon$ & $\fullmoon$ & $\newmoon$ \\ 
		DM Halo    & $\fullmoon$     & $\times$ & $\newmoon$ & $\newmoon$ \\ 
		Black Hole & $\fullmoon$     & $\newmoon$&$\fullmoon$ & $\newmoon$ \\ \hline \hline
	\end{tabular}\\[0.5em]
	\footnotesize
	$\fullmoon$ = Calculated in Newtonian, $V_i^2 = V_{N,i}^2$\\
	$\newmoon$ = Calculated in Yukawa-correction, $V_i^2 = V_{N,i}^2+V_{Y,i}^2$\\
	$\times$ = Not considered
\end{table}

\section{Overview of Yukawa gravity}
\label{yukawa}
In the Newtonian regime, the most common method to probe modified gravity interactions is by considering the possibility of an additional contribution to the standard $1/r$ potential. One of the popular models involves adding the exponential term analogous to the Yukawa correction in electromagnetic potential.  In Yukawa-like gravity (which we refer to as Yukawa gravity afterwards), an exponential deviation called the Yukawa-correction potential $(\Phi_{Y})$ is added to the Newtonian inverse-square law $(\Phi_{N})$. The total gravitational potential of a system becomes
\begin{align}
	\label{bab2:totpotNMG}
	\Phi_{G}(\mathbf{r}) &= \Phi_{N}(\mathbf{r}) + \Phi_{Y}(\mathbf{r})\\
	&= -G\int\frac{\rho(\mathbf{r}')}{|\mathbf{r}-\mathbf{r}'|}(1+\beta e^{-|\mathbf{r}-\mathbf{r}'|/\lambda}) \odif[order=3]{\mathbf{r}'}.
\end{align}
The Yukawa-like correction introduces two parameters: the parameter $\beta$, which controls the total strength of the exponential term, and the parameter $\lambda$, which controls the range of interaction. If $\beta = 0$, the gravitational interaction reduces to Newtonian gravity. This model has been stringently tested on microscopic to cosmic scales \cite{Will2018,deRham2017}.

Historically, Yukawa gravity parameterization has been commonly used to probe the gravity-like \textit{fifth} fundamental force (or simply the fifth force). Any deviation from standard Newtonian predictions (assuming that there are no systematic effects) is interpreted as a new force encoded in the $\Phi_{Y}$ term \cite{Fischbach1986,Gibbons1981}. 

From a theoretical perspective, numerous theories of gravity can be expressed in Yukawa terms on the Newtonian scale. As an early example, Nambu-Goldstone theory \cite{Fujii1971} predicted that dilaton-mediated interactions should follow Yukawa parameterization. This fact triggered the search for the "fifth force" in short-range gravity (see Refs. \cite{Murata2014,Lee2020} for reviews of the literature and recent experiments).  Yukawa gravity also appears in massive gravity, where the parameter $\lambda$ is related to the Compton length of the mass of the graviton $\lambda_g = \hbar / m_g c$ \cite{deRham2017}. Yukawa-like characteristics also appear in the $f(R)$ family (including fourth-order gravity \cite{Schmidt2007}), as it is related to scalar-tensor-based gravity \cite{Sotiriou2006}.

Beyond theoretically motivated investigations, phenomenologically motivated studies also probe gravity modification on a galactic scale using Yukawa gravity. For example, early studies showed that the $\beta<0$ case could replace the contribution of dark matter in galactic rotation curves \cite{Sanders1986}. More recently, Ref. \cite{Jusufi2023} demonstrated that the Yukawa potential could approximate flat rotation curves in galaxies through MOND phenomenology. These examples show that Yukawa gravity might replace the dark matter contribution in the galactic rotation curve.

Another investigation by Refs. \cite{Almeida2018,Henrichs2021} treats the dark matter component differently from other components. In the context of a scalar-field-mediated force scenario, they assumed that the field $\varphi$ couples differently to baryons and DM. In a scalar-field scenario, the Yukawa strength parameter evaluates as
\begin{equation}
	\beta = \alpha_{x}\alpha_{y},
\end{equation}
where, following Ref. \cite{Damour2002}, the dimensionless parameter introduced as
\begin{equation}
	\alpha_A(\varphi) \equiv \pdv{\ln{(m_A)}}{\varphi},
\end{equation}
\sloppy which measures the coupling of a scalar field $\varphi$ to a particle of type $A$. The aforementioned models assume that baryons interact with both baryon-baryon, $\alpha_B^2$, and baryon-DM, $\alpha_B\alpha_{DM}$; however, baryon-baryon interactions can be ignored to pass local gravity constraints. In our study, we relax this assumption in the NTDMC case; hence, we assume that both baryons and DM couple effectively to the fifth force, i.e., baryons exert standard gravity with other baryons and Yukawa-like while interacting with DM.

\section{Galactic structure model}
\label{galacticstruc}

Generally, spiral or disk galaxies can be decomposed into several primary components, i.e., the disk, bulge, DM halo, and supermassive black hole (SMBH).  In several studies, the spiral galaxy model can be approached by applying perturbations to the disk. Here, for the sake of simplicity, we consider a galaxy model with an axisymmetric thin disk, with S\'ersic bulge and NFW dark matter profile, with a point-mass black hole in the center. This section describes these components and the density profiles considered in our study.

\subsection{Disk}
In spiral galaxies, most stars reside within the galactic disk, a predominantly flat and axisymmetric structure within the galactic plane. In the galaxy outside the Milky Way, the distribution of stars can be determined through its observed surface brightness, which is defined as the total stellar luminosity emitted per unit area \cite{Binney2008}. From luminosity observations, the galactic disk structure is found to mostly follow an exponential profile \cite{Freeman1970} as
\begin{equation}
	\label{surfmassDisk}
	\Sigma(r)=\Sigma_0 e^{-r/a_d},
\end{equation}
where $\Sigma_0$ represents the central surface brightness value and $a_d$ represents the scale length of the disk \cite{Sofue2015}. Generally, the galactic disk can be separated into two main components based on spatial, kinematic, or chemical criteria \cite{Rix2013}: a thin disk, aligned closely with the galactic plane, and a thick disk, which extends vertically away from the plane. The thick disk of galaxies is believed to be composed of older, metal-poor stars \cite{Adibekyan2011} compared to the thin disk counterpart. While some authors argue against the existence of a galactic thick disk in the Milky Way \cite{Bovy2012}, the thick disk only takes a few fraction of the total galactic disk mass if it exists. Therefore, for the sake of simplification and generalization in spiral galaxies, we model the galactic disk as a thin disk, which has a matter distribution in the form of
\begin{equation}
	\label{bab2:rhodisk}
	\rho_{Disk}(r,z) = \Sigma(r) \delta(z) = \Sigma_0e^{-r/a_d}\delta(z),
\end{equation}
where $\delta(z)$ represents the delta Dirac function, $r$ is the radial distance from the galactic center, and $z$ is the height measured from the galactic plane \cite{Mannheim2006}. In Ref. \cite{Sofue2009}, the galactic bar and spiral arms are modeled as perturbations to the galactic disk, thereby introducing extra terms into Eq. ($\ref{surfmassDisk}$).

\subsection{Bulge}
The galactic bulge is a spheroidal concentration of stars around the galactic center, forming a significant mass distribution in the inner regions of galaxies. In contrast to the surface brightness of the disk, the bulge brightness is only measurable from its surface $I(r)$ due to the projection of its 3D structure \cite{Mannheim2006}. From empirical observations, the surface matter distribution can be represented by the de Vaucouleurs profile \cite{deVaucouleurs1948}, or equivalent to the more general S\'ersic profile for $n=4$ \cite{Sersic1963}. The S\'ersic profile has the form of
\begin{equation}
	\label{bab2:deVau}
	I(r) = I_{e} \exp\left(-b_n\left[\left(\frac{r}{a_b}\right)^{n}-1\right]\right).
\end{equation}
For the de Vaucouleurs profile, $b_n = 7.6695$, and $I_e$ represents the surface matter distribution at the scale length radius $r = a_b$ \cite{Sofue2015}. For an arbitrary value of $n$, $b_n$ can be estimated by $b_n = 1.992n-0.3271$ for $0.5<n<10$ \cite{Capaccioli1989}. The volume density of the bulge can then be calculated through an Abel transformation as
\begin{equation}
	\label{bab2:rhobul}
	\rho_{Bul}(r) = -\frac{1}{\pi}\int_{r}^\infty\odv{I(x)}{x}\frac{\odif{x}}{\sqrt{x^2-r^2}}.
\end{equation}
Though closed-form solutions of the rotational velocity of bulges have been derived in the literature, such as Refs. \cite{Baes2010,Baes2011}, it cannot be implemented using elementary or special functions due to the analytical complexities involved. Therefore, a more straightforward and practical method involves numerically computing the rotational velocity instead. 

\subsection{DM halo}
The distribution of DM can be modeled as a spherical halo that is concentrated in the galaxy's outer regions to account for observed mass discrepancies. The current standard cosmological model describes that these structures are formed through gravitational collapse due to initial density perturbations in the early universe, see Ref. \cite{Benson2010}. Several density models have been formulated for the DM halo, including the Navarro–Frenk–White (NFW) profile, derived through N-body simulations within the standard CDM model \cite{Navarro1996}. The NFW density profile is explicitly given by
\begin{equation}
	\label{bab2:rhoDM}
	\rho_{DM}(r) = \frac{\rho_{s}}{\left( \frac{r}{r_s}\right)\left(1+\frac{r}{r_s}\right)^2}.
\end{equation}
The parameter $\rho_{s}$ denotes the scale density while $r_{s}$ denotes the scale radius. In the inner regions of the halo ($r \ll r_s$), the profile follows $\rho_{DM} \propto 1/r$, while in the outer regions ($r \gg r_s$), it follows $\rho_{DM} \propto 1/r^3$. The intermediate radii follow a similar proportionality to the isothermal profile, i.e., $\rho_{DM} \propto 1/r^2$ \cite{Navarro1996}. The scale density can be calculated as $\rho_s = \rho_{crit}\delta_c$, where $\rho_{crit}$ represents the critical density of the universe, derived from the Friedmann equation for zero curvature ($k=0$) as
\begin{equation}
	\rho_{crit} = \frac{3H_0^2}{8\pi G},
\end{equation}
with Hubble's constant $H_0$ and characteristic overdensity of the halo $\delta_c$ determined by
\begin{equation}
	\delta_c = \frac{200}{3}\frac{c_{200}^3}{\ln(1+c_{200})-c_{200}/(1+c_{200})}.
\end{equation}
The concentration $c_{200}$ is a dimensionless parameter defined as $c_{200} \equiv r_{200}/r_{s}$, where $r_{200}$ represents the virial radius, i.e., the radius of a sphere within which the mean overdensity is $\bar{\delta} = \Delta_c\rho_{crit} = 200 \rho_{crit}$ \cite{Navarro1996}. Throughout this study, we adopt the value of the overdensity constant at virialization as $\Delta_c = 200$.

\subsection{Black hole}
\label{bab2:BH}
It has been confirmed by observations that there is a supermassive blackhole (SMBH) resides in the center of a galaxy \cite{Ghez1998,GaiaCollaboration2023}, e.g., in our Milky Way center it known as Sagittarius $\text{A}^{\star}$ (Sgr $\text{A}^{\star}$). We follow Ref. \cite{Sofue2020} where the density profile of the SMBH can be modeled as a point mass at the origin of the galaxy, 
\begin{equation}
	\label{bab2:rhoBH}
	\rho_{BH}(r) = \frac{M_{BH}\delta(r)}{4\pi r^2},
\end{equation}
where $M_{BH}$ is the mass of black hole, and $\delta(r)$ is the delta Dirac function. 

\section{Rotational velocity profile in Yukawa gravity}
\label{rotvel}

The circular rotational velocity is derived from the gravitational potential as
\begin{equation}
	V_{circ}^2(r) = r\frac{d\Phi(r)}{dr}.
\end{equation}
In general, we model the total rotational velocity of the galaxy as a superposition of components explained above, i.e.,
\begin{equation}
	V_{Tot}^2(r) = \sum V_{i}^2(r)
\end{equation}
where $V_{i}$ is the velocity contribution from each component $i$. In this section, we will briefly describe how we derive the rotational velocity contribution of the galactic components with Yukawa corrections and their numerical results. The more detailed mathematical description of the derivation can be found in  \ref{app:derivation}.
\subsection{Disk}

The velocity distribution of the galactic disk is typically derived using cylindrical coordinates due to its flat structure. Here, the Newtonian disk potential calculation follows the derivation done by Ref. \cite{Mannheim2006}. In contrast, we replace the form of the Green's function and derive it analytically and numerically for the Yukawa-correction potential. As it turns out, the velocity distribution of the galactic disk is significant in the mid-range portion of the galaxies.

The disk component's gravitational potential and circular velocity, both for the Newtonian and Yukawa-correction potential, are derived using the assumption of an axisymmetrical cylindrical matter distribution, evaluated at the $z=0$ plane. Starting with the Green's function in cylindrical coordinates, we derive the Newtonian potential for the disk following the procedure from Ref. \cite{Mannheim2006}, where the assumption of an axisymmetrical distribution simplifies the integral in the gravitational potential equation, from which we can then substitute the matter distribution and solve for the circular velocity. The final Newtonian circular velocity equation is expressed in modified Bessel functions as,
\begin{align}
	V_{Disk,N}^2 =& \frac{G M_d r^2}{2a_d^3} 
	\left[I_0\left(\frac{r}{2a_d}\right)K_0\left(\frac{r}{2a_d}\right)\right. \notag \\
    &\qquad \qquad \left. - I_1\left(\frac{r}{2a_d}\right)K_1\left(\frac{r}{2a_d}\right) \right].
\end{align}
For the Yukawa-correction, we derive it upwards to the gravitational potential, where we substitute in the Green's function for the modified Helmholtz case to obtain
\begin{equation}
	\Phi_{Disk,Y}(r,z=0) = - \frac{G\beta M_d}{a_d^2} \int_0^\infty e^{- r'/a_d} G_H^0(\lambda,r,r') r' \odif{r'},
\end{equation}
where we take 
\begin{equation}
	G_H^0(\lambda,r,r')= \frac{1}{\pi}\int_0^\pi\frac{\exp\left(-\frac{1}{\lambda}\sqrt{r^2+{r'}^2-2rr'\cos\psi}\right)}{\sqrt{r^2+{r'}^2-2rr'\cos\psi}}\odif{\psi},
\end{equation}
in the near-field case and 
\begin{equation}
	G_H^0(\gamma,\mathcal{K},r,r')= \frac{\mathcal{K}}{2\sqrt{rr'}}\exp{[\xi(1-\mathcal{K}^2/4)]}I_0(-\xi \mathcal{K} ^2/4),
\end{equation}
with $\mathcal{K} = 2\sqrt{rR}/|r+R|$ and $\xi = \frac{i}{\lambda}|r+R|$
for the far-field case. Subsequent calculations are carried out numerically, especially for the Yukawa-correction circular velocity. This process involves generating logarithmically spaced radial points, computing the Helmholtz Green's function for either near-field or far-field expression, and integrating it using its logarithmic transformation. The gravitational potential is then computed using a quadrature function, and the velocity is numerically approximated using a five-point central difference method. A combination of cubic and linear interpolation is used to avoid overshoot close to zero to determine the final velocity function.

\subsection{Bulge}
We follow the derivation for the Newtonian potential from Ref. \cite{Mannheim2006} and adapt it to our derivation for the Yukawa-correction potential. To calculate a bulge rotational velocity component, we derive the generic form of the gravitational potential in a spherically symmetric matter for both the Newtonian and Yukawa gravity cases. In these two cases, the equations of Green's functions are significantly different.  We then solve this to calculate the circular velocity for each case, where we derive it analytically to obtain
\begin{equation}
	\label{Eq:vbulnewt}
	V_{Bul,N}^2(r) = \frac{b_n e^{b_n}M_b G}{\eta a_b^{\frac{2n+1}{n}}r} \int_0^r \int_{r'}^\infty\frac{e^{-b_n (x/a_b)^{1/n}}}{x^{\frac{n-1}{n}}\sqrt{x^2-{r'}^2}} {r'}^2 \odif{x,r'}.
\end{equation}
for the Newtonian gravity case and
\begin{align}
	V_{Bul,Y}^{2}(r) &=  \frac{b_n e^{b_n} M_b G\beta\lambda }{\eta a_b^{\frac{2n+1}{n}}r} \left[ e^{-r/\lambda}\left(1+\frac{r}{\lambda}\right)\mathcal{I}_{Bul,1} \right. \notag\\
	&\qquad \left. + \left\{\sinh{\left(\frac{r}{\lambda}\right)} -\frac{r}{\lambda}\cosh{\left(\frac{r}{\lambda}\right)}\right\}\mathcal{I}_{Bul,2} \right],
\end{align}
for the Yukawa gravity, where
\begin{equation}
	\label{Eq:bulfirstint}
	\mathcal{I}_{Bul,1} = \int_0^r \int_{r'}^\infty\frac{e^{-b_n (x/a_b)^{1/4}}}{x^{3/4}\sqrt{x^2-{r'}^2}}r'\sinh{\left(\frac{r'}{\lambda}\right)}\odif{x,r'},
\end{equation}
and
\begin{equation}
	\label{Eq:bulsecint}
	\mathcal{I}_{Bul,2} = \int_r^\infty \int_{r'}^\infty\frac{e^{-b_n (x/a_b)^{1/4}}}{x^{3/4}\sqrt{x^2-{r'}^2}}r' e^{-r'/\lambda}\odif{x,r'}.
\end{equation}

Since the volume mass density profile of the bulge cannot be calculated with elementary functions, subsequent calculations from our derivations will be carried out numerically. The numerical method is similar to the disk case, which involves creating logarithmically spaced radial arrays and integrating the integrands in Eqs. (\ref{Eq:vbulnewt}-\ref{Eq:bulsecint}) using the quadrature method. Finally, the circular velocity function is constructed by interpolating between integrated points. Detailed steps to calculate the rotational velocity solution are explained at \ref{app:Bulge}.

\subsection{DM halo}
\begin{figure}[t]
	\includegraphics[width=0.5\textwidth]{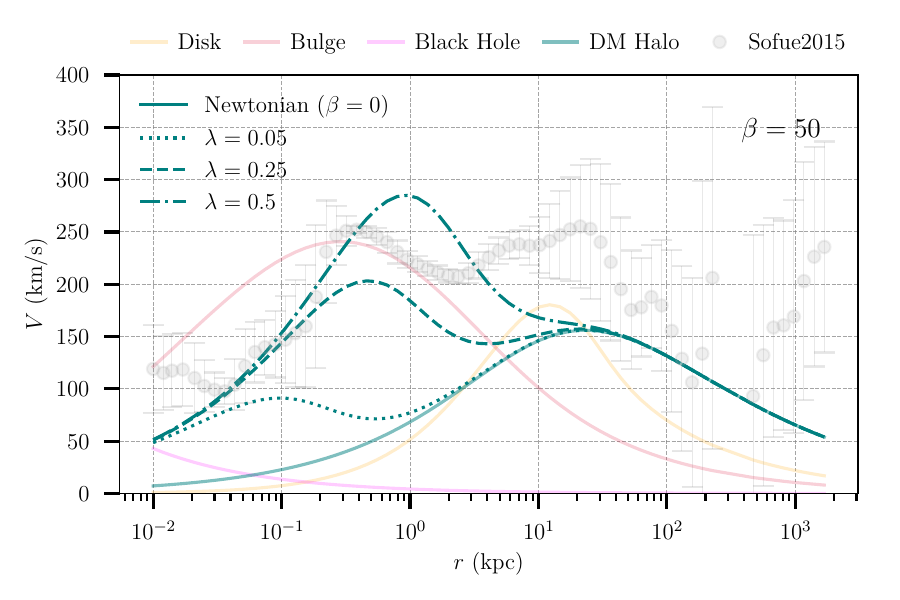}
	\caption{\label{fig:varyuk_DM}Radial velocity distribution in the DM halo affected by Yukawa-correction with variations of $\lambda$ values. The $\beta$ parameter is held constant at $\beta=50$, while the $\lambda$ parameters are varied; $\beta=0$ correspond to the Newtonian case (solid line), $\lambda=0.05$ (dotted), $\lambda = 0.25$ (dashed), and $\lambda = 0.5$ (dash and dot). The opaque lines show velocity components of bulge, disk, and black hole from prior measurements, and the opaque points with vertical bars are measurements from Ref. \cite{Sofue2015}.}
\end{figure}

The rotational velocity estimation for the DM halo follows the derivation procedure from literature for the standard Newtonian potential (i.e., Ref. \cite{Navarro1996}) and the Yukawa-correction potential (i.e., Ref. \cite{Almeida2018}). The dark matter halo is modeled as a spherical symmetric matter distribution with the NFW profile. Similar to the bulge derivation, the Newtonian and Yukawa-correction potentials have different Green's functions. We use the standard derivation of the NFW velocity to obtain the Newtonian velocity,
\begin{equation}
	\label{v DM Newt}
	V_{DM,N}^2 = \frac{4\pi G \rho_s r_s^3}{r} \left[\ln\left(1+\frac{r}{r_s}\right) - \frac{r/r_s}{1+\frac{r}{r_s}}\right].
\end{equation}
and we follow the derivation done by Ref. \cite{Almeida2018} to obtain the Yukawa-correction velocity,
\begin{align}
	V_{DM,Y}^2 &= -\frac{2\pi G \beta\rho_{s}r_{s}^{3}}{r} \left[\frac{2r}{r+r_{s}} \right. \\ 
    & \left. \qquad +e^{-\frac{r+r_{s}}{\lambda}}\left(1+\frac{r}{\lambda}\right) \left\{    \text{Ei}\left(\frac{r_{s}}{\lambda}\right)+ e^{\frac{2r_s}{\lambda}} \text{Ei}\left(-\frac{r_{s}}{\lambda}\right) \right.\right.\notag\\ 
	&\qquad\left. \left.  - \text{Ei}\left(\frac{r+r_{s}}{\lambda}\right) \right\}  + e^{ \frac{r+r_{s}}{\lambda}}\left(\frac{r}{\lambda}-1\right)  \text{Ei}\left(-\frac{r+r_{s}}{\lambda}\right) \right],
\end{align}
where $\text{Ei}(x)$ stands for the exponential integral function. The above formula is then can be calculated numerically by solving $\text{Ei}(x)$ using numerical special function solvers.

\subsection{Black hole}
\begin{figure}[t]
	\includegraphics[width=0.5\textwidth]{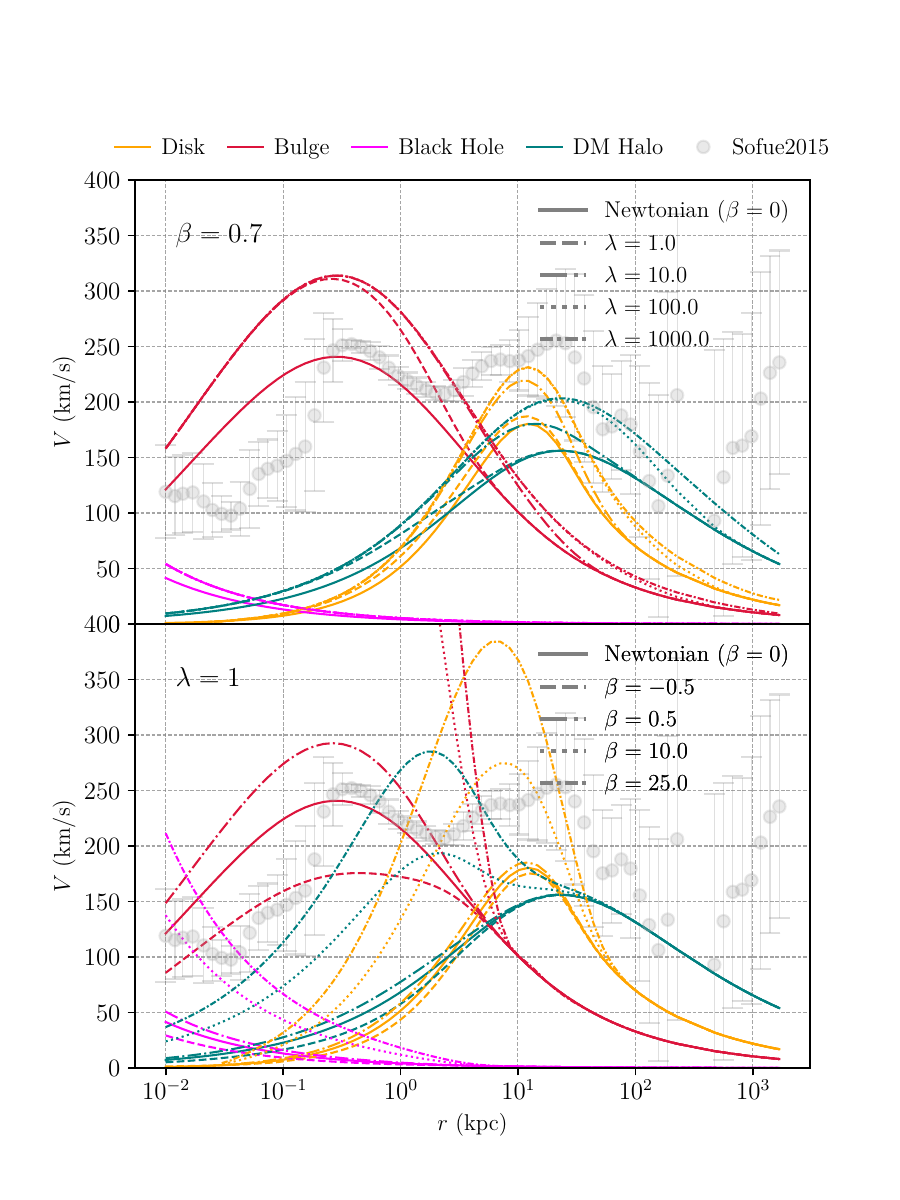}
	\caption{\label{fig:varyuk_all} Solutions of the galactic rotation curve in Yukawa gravity with variations of $\beta$ and $\lambda$ values. \textit{Top:} The $\beta$ parameter is held constant at $\beta=0.7$, while we vary the $\lambda$ parameters for $\lambda=0$ or the Newtonian case (solid line), $\lambda=1$ (double dash and dot), $\lambda = 10$ (dash and double dot), $\lambda = 100$ (dotted), and $\lambda=1000$ (dash and dot). \textit{Bottom:} Same as top graph, but we hold the $\lambda$ parameter constant at $\lambda = 1$ and vary the $\beta$ parameters for $\beta=0,-0.5,0.5,10,25$ with the same order of legends as before. Note that the peak of bulge components in $\beta=10,25$ is outside the range of the plot. In the background, the Milky Way data from \cite{Sofue2015} are shown in opaque scattered points with error bars as a visual comparison.}
\end{figure}

The velocity distribution of the SMBH can be calculated analytically due to the simplicity of its assumption. Since we assume a point-mass distribution of the black hole, the integral falls off as we integrate over spherical coordinates, and we obtain
\begin{align}
	V_{BH,N}^2 = \frac{GM_{BH}}{r},
\end{align}
for the Newtonian velocity and
\begin{equation}
	V_{BH,Y}^2 = \frac{GM_{BH}\beta e^{-\frac{r}{\lambda}}(r+\lambda)}{r \lambda},
\end{equation}
for the Yukawa correction of velocity.

\section{Rotation curve decomposition}
\label{rotcurdecompose}

In our study, as shown beforehand, we decompose the total rotation curve of the disk-like galaxies as a superposition of the galactic components, e.g.,
\begin{equation}
	V_{Tot}^2 = V_{Disk}^2+V_{Bulge}^2+V_{BH}^2 + V_{DM}^2.
\end{equation}
We factorize the contribution of each component in the cases considered in Table \ref{tab:cases} through Bayesian statistics to fit the models into our data to obtain the probability distribution of each parameter from the data set. This way also enables the calculation of the Bayes factor to calculate the statistical preference from one model to another.

\subsection{Posterior and evidence estimation}
In Bayesian statistics, the probability distribution of parameters given available data is determined by utilizing Bayes' theorem to calculate the following probability, i.e.,
\begin{equation}
	\label{bab2:bayestheorem}
	P(\Theta|\mathbf{D},\mathcal{M}_i)=\frac{P(\mathbf{D}|\Theta,\mathcal{M}_i)P(\Theta|\mathcal{M}_i)}{P(\mathbf{D}|\mathcal{M}_i)} \equiv \frac{\mathcal{L}(\Theta)\pi(\Theta)}{\mathcal{Z}}.
\end{equation}
\sloppy Here, we adopt the notation from Ref. \cite{Speagle2020}, where $P(x)$ denotes the probability of a random variable, $\mathcal{M}_i$ denotes a specific model under consideration. At the same time, $\mathbf{D}$ represents sample data. The parameters are represented by $\Theta \equiv \theta_{1}, \theta_2,\dots,\theta_n$. The resulting output of Bayes' theorem is the posterior $P(\Theta|\mathbf{D},\mathcal{M}_i)$, which is the probability distribution of various parameter values given the data. $P(\mathbf{D}|\Theta,\mathcal{M}_i) \equiv \mathcal{L}(\Theta)$ denotes the likelihood, which represents the probability of observing data given specific values for the parameters. Meanwhile, $P(\Theta|\mathcal{M}_i) \equiv \pi(\Theta)$ represents the prior, which describes the prior belief of every parameter value $\theta_n$. The marginal likelihood, $P(\mathbf{D}|\mathcal{M}_i) \equiv \mathcal{Z}$, is commonly referred to as evidence and represents the probability of observing data given a model, which is independent of the parameters. The evidence is defined as
\begin{equation}
	\label{bab2:evidence}
	\mathcal{Z}=\int \mathcal{L}(\Theta)\pi(\Theta)d\Theta,
\end{equation}
where the integral is carried over all possible parameter combinations, this equation also shows that, even through estimation, the prior choice is an important aspect of the evidence calculation. We can also use Bayes' theorem to write the posterior probability of the $k$-th model as
\begin{equation}
	P(\mathcal{M}_k|\mathbf{D}) = \frac{P(\mathbf{D}|\mathcal{M}_k) P(\mathcal{M}_k)}{P(\mathbf{D})}.
\end{equation}
To compare a model $\mathcal{M}_1$ to another model $\mathcal{M}_2$, we can use the Bayes factor (BF) which is the ratio of each model's posterior probabilities,
\begin{align}
\label{bab2:BF12}
BF_{12} &= \frac{P(\mathcal{M}_1|\mathbf{D})}{P(\mathcal{M}_2|\mathbf{D})} \\
&= \frac{P(\mathbf{D}|\mathcal{M}_1)P(\mathcal{M}_1)}{P(\mathbf{D}|\mathcal{M}_2)P(\mathcal{M}_2)} \\
&= \frac{\mathcal{Z}_1}{\mathcal{Z}_2}
\end{align}
where we set $P(\mathcal{M}_1) = P(\mathcal{M}_2)$ since we have no preference for either model. The range of BF value can be interpreted with the BF interpretation table in Ref. \cite{Kass1995}, where the base-10 logarithm is often used in comparisons. This quantity can be calculated through a logarithmic transformation of Eq. (\ref{bab2:BF12}) into
\begin{equation}
	\label{bab2:logBF12}
	\log_{10}(BF_{12}) = \log{\mathcal{Z}_1} - \log{\mathcal{Z}_2}.
\end{equation}
Our approach assumes a Gaussian form for the likelihood probability density, i.e.,
\begin{equation}
	\label{loglikelihood}
	\mathcal{L}(\Theta) = \prod_i \frac{1}{\sqrt{2\pi}\sigma_i}\exp{-\frac{\left(v_{obs}(r_i)-v_{pred}(r_i)\right)^2}{2\sigma_i^2}}.
\end{equation}
Here, $v_{obs}(r_i)$ represents the observed rotational velocity at radius $r_i$, while $v_{pred}(r_i)$ represents the rotational velocity predicted by a model $\mathcal{M}_i$ with the parameters $\Theta$ at $r_i$. The errors in the observed data are represented by $\sigma_i$. 

We used nested sampling \cite{Skilling2006} to estimate both posteriors of each parameter and evidence for each model.  The evidence calculation is reformulated in the nested sampling algorithm by converting the integral in Eqn. (\ref{bab2:evidence}) into
\begin{equation}
	\mathcal{Z} = \int_0^1 \mathcal{L}(\mathcal{X}) \odif{\mathcal{X}}, 
\end{equation}
where $\mathcal{X}$ is a fraction of the prior volume of the enclosed parameter space, the algorithm derives a set of samples called live points from the prior distribution, which are progressively replaced by new points from higher-likelihood areas, with the live point with the least likelihood added to a set of samples known as dead points \cite{Higson2018}. This process is continued until the remaining evidence to be integrated is below a specified tolerance level ($< 1\%$ of the evidence). Here, we use the static nested sampling method implemented through the \texttt{dynesty} Python library, chosen for its effectiveness in calculating the evidence and the posterior distributions simultaneously \cite{Speagle2020}. By default, \texttt{dynesty} calculates the posterior and evidence from generated samples in an 80:20 ratio.

\subsection{Data and priors}
\label{subsect:dataprior}

There are several options to combine the data from different kinds of observation to cover a wide range of scales while at the same time maintaining meaningful and consistent measurement errors. The most common approach is to use a combined data set such as the primary RC source used in this study, i.e., Refs. \cite{Sofue2009,Sofue2015,Sofue2016,Sofue2020}, which combined rotational velocity from multiple sources using running averaged rotational velocity, resulting in the equally-(log)spaced data at a wide range of $r$ (from sub-kpc to several thousands kpc). Here, we also combined information derived from other observations, such as GRAVITY and Gaia data for MW and luminosity profile measurement for M31. In this subsection, we will explain how we select and combine the datasets and prior assumptions to represent cases in this study. 

\textit{Milky Way data.}
The Milky Way RC data is mainly derived from survey observations of stellar and gas movements, measuring their relative velocity to the Solar System \cite{Sofue2016}.  The derivation of the Milky Way RC differs from other galaxies due to our position within the Galaxy. Thus, several methods were used to derive the rotational velocities in the inner and outer regions. For example, where tangential velocity is measurable in the inner regions, the rotational velocity can be determined using HI tangential velocity \cite{Burton1978, Fich1989} and velocity from CO emissions \cite{Clemens1985}. However, no tangential point exists along the line of sight in the outer regions \cite{Sofue2009}, thus requiring another approach, such as analyzing the geometry of the HI disk \cite{Honma1997, Petrovskaia1986,Merrifield1992} or from known distances to galactic objects, for example, the HII region \cite{Brand1993}. The astrometric information of Galactic stars can accurately measure RC around the solar neighborhood, such as using recent Gaia DR3 data \cite{GaiaCollaboration2023}. The above approaches enable precise velocity measurement in the wide range of $r$ around the Solar System. However, the current observation of the Milky Way may lack robust measurements of global features, i.e., there are no direct observations to probe, for example, the S\'ersic profile accurately. 

We analyze the MW data using a combination of several sources, mostly from Refs. \cite{Sofue2015,Sofue2016,Sofue2020} which ranges from $1 \text{ pc} - 1 \text{ Mpc}$ to the Galactic center. The Sofue 2015 and Sofue 2017 RC feature relatively large error bars in the inner and regions of the galaxies, mostly at $\sigma_V \sim 10-30$ km/s in the inner regions ($r < 0.2$ kpc) and outwards to $\sigma_V > 100 \sim$ km/s in the outer regions ($r>100$ kpc), while the middle regions ($1<r<20$ kpc) features relatively more precise data with $\sigma_V < 10$ km/s. In contrast, the 2020 RC features more precise data with $\sigma_V \lesssim 20$ km/s. 

Additionally, we include the precise RC data around the Solar neighborhood derived from Gaia measurements. The RC data are derived from the latest Gaia DR3 dataset from Ref. \cite{Wang2022} and Ref. \cite{Zhou2023} in the $5 - 30$ kpc region. The RC derived with this method features a minimal uncertainty compared to other methods, with errors of about $\sigma_V \lesssim 5$ km/s. However, in this region, the RC is in tension with the \cite{Sofue2020} result, making the averaged rotational velocity less precise than the above error.

To probe the RC at a short distance, that is, at $r= 0.38$ AU from the Galactic center, we derive the equivalent rotational circular velocity quantity from GRAVITY data \cite{Abuter2023} to cover the area near the SMBH. This data will give stringent constraints to the SMBH component, which now depends on the Yukawa gravity parameters ($\beta,\lambda$). Details about data derivation are available in the  \ref{constrainingMBH}, from the data provided in Table 2 in \cite{Abuter2023}. 

The prior distributions chosen for all parameters in the Milky Way are uniform, as we would like to probe a wide range of possibilities. Several parameters, $\lambda$, $\rho_s$, $r_s$, and $M_{BH}$, are transformed into a logarithmic scale to improve the localization of the best-fit values. Table \ref{table:priorsCombined} summarizes the parameters and their corresponding prior ranges. 

\begin{table}[tb!]
	\centering
	\caption{Prior choices for parameters were used to model the Milky Way (MW) and Andromeda (M31) galaxies and their respective units. For the Milky Way, all priors are uniform distributions $\mathcal{U}$(lower, upper). For Andromeda, parameters $a_b$, $a_d$, and $n$ use normal distributions $\mathcal{N}(\mu, \sigma)$, while others are uniform. Units marked with (--) are dimensionless. The SR stands for "short range" case, while LR stands for "long range".}
	\label{table:priorsCombined}
	\begin{tabular}{c c c c}
		\hline\hline
		\textbf{Parameter} & \textbf{Units} & \textbf{MW Prior} & \textbf{M31 Prior} \\ [0.5ex]
		\hline
		$\beta$ & -- & $\mathcal{U}(-150, 150)$ & $\mathcal{U}(-150, 150)$ \\
		\multirow{2}{*}{$\ln \lambda$} & $\ln[\mathrm{kpc}]$ & SR: $\mathcal{U}(-4, 11)$  & $\mathcal{U}(-4, 10)$  \\
		& $\ln[\mathrm{kpc}]$ & LR: $\mathcal{U}(2, 11)$  & $\mathcal{U}(2, 10)$  \\
		$M_b$ & $M_\odot$ & $\mathcal{U}(10^8, 10^{11})$ & $\mathcal{U}(10^8, 10^{11})$ \\
		$a_b$ & kpc & $\mathcal{U}(0.01, 10)$ & $\mathcal{N}(1, 0.2)$ \\
		$M_d$ & $M_\odot$ & $\mathcal{U}(10^{10}, 3{\times}10^{13})$ & $\mathcal{U}(10^7, 3{\times}10^{13})$ \\
		$a_d$ & kpc & $\mathcal{U}(1.5, 20)$ & $\mathcal{N}(5.3, 0.5)$ \\
		$n$ & -- & 4 & $\mathcal{N}(2.2, 0.3)$ \\
		$\ln \rho_s$ & $\ln[M_\odot\,\mathrm{kpc}^{-3}]$ & $\mathcal{U}(-5, 20)$ & $\mathcal{U}(0.01, 20)$ \\
		$r_s$ & kpc & $\mathcal{U}(0.1, 385)$ & $\mathcal{U}(0.1, 385)$ \\
		$\ln M_{\mathrm{BH}}$ & $\ln[M_\odot]$ & $\mathcal{U}(0.01, 20)$ & $\mathcal{U}(0.01, 40)$ \\ [1ex]
		\hline\hline
	\end{tabular}
\end{table}

\begin{figure*}[!htbp]
	\captionsetup[subfloat]{farskip=0pt,captionskip=0.5pt}
	\subfloat[Newtonian model]{%
		\includegraphics[width=0.33\textwidth]{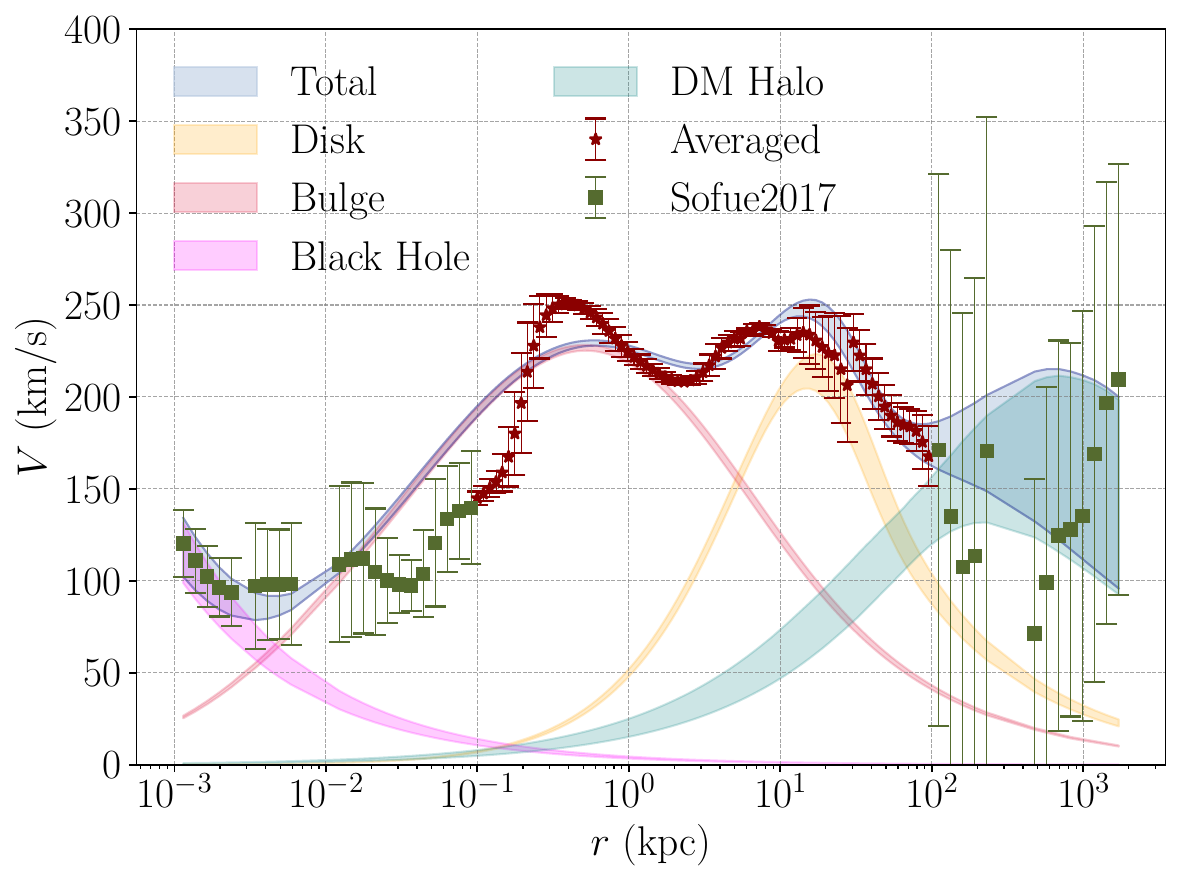}}%
	\subfloat[NTDMC (short-range $\lambda$)]{%
		\includegraphics[width=0.29\textwidth]{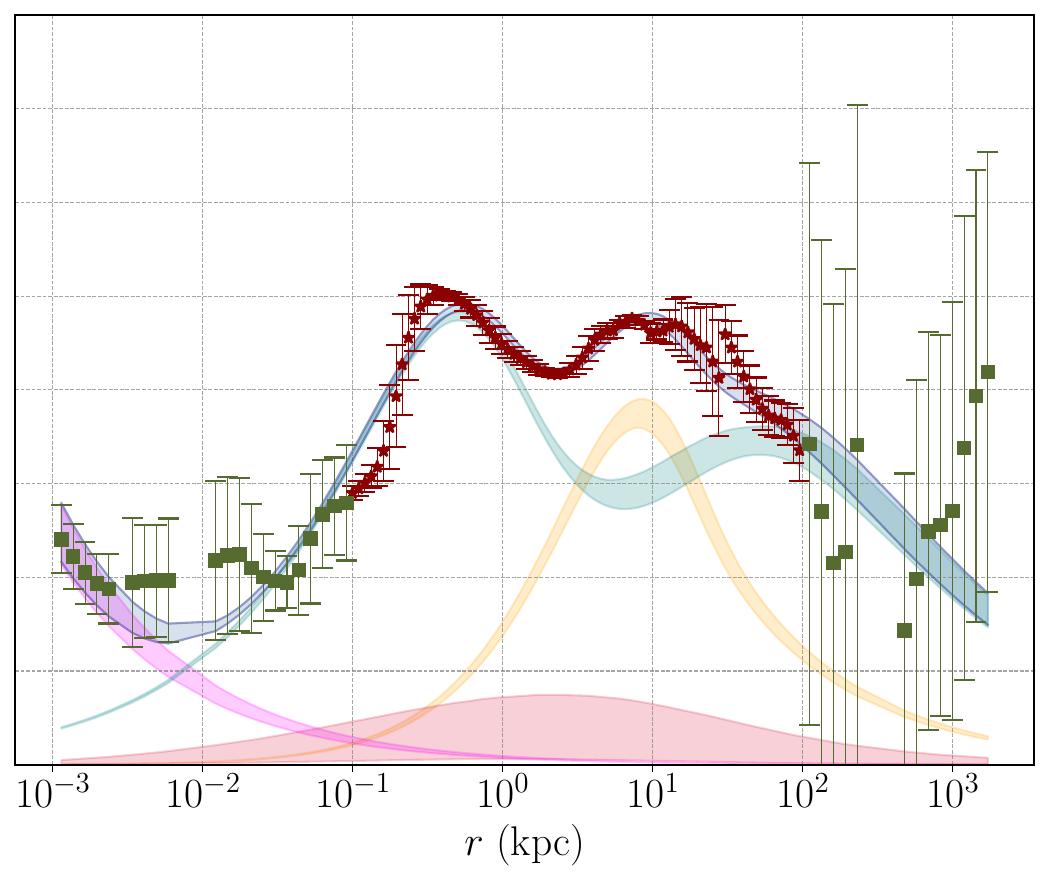}}%
	\subfloat[NTDMC (long-range $\lambda$)]{%
		\includegraphics[width=0.29\textwidth]{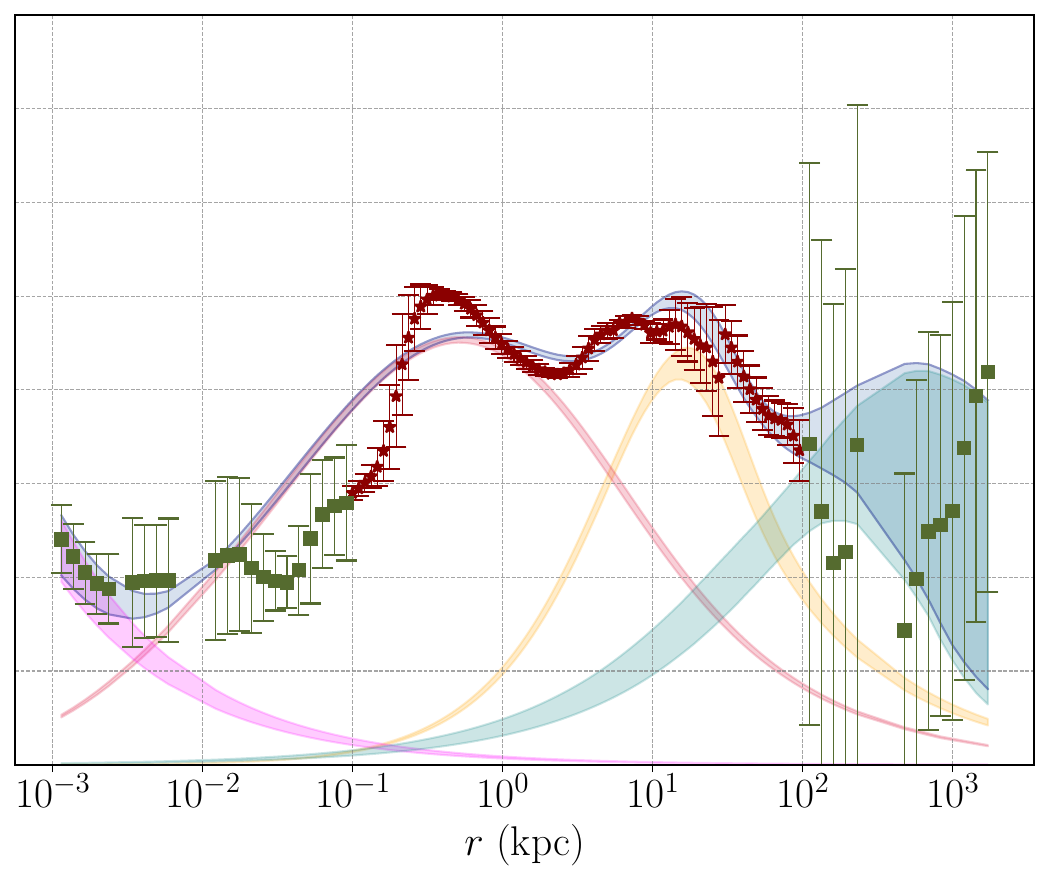}}\\
	\subfloat[MG model (short-range $\lambda$)]{%
		\includegraphics[width=0.324\textwidth]{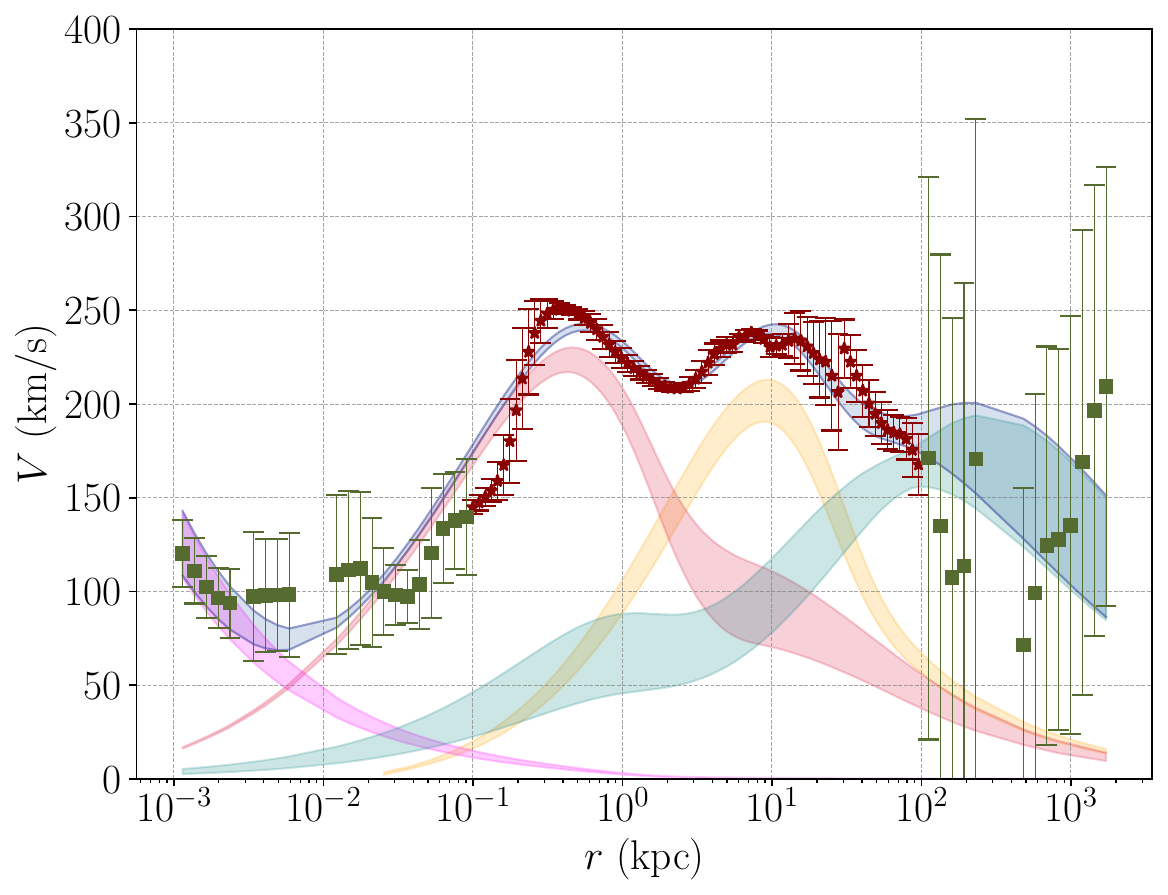}}%
	\subfloat[MG model (long-range $\lambda$)]{%
		\includegraphics[width=0.29\textwidth]{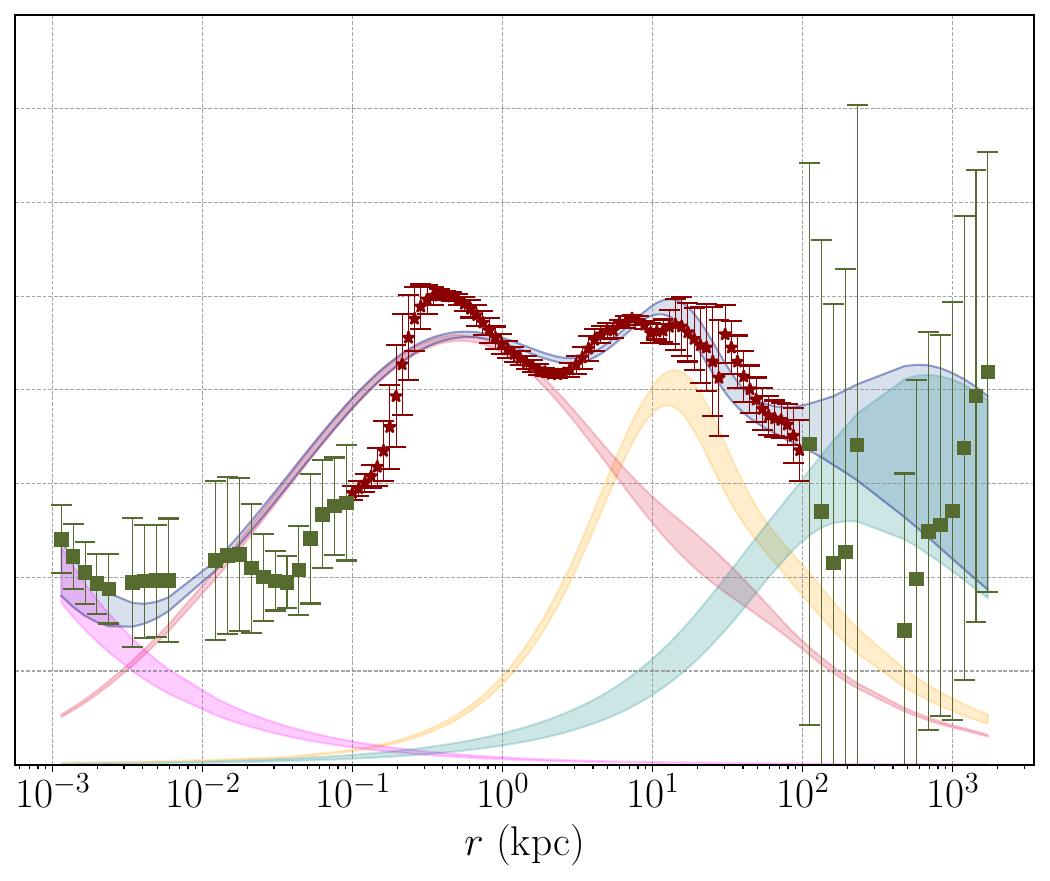}}%
	\subfloat[Y-No-DM model (short-range $\lambda$)]{%
		\includegraphics[width=0.29\textwidth]{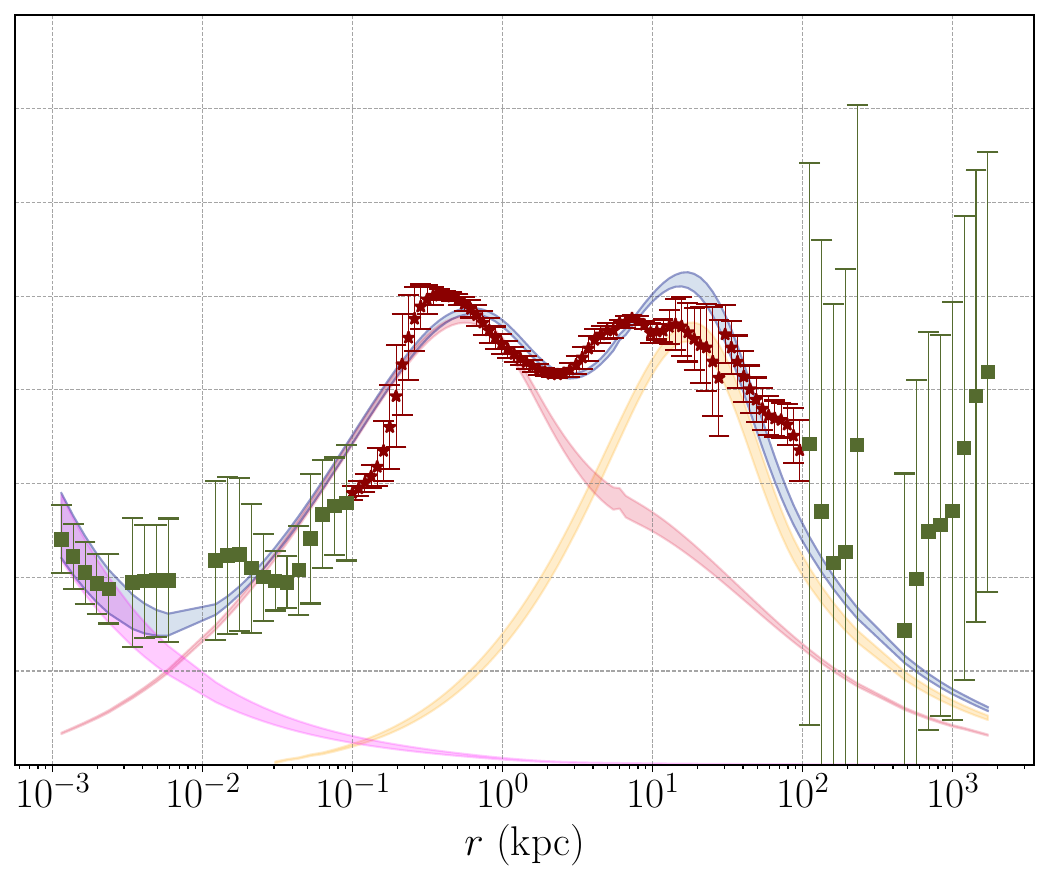}}\\%
	\subfloat[Y-No-DM model (long-range $\lambda$)]{%
		\includegraphics[width=0.33\textwidth]{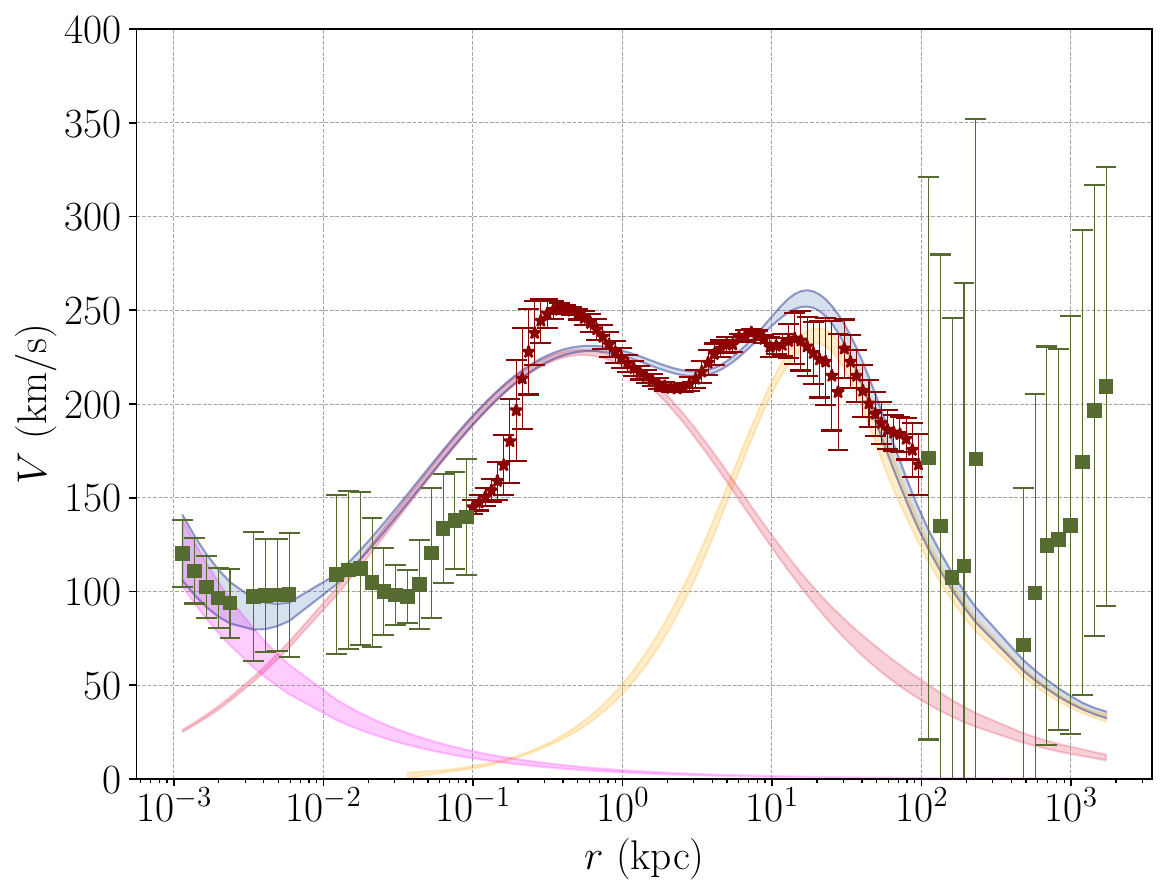}}%
	\caption{Rotation curves for the MW using the Sofue (2017), the averaged Sofue (2020), and the Gaia dataset. The GRAVITY data has
		been omitted from the graph to conserve space on the axis. The shaded regions show the 95\% CIs. (a) Newtonian model, (b) NTDMC model in the short-range case, (c) NTDMC model in the long-range case, (d) MG model in the short-range case, (e) MG model in the long-range case, along with Yukawa-No-DM models in the (f) short-range and (g) long-range case.}
	\label{fig:mw_ppc}
\end{figure*}

\begin{figure*}[!htbp]
	\captionsetup[subfloat]{farskip=0pt,captionskip=0.5pt}
	\subfloat[Newtonian model]{%
		\includegraphics[width=0.33\textwidth]{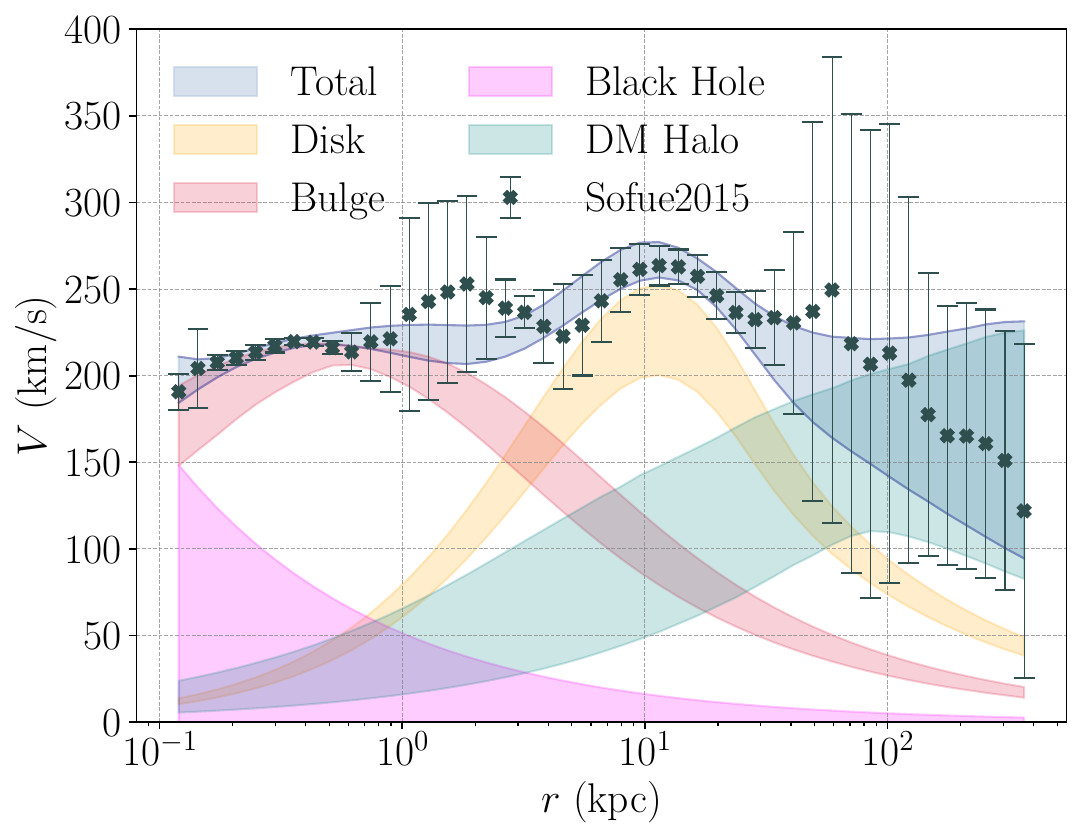}}%
	\subfloat[NTDMC (short-range $\lambda$)]{%
		\includegraphics[width=0.3\textwidth]{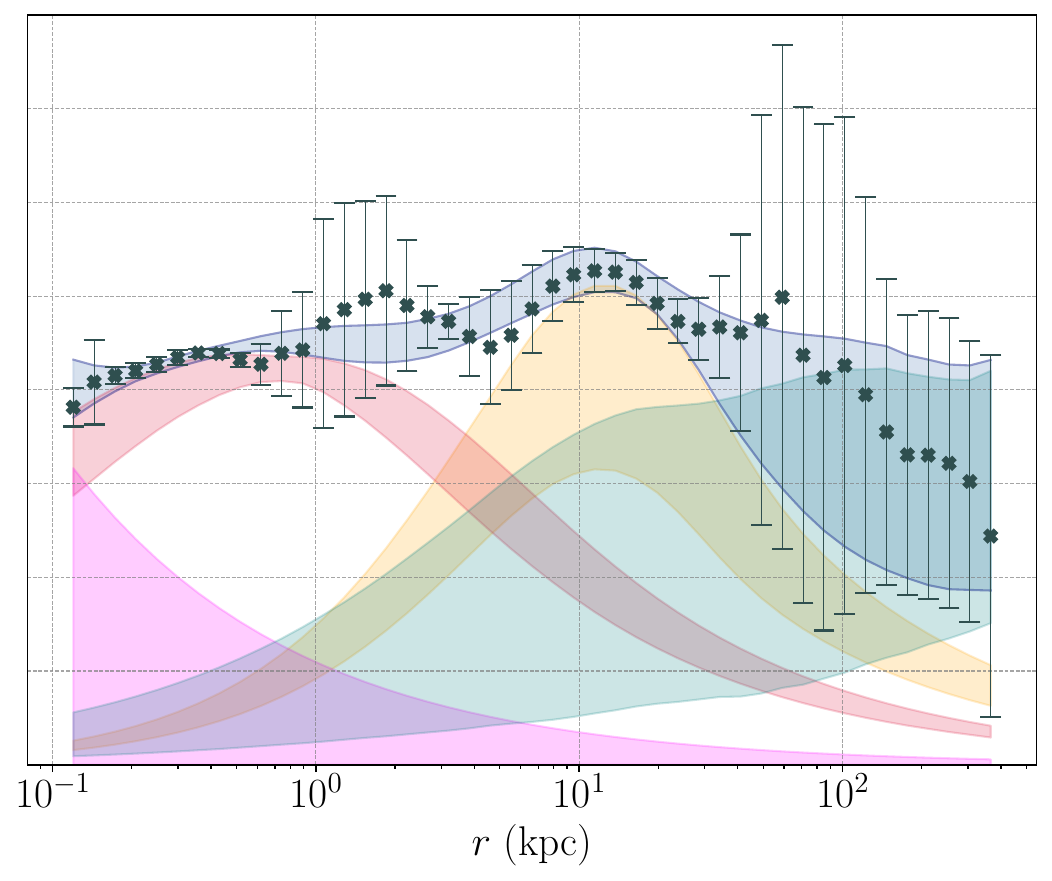}}%
	\subfloat[NTDMC (long-range $\lambda$)]{%
		\includegraphics[width=0.3\textwidth]{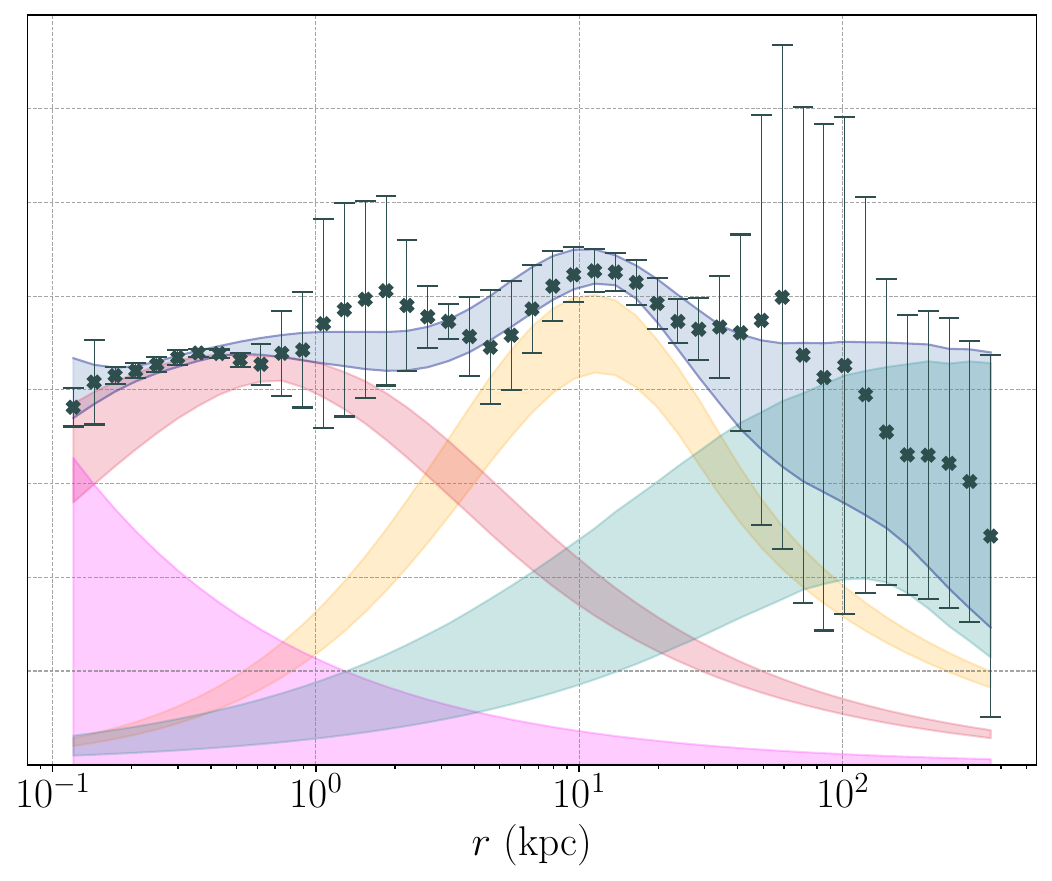}}\\
	\subfloat[MG model (short-range $\lambda$)]{%
		\includegraphics[width=0.33\textwidth]{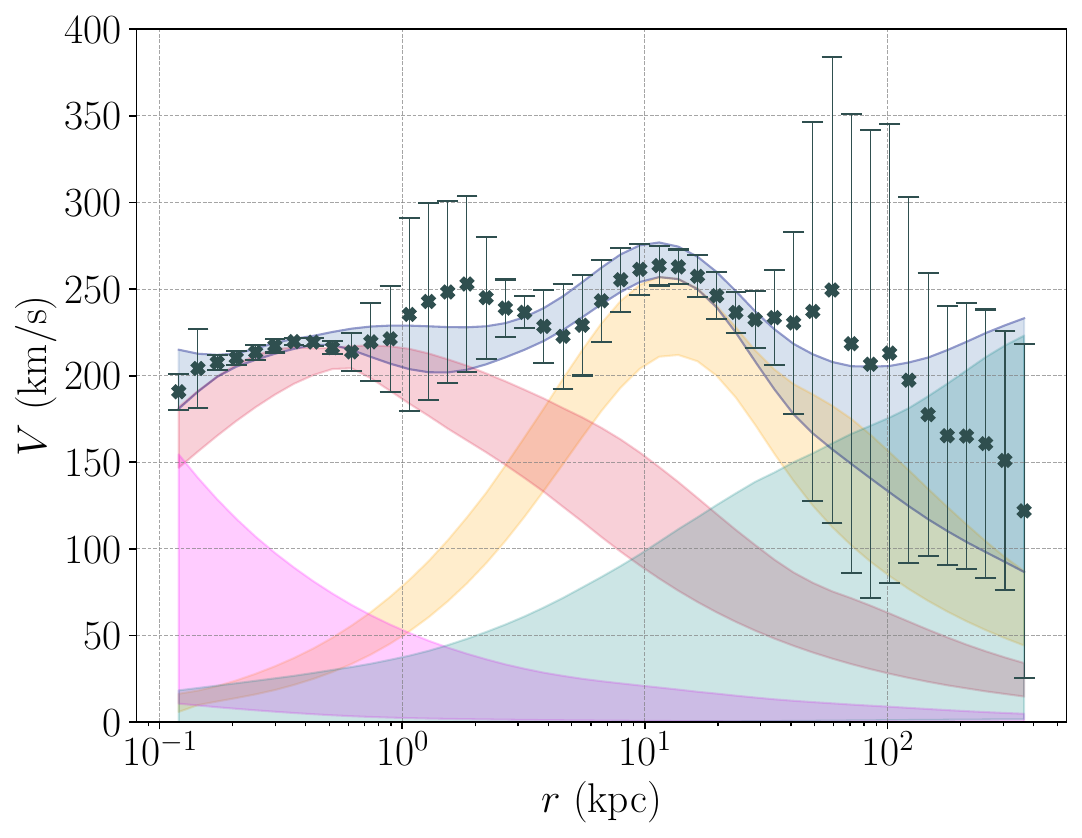}}%
	\subfloat[MG model (long-range $\lambda$)]{%
		\includegraphics[width=0.3\textwidth]{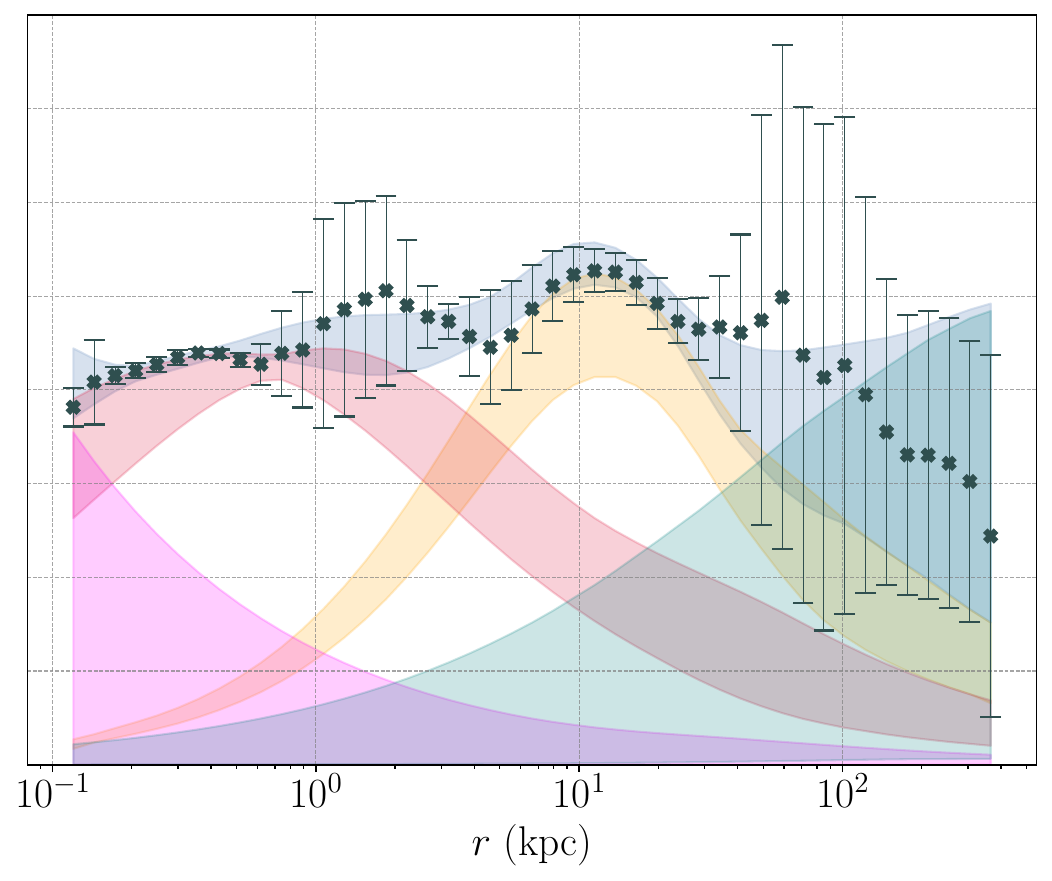}}%
	\subfloat[No-DM model (short-range $\lambda$)]{%
		\includegraphics[width=0.3\textwidth]{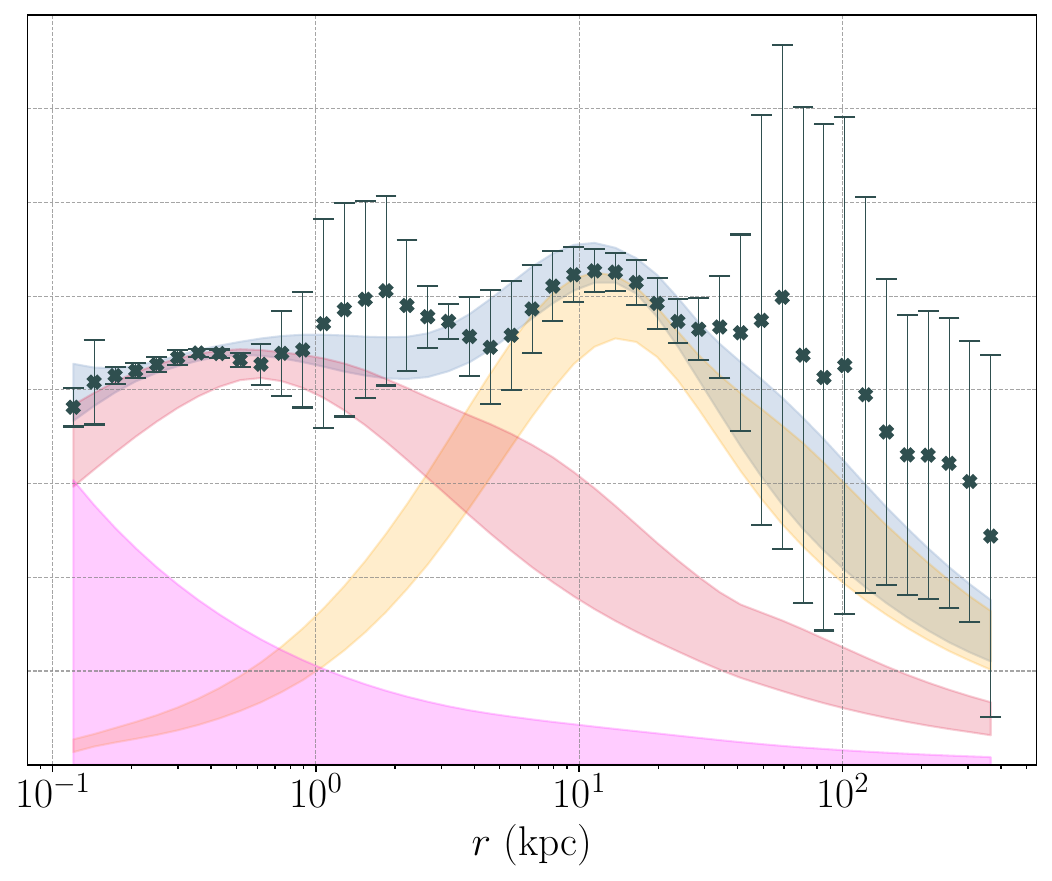}}\\%
	\subfloat[No-DM model (long-range $\lambda$)]{%
		\includegraphics[width=0.33\textwidth]{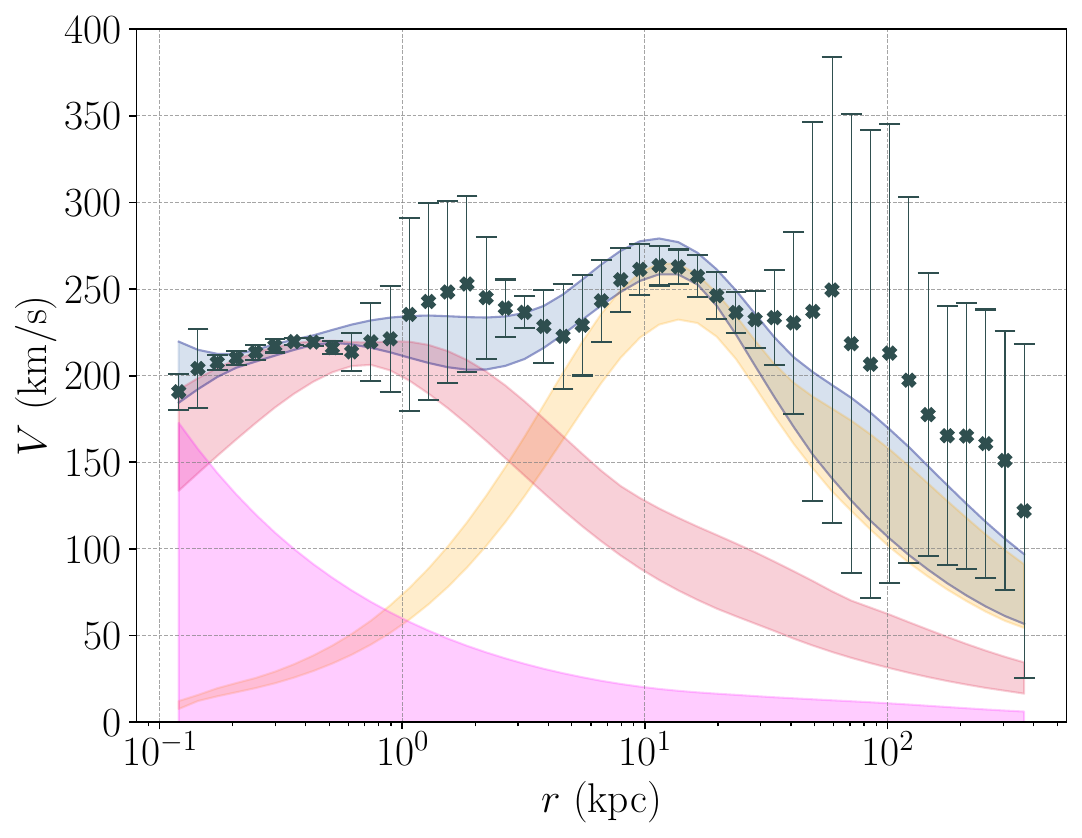}}%
	\caption{Rotation curves for the M31 using the Sofue (2015) dataset. The shaded regions show the 95\% CIs. (a) Newtonian model, (b) NTDMC model in the short-range case, (c) NTDMC model in the long-range case, (d) MG model in the short-range case, (e) MG model in the long-range case, along with No-DM models in the (f) short-range and (g) long-range case.}
	\label{fig:m31_ppc}
\end{figure*}

\begin{figure}[t]
	\includegraphics[width=0.5\textwidth]{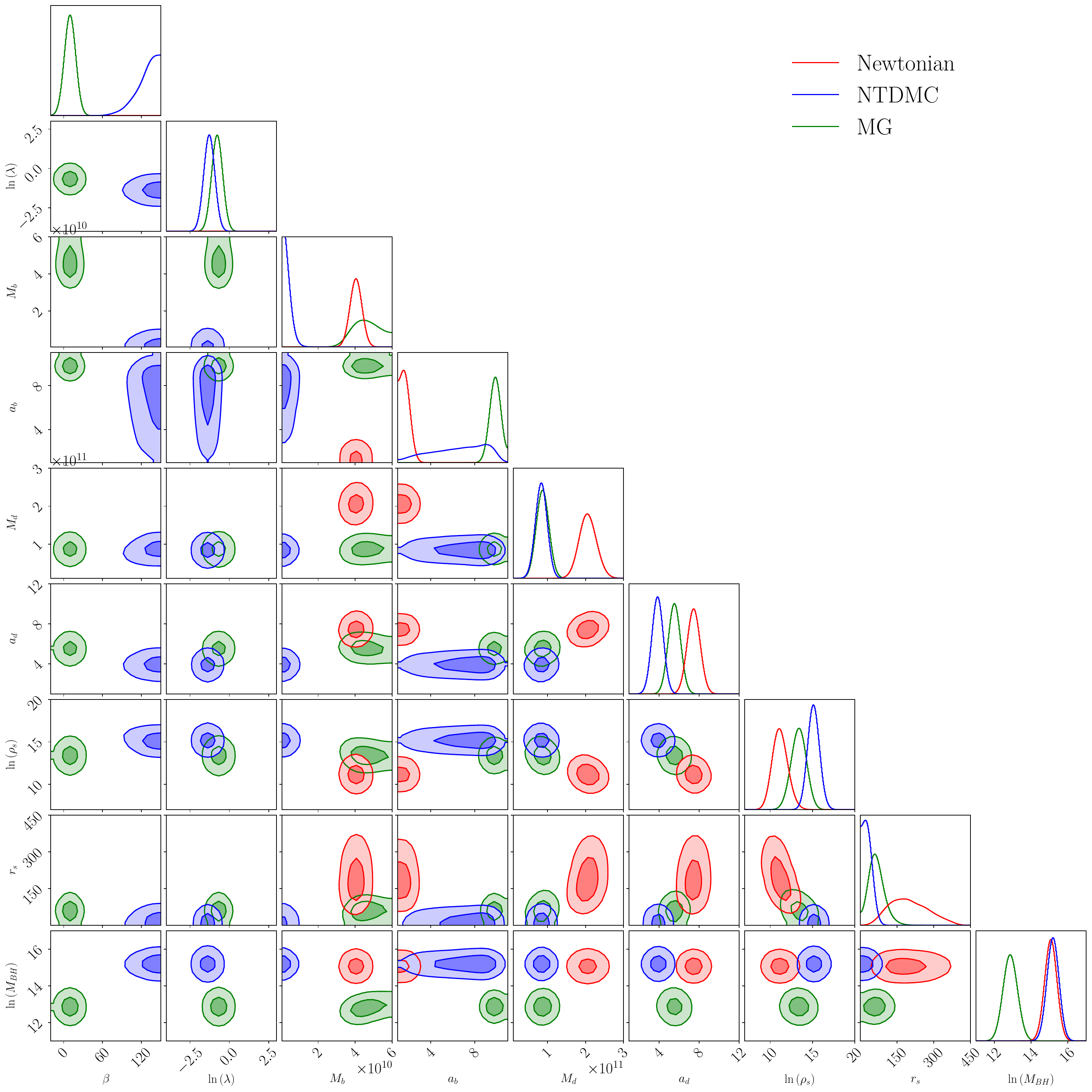}
	\caption{\label{fig:mw_shortrange} Marginal posterior distributions for the Newtonian (red), NTDMC (blue), and MG (green) models for the MW using the averaged data between Sofue (2017), Sofue (2020), and the Gaia dataset. The short-range case is investigated with a prior distribution implemented on the length scale parameter with $\ln{\lambda} \in [-4,10]$.
		The darker regions and their surrounding lighter areas correspond to the 1$\sigma$ and 2$\sigma$ confidence levels
		regions, respectively.}
\end{figure}

\begin{figure}[t]
	\includegraphics[width=0.5\textwidth]{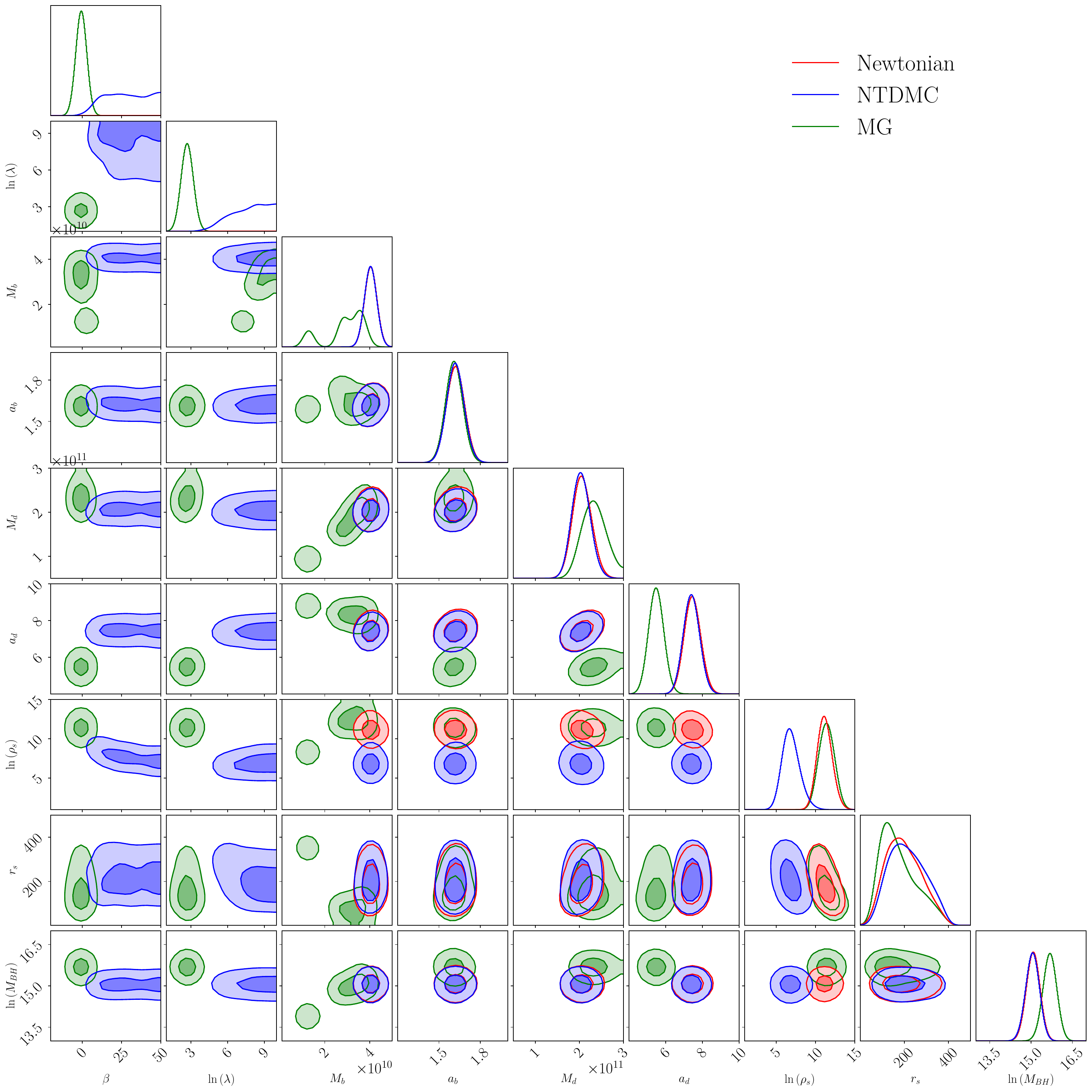}
	\caption{\label{fig:mw_longrange} Same as Fig. \ref{fig:mw_longrange}, but for the long-range case with $\lambda = [2,10]$ as the priors.}
\end{figure}

\begin{figure}[t]
	\includegraphics[width=0.5\textwidth]{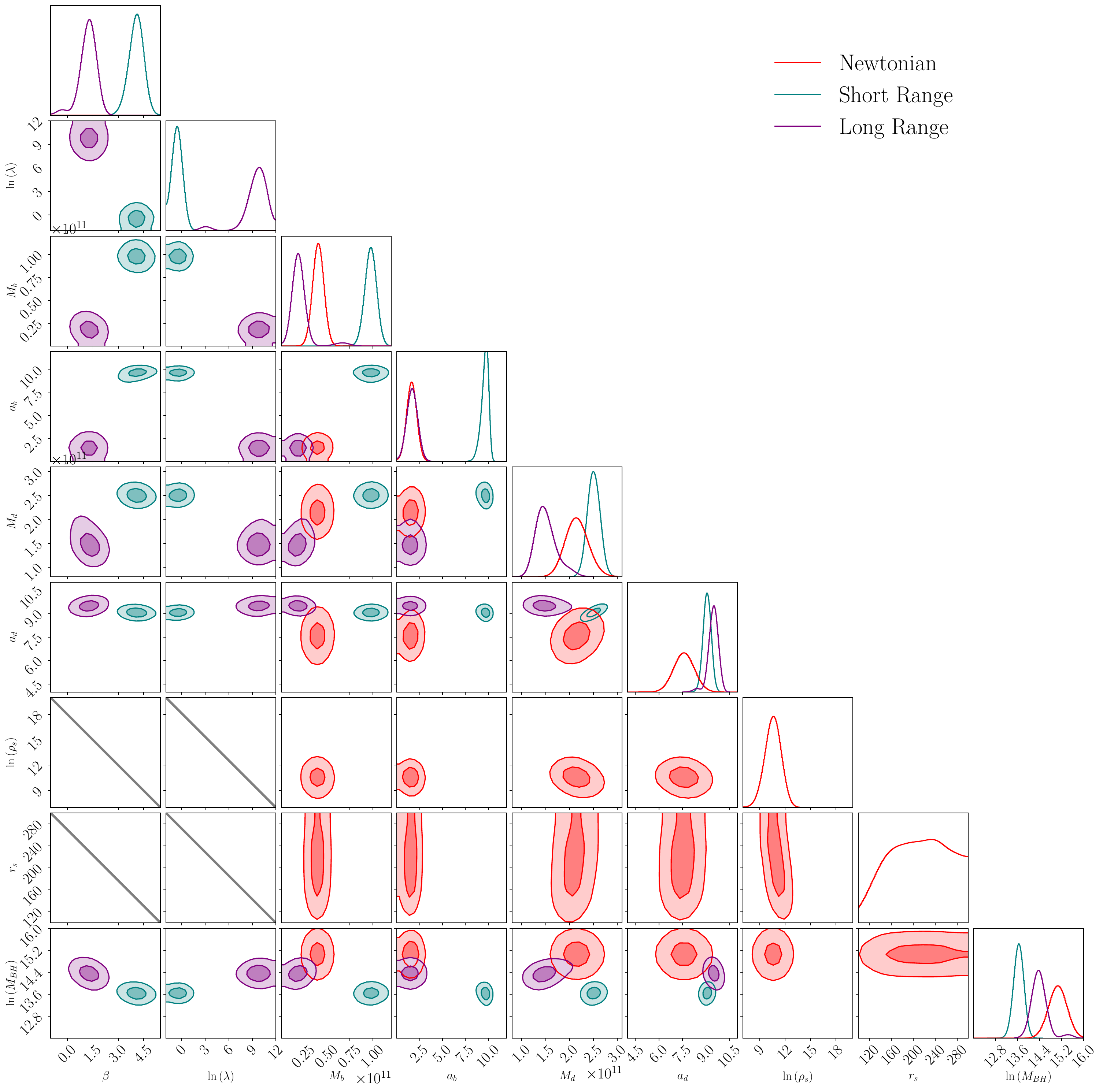}
	\caption{\label{fig:mw_nodm} Same as Fig. \ref{fig:mw_longrange}, but for the Newtonian with dark matter (red) and No DM model for both short-range (teal; $\ln{\lambda} \in [-4,10]$) and long-range (purple; $\ln{\lambda} \in [2,10]$) cases.}
\end{figure}

\begin{figure}[t]
	\includegraphics[width=0.5\textwidth]{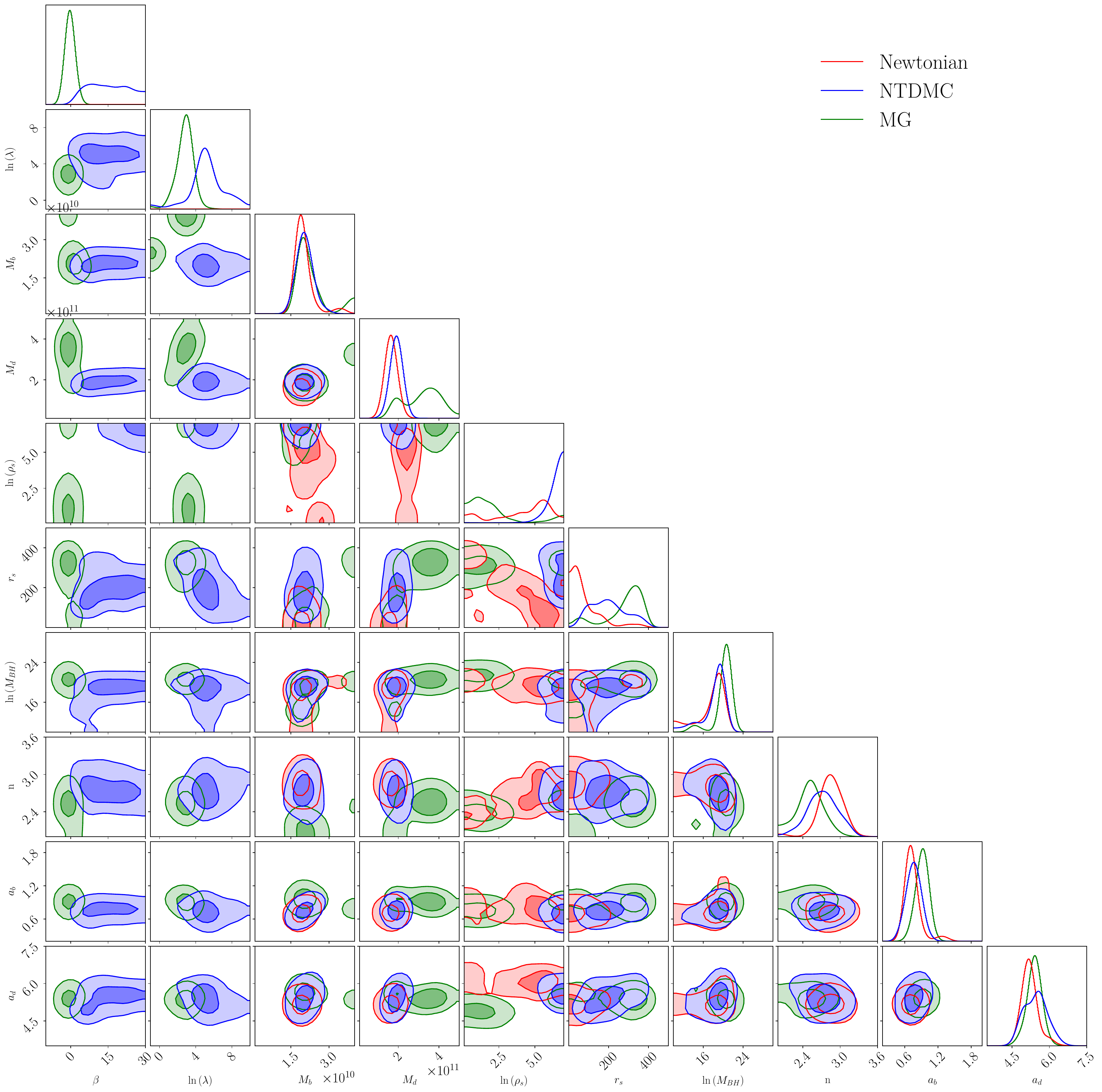}
	\caption{\label{fig:m31_shortrange} Marginal posterior distributions for the Newtonian (red), NTDMC (blue) and MG (green) model for the M31 using the Sofue (2015) dataset. The short-range case is investigated with a prior distribution implemented on the length scale parameter with $\ln{\lambda} \in [-4,10]$.
		The darker regions and its surrounding lighter areas correspond to the 1$\sigma$ and 2$\sigma$ confidence
		regions, respectively.}
\end{figure}

\begin{figure}[t]
	\includegraphics[width=0.5\textwidth]{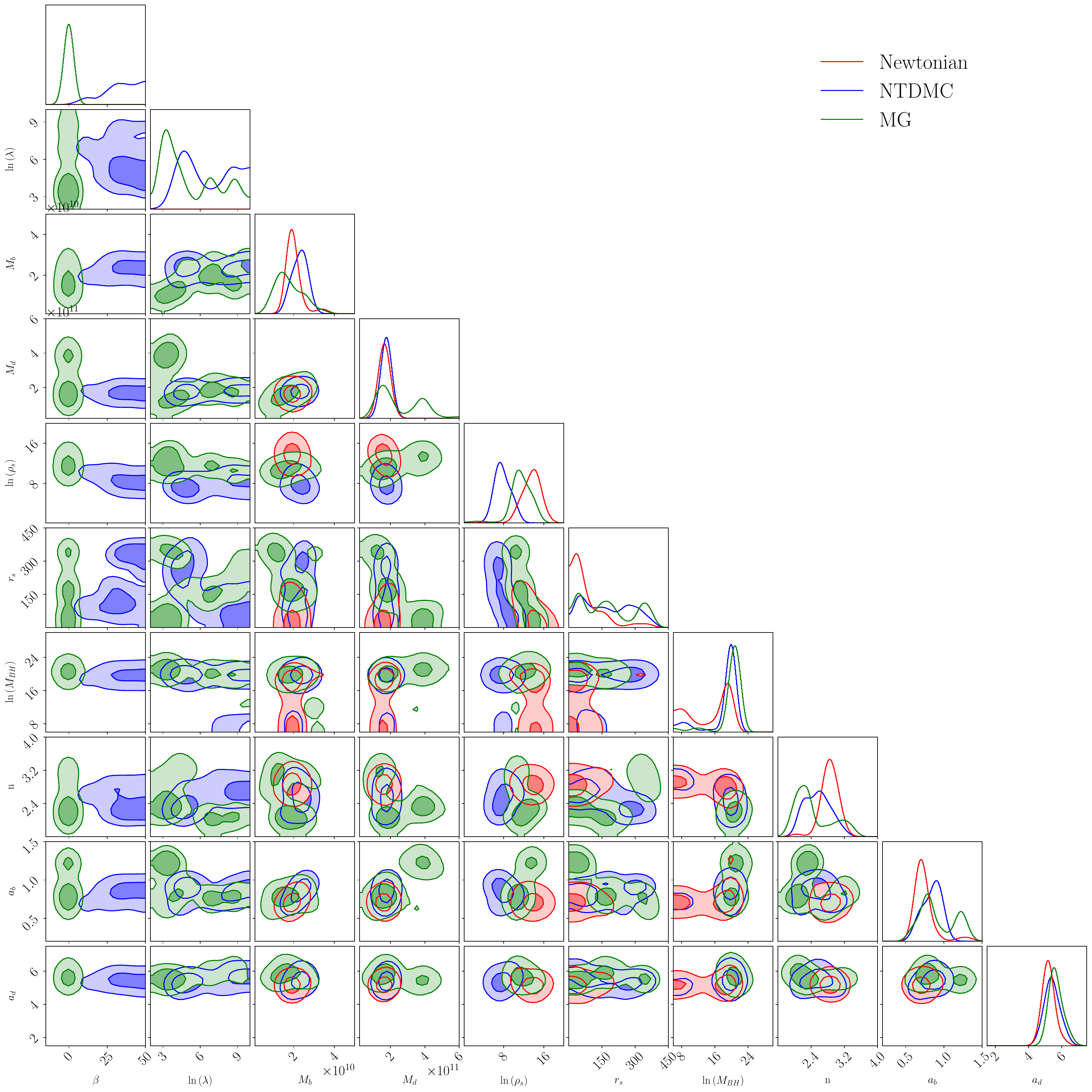}
	\caption{\label{fig:m31_longrange} Same as Fig. \ref{fig:m31_shortrange}, but for the long-range case with $\lambda = [2,10]$ as the priors.}
\end{figure}

\begin{figure}[t]
	\includegraphics[width=0.5\textwidth]{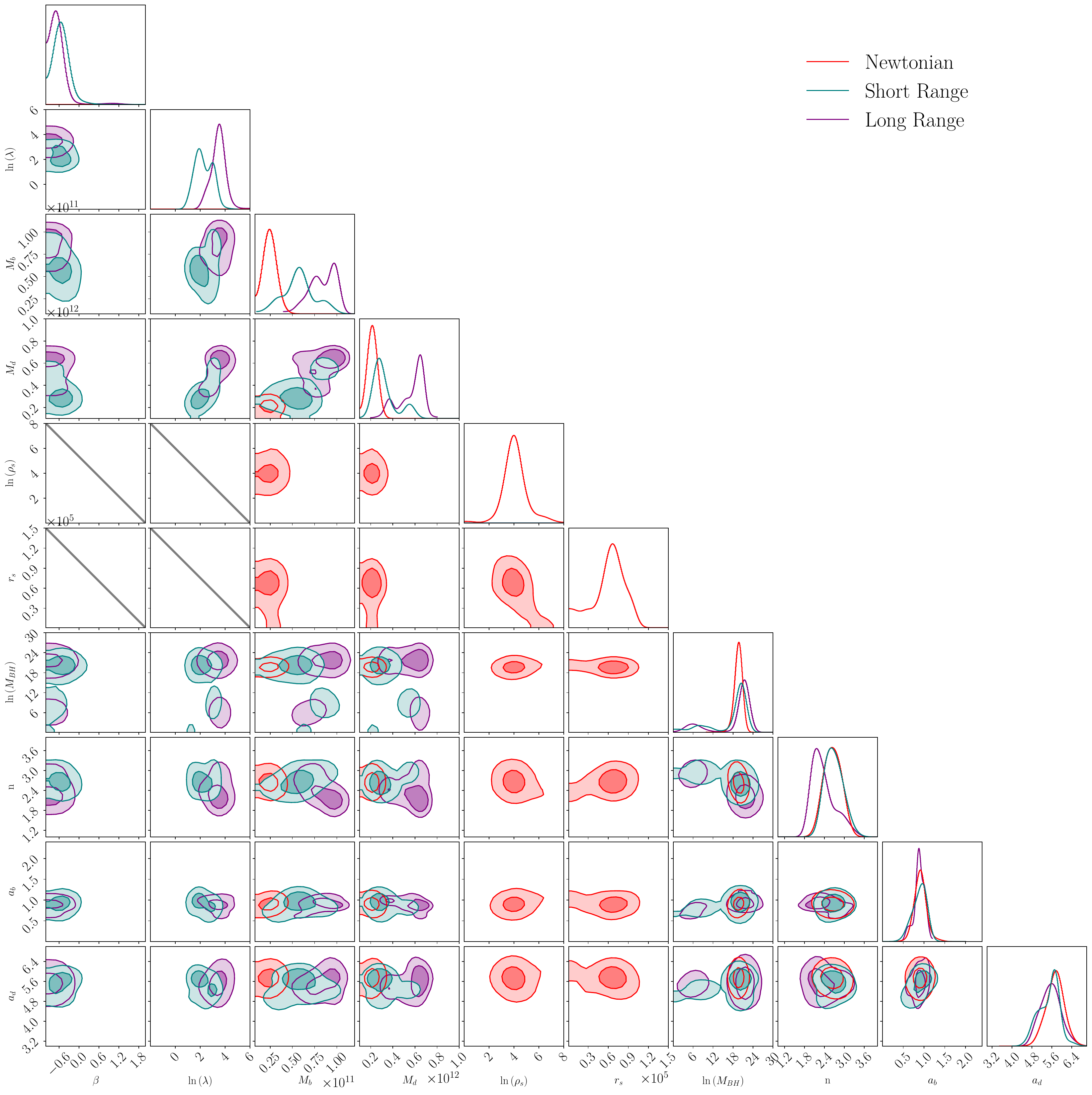}
	\caption{\label{fig:m31_nodm} Same as Fig. \ref{fig:m31_shortrange}, but for the Newtonian with dark matter (red) and No DM model for both short-range (teal; $\ln{\lambda} \in [-4,10]$) and long-range (purple; $\ln{\lambda} \in [2,10]$) cases.	
	}
\end{figure}

\textit{M31 data.} Unlike the MW, the RC measurement of M31 is more straightforward since it only needs a single direction to observe the kinematics of objects around the galactic plane. Furthermore, photometry analysis of M31 enables the measurements of global features such as luminosity structure profile, which is more challenging to determine in the MW. 

We analyze the M31 rotation curve from Ref. \cite{Sofue2015}, which is a combination of observations from several sources, such as HI, CO, and H$\alpha$, including observations of globular clusters and galactic satellites to probe the RC in the halo region. This data ranges from $0.1 < r < 2000$ kpc, with a slightly significant error of $\sigma_V \sim 10-50$ km/s in the midregions ($1<r<40$ kpc), and relatively high error bars of $\sigma_V \gtrsim 100$ km/s at the outer regions $r>40$ kpc compared to the MW. Hence, we truncate the rotation curve up to $r = 400$ kpc due to the large error bars at $r>400$ kpc. Additionally, the photometry observations of M31 galactic structure (e.g., Ref. \cite{Courteau2011}) enable direct measurements of the luminosity profile and give precise prior information on bulge and disk parameters.  The measured global structure information, such as S\'ersic parameter $n$ and scale lengths of the bulge ($R_b$) and disk ($R_d$), enters through the prior in our inference. For that reason, we assume Gaussian priors of $n, R_b$ and $R_d$ based on the observational values from Ref. \cite{Courteau2011} as shown in Table \ref{table:priorsCombined}. In this case, unlike in MW, the S\'ersic parameter $n$ is treated as an additional free parameter.

\subsection{Model comparison}

In order to compare models that resemble the observed dynamics of a galaxy, the previously computed Bayesian evidences are used to calculate the Bayes factor to determine the statistical preference of multiple competing models. We compute the Bayes factors for every pair of models considered here. The models being compared are (the parameters for each component are omitted in this notation): 
\begin{enumerate}
	\item Newtonian gravity with 7 free parameters (MW) and 8 free parameters (M31). This model represents the standard description of gravity,
	\begin{equation}
		V_{Tot,N}^2(r) = V_{Disk,N}^2(r) + V_{Bul,N}^2(r) +  V_{BH,N}^2(r)  +  V_{DM,N}^2(r).
	\end{equation}
	\item Newtonian gravity with Yukawa correction on components without dark matter model (Y-noDM). This model represents the implementation of the Yukawa model as an alternative description to DM, 
	\begin{equation}
		V_{Tot,Y-noDM}^2(r) = V_{Disk,N+Y}^2(r) + V_{Bul,N+Y}^2(r) +  V_{BH,N+Y}^2(r).
	\end{equation}
	\item Newtonian gravity with Yukawa correction in the DM halo as in Ref. \cite{Almeida2018} with 9 free parameters (MW) and 10 free parameters (M31). This model represents the assumption of the DM model with alternative coupling behavior,
	\begin{align}
		V_{Tot,NTDMC}^2(r) &= V_{Disk,N}^2(r) + V_{Bul,N}^2(r) +  V_{BH,N}^2(r) \notag \\
		&\quad +  V_{DM,N}^2(r) + V_{DM,Y}^2(\beta,\lambda,r).
	\end{align}
	\item Newtonian gravity with Yukawa correction in both baryonic components and the DM halo with 9 free parameters (MW) and 10 free parameters (M31). This model represents the gravity with Yukawa interaction on all components,
	\begin{align}
		V_{Tot,MG}^2(r) &= V_{Disk,N+Y}^2(\beta,\lambda,r) + V_{Bul,N+Y}^2(\beta,\lambda,r)  \notag \\ 
		&\quad +  V_{BH,N+Y}^2(\beta,\lambda,r) +  V_{DM,N+Y}^2(\beta,\lambda,r).
	\end{align}
\end{enumerate}
We refer to each as the Newtonian, Y-noDM, NTDMC, and modified gravity (MG) case, respectively.

\begin{table*}
	\label{tab:betalamposterior}
	\centering
	\caption{Summary of $\beta$ and $\lambda$ posteriors for the Milky Way (MW) and M31 galaxies. Values are presented with 95\% confidence intervals. Parameters unconstrained by data (i.e., constrained by priors) are indicated with inequalities ($\gtrsim$ or $\lesssim$), where in that case, is determined by the lower/upper percentile of the posterior, which corresponds to the opposite bound imposed by the prior. The SR stands for "short range" case, while LR stands for "long range".}
	\begin{tabular}{l l l l l c c}
		\hline\hline
		\textbf{Galaxy} & \textbf{Model} & \textbf{$\beta$} & \textbf{$\lambda$ [kpc]} \\ 
		\hline
		\multirow{6}{*}{MW} 
		& NTDMC (SR) & $\gtrsim 87$ & $0.28^{+0.01}_{-0.01}$ \\
		& NTDMC (LR) & $84^{+62}_{-74}$ & $\gtrsim 140$ \\
		& MG (SR) & $9.9^{+3.7}_{-3.8}$ & $0.46^{+0.03}_{-0.02}$ \\
		& MG (LR) & $-0.58^{+0.05}_{-0.03}$ & $14.7^{+9.5}_{-4.2}$ \\
		& No DM (SR) & $4.1^{+0.5}_{-0.8}$ & $0.55^{+0.01}_{-0.04}$ \\
		& No DM (LR) & $1.3^{+0.5}_{-1.6}$ & $\gtrsim 23$ \\
		\hline
		\multirow{6}{*}{M31} 
		& NTDMC (SR) & $34.5^{+100}_{-31.2}$ & $155^{+423}_{-153}$ \\
		& NTDMC (LR) & $\gtrsim 24$ & $\gtrsim 35$ \\
		& MG (SR) & $-0.54^{+2.6}_{-0.25}$ & $16.6^{+46.2}_{-16.4}$ \\
		& MG (LR) & $-0.02^{+1.2}_{-0.7}$ & $\gtrsim 13$ \\
		& No DM (SR) & $-0.53^{+3.3}_{-0.23}$ & $8.8^{+794}_{-6.6}$ \\
		& No DM (LR) & $-0.69^{+0.7}_{-0.06}$ & $33.5^{+539}_{-22}$ \\
		\hline\hline
	\end{tabular}
\end{table*}

\begin{table*}
	\caption{Logarithmic evidence ($\log_{10} Z$) of the Newtonian model, and Bayes factors for the three alternative models: a) NTDMC, b) Modified Gravity (MG), and c) Yukawa - No Dark Matter (Y-No-DM), with the Newtonian model as the reference. Both short-range (SR) and long-range (LR) cases are presented for the MG and Y-No-DM models.}
	\label{tab:BayesF_log10}
	\centering
	\begin{tabular}{lcccccccc}
		\hline\hline
		& & \multicolumn{6}{c}{$\log_{10} BF_{i,N} = \log_{10} Z_{i} - \log_{10} Z_{N}$} \\
		\cline{3-8}
		Data & $\log_{10} Z$ (Newtonian) & \multicolumn{2}{c}{NTDMC} & \multicolumn{2}{c}{MG} & \multicolumn{2}{c}{Y-NO-DM} \\
		\cline{3-8}
		&  & SR & LR & SR & LR & SR & LR \\
		\hline
		MW  & $-426.223$ & $183.032$ & $-0.547$ & $143.086$ & $3.800$  & $106.896$ & $-9.342$ \\
		M31 & $-92.615$  & $-0.511$  & $-0.529$ & $-2.210$  & $-2.873$ & $-3.347$  & $-3.127$ \\
		\hline\hline
	\end{tabular}
\end{table*}

\section{Discussions}
\label{discuss}

\subsection{Phenomenological consequences of Yukawa correction}
The computed profiles of the galactic rotation curve are presented in Figs. \ref{fig:varyuk_DM} and \ref{fig:varyuk_all}. Generally, the strength parameter $\beta$ adjusts the magnitude of the velocity in the galactic components. A positive $\beta$ introduces an attractive "fifth force" component by increasing the velocity, effectively enhancing the gravitational potential at a point in space. Meanwhile, a negative $\beta$ has the opposite effect. On the other hand, the interaction distance parameter $\lambda$ determines the location where the Yukawa correction is applied effectively. For example, a small $\lambda$ only modifies the region close to the galactic center (small $r$), while a large $\lambda$ results in a more uniform modification of the velocity across both small and large distances from the galactic center.

The dominant peak of the rotational velocity of a galactic component can shift depending on the value of $\lambda$ and the sign of $\beta$. For instance, a significant positive value of $\beta$ combined with a small $\lambda$ can increase the velocity at shorter distances from the galactic center and shift the peak towards smaller $r$. As a result, galactic components that dominate at large $r$ in the Newtonian framework, such as dark matter, can shift to smaller distances for small $\lambda$, where the bulge and disk typically dominate. In such cases, the velocity of the dark matter component with a large positive $\beta$ can replace the dominance of the bulge or disk at smaller $r$ (see Fig. \ref{fig:varyuk_DM}). This behaviour would affect the decomposition results, as the dark matter component could now mimic or replace the contribution of the bulge or disk at shorter distances (i.e., from sub- to several kpc).

The actual impact of the Yukawa parameters also depends on screening mechanisms (e.g., Vainshtein and chameleon mechanisms), which suppress Yukawa corrections in the inner regions of galaxies while allowing fifth-force effects to manifest only in the outer regions. Typically, the fifth force is screened in high-density regions similar to the Solar System. Near the screening radius, an observable "upturn" emerges in the rotation curve where the fifth force begins to contribute significantly \cite{Naik2018}.

A negative value of $\beta$ yields the opposite effect on the rotation curve as it has repulsive characteristics. A small $\lambda$ case with negative $\beta$ will give lower velocity at short distances, which can shift the velocity distribution peak to larger $r$. The more substantial magnitude of $\beta$ in the negative case would imply weaker gravity or even repulsion if the effect from the Yukawa term is larger than that from the Newtonian one. This situation can be effortlessly ruled out, e.g., from the orbital stability, as the total velocity value must be positive everywhere. From this, we can expect that $ \ beta$'s lower (negative) bound of $\beta$ is closer to zero than the upper (positive) one. Recent observations, particularly from S-star orbits around the Galactic Center, constrain the strength parameter to be $|\beta| < 0.009$ for $\lambda \sim 150 \text{AU}$ (around $7 \times 10^{-7}$ kpc from the Galactic Center), which implies that if $\beta$ and $\lambda$ values are universal, large negative values of $\beta$ are disfavored by existing data \cite{Tan2024}.

\subsection{Statistical bounds of models}
From the results explained in Sec. \ref{rotcurdecompose}, we primarily focus on extracting the information of the model's bound from the posterior probability distribution and determining the statistical preference between models and assumptions using the calculated evidence.  
In this section, the posterior probability of each model is discussed, followed by an analysis of the statistical preferences of these models in the next section.  
The summary of the posterior bounds can be found in Table \ref{tab:Posterior_MW} and Table \ref{tab:Posterior_M31}.  
In this discussion, the parameter posterior distributions, which are limited by their priors, are treated as unconstrained, except for the priors derived from observational information.

\subsubsection{Newtonian model}
With this model, our primary goal is to compare our fit results to those in the literature. In the Milky Way, the central values of all parameters are generally of the same order of magnitude as the results of \cite{Sofue2015}. However, the bulge and disk parameters ($a_b,M_b,a_d,M_d$) and scale density $\rho_s$ in our study are systematically larger than those from \cite{Sofue2015}.  Notably, our result shows a significantly larger scale radius compared to Ref. \cite{Sofue2015}, with $r_s = 192^{+162}_{-117}$ kpc. This discrepancy arises from differences in variable selection, dataset scope, and the fitting methodology, as our approach simultaneously accounts for all galactic component contributions using multiple observables spanning a broad radial range.

In M31, we find slightly smaller bulge parameters than Ref. \cite{Sofue2015}, which did not include the central black hole as an additional mass source in the inner galactic region. We emphasize that in M31, the morphological parameters (i.e., $n, a_b, a_d$) are observationally constrained through photometric measurements. 

The derived supermassive black hole mass aligns with established values. For the MW, we obtain $M_{BH} \in (2.65,4.71)\times 10^6 M_\odot$, consistent with the measurement from Refs. \cite{Ghez1998,Abuter2023}. In contrast, for the M31, we derive a lower bound limit of $M_{BH} \gtrsim 3.7 \times 10^5 M_\odot $ based on the minimum mass required to explain the velocity curve at small $r$. While this only represents the lower bound, it remains consistent with the independent measurements from literature (see Ref. \cite{AlBaidhany2020} for a summary). We should note that this M31 black hole mass determination relies exclusively on rotation curve data.

\subsubsection{Yukawa without dark matter model (Y-noDM)}
This model provides a phenomenological description of the fifth force as an alternative to dark matter. While Refs. \cite{Sanders1984,Sanders1986} found that $\beta$ tends to have negative values to reproduce galactic rotation curves at around 5 to 50 kpc; our results show contrasting trends between the MW and M31.

For the MW, $\beta$ converges to a positive value in both the SR and LR cases. Specifically, in the SR case, the Yukawa correction enhances the total rotational velocity to fine-tune the data by slightly increasing the contribution from the bulge and disk at around $r \sim 10-100$ kpc (see Fig. \ref{fig:mw_ppc}f). In the LR case, $\beta$ converges to a weaker value with a relatively long range ($\lambda \gtrsim 23$ kpc), subtly modifying the disk and bulge components to maintain the flatness of the rotation curve in the outer region, $r \gtrsim 10^2$ kpc (see Fig. \ref{fig:mw_ppc}g).

For M31, $\beta$ is constrained to a negative value in both SR and LR scenarios. This repulsive Yukawa interaction reduces velocity at small radii ($r\sim 10$kpc), shifting the peak of the bulge velocity component outward. To compensate, the inferred bulge and disk masses increase as a counterbalance to match observational constraints. Critically, the requirement of positive $v^2$ restricts $\beta$ to have a significantly positively skewed distribution. Unlike MW, the posteriors of M31 are more realistic, benefiting from observationally informed priors. This fact reduces parameter degeneracies, yielding more physically realistic posteriors.

\subsubsection{Non-trivial dark matter -- matter coupling model (NTDMC)}
In this model, the Yukawa correction is assumed to affect only the DM component due to the non-trivial coupling between DM and baryonic matter. For this scenario, $\beta$ converges to relatively large values (typically $\beta \sim 10-10^2$) to influence the rotation curve significantly. Our results show substantial uncertainty in the posterior distributions of both $\beta$ and $\lambda$, which arises because the Yukawa modification exclusively affects the DM component. This component dominates at large $r$, where the uncertainty in the observational data is more pronounced.

For the SR scenario in both the MW and M31, $\lambda$ converges to small values while $\beta$ converges to large values. We find the peak probability around $\lambda \sim 0.3$ kpc in the MW with $\beta \gtrsim 87$. This fact occurs because the Yukawa corrections on the DM component affect the gravitational potential at low radii, altering the rotation curve. As a result, the DM component suppresses the bulge contribution and becomes the dominant potential component at small radii ($r \sim 1$ kpc). In contrast, the bulge and disk structure in M31 are tightly constrained by photometric data, preventing the DM component from dominating at smaller radii. Nevertheless, the posterior distribution of Yukawa parameters remains poorly constrained in M31, indicating that current data do not robustly support the SR NTDMC model.

In the LR case, the upper limits of $\lambda$ for both galaxies remain unconstrained. This finding indicates that the prior assumption forces modifications to dominate at large $r$, where rotation curve data have relatively poor precision. Notably, this regime's constraints on $\beta$ are far weaker than those reported in earlier studies (i.e., Refs. \cite{Almeida2018,Henrichs2021}). This discrepancy arises because, in our analysis, the prior limit is intentionally set to be wide to avoid artificially restricting the parameter space, ensuring that the Yukawa parameters are bounded by the data and model rather than the prior choice.

\subsubsection{Completely modified gravity model (MG)}
In this model, the galactic structure is similar to the standard $\Lambda$CDM description, incorporating universal Yukawa effects (i.e., a fifth force) affecting all mass components (bulge, disk, DM halo, SMBH). This model demonstrates the extent to which galactic rotation velocity data can reveal deviations from the current standard understanding of gravity in the galactic rotation curve, leveraging the degrees of freedom introduced by the Yukawa term.

In the Milky Way analysis, $\beta$ and $\lambda$ are constrained. The SR model converged at small $\lambda$ with large $\beta$ ($\beta \sim 10$), favoring modifications around $r\sim1$ kpc by adding new degrees of freedom in the bulge and disk distribution. Conversely, the LR case converges at $\lambda \sim 15$ kpc with a slightly repulsive $\beta \sim -0.6$, modifying the disk-dominated region's bulge potential.

For M31, both SR and LR are more homogeneous than the MW case, and the distribution coincides with Newtonian gravity within $2\sigma$. The $\lambda$ parameter remains unconstrained in the LR case, while the Bayes factor shows less preference than the Newtonian case. This fact indicates that the M31 data used here are insufficient, either in terms of precision or coverage, to extract the signal of gravity modification.

The completeness of data coverage may influence our results: the actual value of $\lambda$ could lie outside the data coverage, or the data may lack sufficient precision and accuracy to extract information about $\lambda$. For instance, recent observations by Ref. \cite{GRAVITYCollaboration2025} using S-stars around the galactic center reveal that the bound on $\beta$ is tightly constrained in the middle of the data range, $|\beta| \lesssim 0.003$, but remains unconstrained at the edge of the S-stars' orbital radius coverage, $|\beta| \gtrsim 10^3$. This finding highlights the importance of data coverage in determining $\beta$.

There is also a scenario in which the value of $\lambda$ could vary across different scales. In this context, the aforementioned tension may be reconciled through the perspective of screened gravity, where $\lambda$ becomes dependent on the environment. For instance, in chameleon screening gravity, as discussed in Ref. \cite{Khoury2003}, the value of the graviton mass (or equivalently, $\lambda$) varies with local density. Under the background cosmological density, this scenario favors a larger value of $\lambda$, on the order of kpc, while maintaining consistency with solar system tests.

\subsection{Statistical preferences of models}
Statistical preference is evaluated by comparing Bayes factors across models using the same dataset. Our findings demonstrate mixed behavior between MW and M31 (Table \ref{tab:BayesF_log10}).

In the MW, SR models exhibit extremely large Bayes Factors (BF) compared to Newtonian or their LR counterparts, indicating strong statistical favour for SR models. In the MW SR models, the best fit of the rotation curve converges to a larger value of $\beta$, up to multiple times that of their LR counterparts with small $\lambda$. This behaviour suggests that SR models favor a stronger Yukawa force at smaller distances, introducing a new degree of freedom with characteristics comparable to those of the bulge/disk components. For example, in the NTDMC case [Fig. \ref{fig:mw_ppc}(b)], the DM component significantly increases the velocity at $r\sim1$ kpc due to the Yukawa effect, while simultaneously suppressing the bulge component. In this case, the DM component acts as an additional degree of freedom around the bulge, finely tuning the overall curve to better fit the data. Therefore, the BF in this case is higher than in other cases. Similarly, for the MG and Y-NO-DM cases with SR, the Yukawa effect fine-tunes the velocity peak in the bulge-dominant ($r \sim 0.1-1$ kpc) and disk-dominant ($r \sim 10$ kpc) regions. In these cases, the Yukawa correction also affects the bulge around the disk-dominant region, fine-tuning the rotational velocity at that distance while simultaneously suppressing the disk velocity component.

While providing better BF values, the above cases are prone to overfitting. For example, they tend to produce false positive errors, suggesting that the modified gravity model is favored at non-physical values in the parameter space, such as extreme values of $\beta$ or significant differences in morphological parameters. In the SR cases of the Milky Way (MW), the bulge and disks are not constrained by observational data other than rotational velocity (which is a crude combination of galactic components), and their prior values do not include any additional constraints: any possible values are considered as valid configurations. Therefore, additional constraints such as conjugate observational data, e.g., independent S\'ersic parameter measurements, are needed to ensure that the overall posterior and BF converge in the physical region. This fact would constrain the morphological properties of the MW rather than relying solely on the rotation curve and its BF. For example, in the case of M31 (discussed later in this section), prior constraints from photometric measurements enable a more consistent posterior and BF between models.

Excluding SR cases, the MG LR strongly prefers the MW data over the Newtonian model. The value converges to the Newtonian case for $\lambda \gtrsim 13$ kpc, which slightly modifies the velocities around the disk at $r\sim 10^{1-2}$ kpc. This result better fits the curve at around 10 kpc by introducing additional degrees of freedom to the bulge component. Compared to the Newtonian model, the LR case of NTDMC is substantially disfavored, while the Y-NO-DM case is extremely disfavored.

Compared to MW data, M31 data exhibit a more homogeneous value of $\log_{10} z$, with the highest preference for the standard Newtonian model, followed by the NTMDC, MG, and Y-NO-DM models. There are no significant differences between the SR and LR models, although their posteriors vary. The homogeneity of $z$ indicates that the models predict and converge more similarly than they do for the MW. This behaviour could result from the strict constraints on morphological parameters ($a_b$, $a_d$, $n$), which enforce consistency with photometric observations and prevent overfitting from modifications. Meanwhile, the simplest model is the most preferred, which aligns with Occam's razor: the data lack sufficient signal (potentially in terms of coverage and precision) to extract information about additional parameters from modified gravity, and thus the simplest model is favored.

\subsection{Importance of the choice of prior}

Our results reveal various characteristics of the posterior distribution, highly sensitive to the model and assumptions (e.g., the $\lambda$ prior). A wide range of Yukawa parameter priors was selected. However, in several cases, the posterior distribution remains constrained by the prior, even when the prior's bounds are set very high.

As our goal is to infer the bounds of the Yukawa parameters from observational data, we choose to set a wide range of flat prior distributions for the Yukawa parameters to cover various possible values, avoiding \textit{ad-hoc} assumption. Here, we ensure that the prior does not truncate the complete distribution unless additional information is from other constraints, such as observational or theoretical bounds. If we set the prior excessively tight, any constraint will be dictated by that prior. Consequently, the posterior will be prone to providing false information that the parameters are tightly constrained.

That is because an overly narrow prior distribution (compared to the "true" posterior distribution purely dictated by the likelihood, e.g., if we set a flat prior without bounds) would misleadingly lead to an interpretation that the real distribution is centered around that value. For example, consider a narrow flat prior distribution located at the tail of a real distribution. In that case, the generated sample distribution should be approximately flat with minor ripples caused by numerical artifacts. This result can lead to a misleading interpretation that the inferred distribution is centered around the prior’s midpoint, even though the actual maximum likelihood estimate lies outside the prior range. 

The issue may worsen with larger datasets. Critically misleading narrow posterior distributions will arise when unconstrained data (e.g., $\beta$ uniformly distributed within priors) are assumed to follow Gaussian posteriors per galaxy. The product of these Gaussians yields an artificially narrow cumulative distribution centered near the prior midpoint, even as the actual data distribution remains unconstrained and random. This issue highlights how overly restrictive priors can generate false confidence in parameter estimates and distort model comparison. Such practices should be avoided in future studies, particularly when interpreting aggregated posteriors from weakly informative datasets.

\section{Conclusions}
\label{conclu}

We have investigated the rotational dynamics of the Milky Way (MW) and Andromeda (M31) galaxies within the framework of Yukawa gravity. This study focuses on two foundations: (i) deriving mathematical solutions and developing a numerical framework for squared rotational velocity ($V^2$) of individual galactic components under Yukawa gravity and studying their phenomenological implications, and (ii) unraveling theoretical consequences of these solutions for galactic-scale modified gravity theories via a Bayesian analysis framework towards rotation curve data.

We derived closed-form and numerical solutions of $V^2$ for all galactic components, which, to our knowledge, have not been explicitly formulated in this form in previous literature. Our calculations reveal that the negative $\beta$ values, which indicate an additional repulsive force from Yukawa gravity, are constrained by the condition  $V^2(r) > 0$ to ensure orbital stability. The $V^2$ curves also reveals that when $\lambda$ is relatively small (compared to galactic size, e.g., around one kpc) and combined with profoundly significant $\beta$ values (on the order of $\sim \mathcal{O}(10^1)$), the velocity component of the dark matter halo is dominant at mid-region, hence can mimic the velocity distribution that is typically observed in bulge or disk components. Dark matter halo within this parameter combination may even suppress effects from other components in these regions.

The implications of the above situation are revealed through Bayesian analysis. While previous studies focus only on parameter estimation, our Bayesian analysis through model comparison reveals that the dataset's characteristics determine the reliability of the parameter constraints. There exist distinct posterior distributions between galaxies and different Bayesian evidences ($Z$) across models. A notable consequence is, for the MW, the short-range $\lambda$ assumption yields a huge Bayes factor ($BF=Z_{model}/Z_{Newtonian}$) in comparison to other scenarios. In these short-range cases, Yukawa gravity introduces additional degrees of freedom, allowing for the "fine-tuning" of predictions, resulting in higher Bayes factor values. For instance, in dark matter models, the DM halo velocity component dominates the velocity around the bulge or disk region, reducing the estimated bulge values and disk masses or radii. However, as the values of the bulge or disk parameters become exceedingly small, short-range models are susceptible to type I errors (false positives). This result indicates a potential for overfitting within this model. One way to overcome this issue is by injecting Bayesian inference with complementary observational data, such as galactic morphological measurements (including S\'ersic parameters), to restrict the posterior to realistic values, as demonstrated in M31. Excluding short-range models, the statistical preference for the MW within long-range (LR) models indicates a preference for the full Modified Gravity (MG) model. In contrast, other models are less favored than the standard Newtonian model.

The above situation in short-range models does not apply to M31, which enables robust constraints from galactic morphology via S\'ersic parameters. The short-range case shows more homogeneous Bayes factors without extreme differences between models, as it converged around the distribution obtained in Newtonian. In M31, the Bayes factor result shows a preference for the Newtonian theory due to Occam's razor, where the data contains insufficient information to justify the Yukawa correction with current precision, and therefore the simplest model is preferred. This finding contrasts with the MW result, which shows a preference for the short-ranged model. This result indicates that high-precision $V^2$ measurements, capable of resolving finer features of galactic structure, are necessary to assess the statistical preference of more complex models.

To illustrate the complementary nature of observational data, we compare two representative cases: MW and M31 here represent two contrasting characteristics of data, one characterized by high-precision, wide-coverage of rotational velocity measurements (MW) and the other one with extensive global morphological data, enabling independent constraints on the structural parameters of the bulge and disk components (M31). Our study shows the importance of both aspects to test modified gravity with robust galactic rotation curves: precise $V(r)$ throughout wide-range coverage and conjugate data representing a global picture of a galaxy.

We also suggest that future studies should carefully handle the $\beta$ prior, as it depends on the underlying theory or assumption. For example, prior assumptions incorporated from solar measurements may be helpful in some cases, but those assumptions are no longer relevant for non-local theories or screened gravity models. Overly tight prior ranges may lead to biased inferences, making the parameter appear tightly constrained when it is, in fact, not.

As a final caveat, methodologies developed here provide a step-by-step template for testing gravity across a larger database of galaxies, such as SPARC \cite{Lelli2016} or its successor BIG-SPARC \cite{Haubner2024}. To improve the constraint, three critical regions must be precisely measured: around the galactic center, where black hole effects dominate, in the middle regions, where the bulge and disk components dominate, and in the flattened areas, where DM/fifth-force components dominate. Concurrent measurements, such as independent constraints from black hole mass and galaxy-wide structural parameter (e.g., S\'ersic indices), are also important to maintain the robustness of the inference.
 

\begin{acknowledgements}
We thank M. Ikbal Arifyanto for the fruitful discussions. The computation in this work has been done using the facilities of MAHAMERU BRIN HPC, National Research and Innovation Agency of Indonesia (BRIN). A. Sulaksono acknowledges financial support from the PUTI Grant for the period 2024–2025, under Grant No. NKB-380/UN2.RST/HKP.05.00/2024.
\end{acknowledgements}

\appendix
\section{Derivation of galaxy rotational velocity in Yukawa gravity}
\label{app:derivation}

For non-trivial DM coupling, we start by deriving the total gravitational potential of each galactic component seen in Eq. $(\ref{bab2:totpotNMG})$ by using the density profiles outlined in Chapter \ref{galacticstruc}. We compute the rotational velocities analytically possible, using numerical methods for more complex cases. Our implementation uses \texttt{numpy} \cite{Harris2020} and \texttt{scipy} \cite{Virtanen2020} libraries to represent scientific functions, \texttt{mpmath} library \cite{mpmathdevelopmentteam2023} for high-precision operations, and \texttt{@jit} decorator from the \texttt{numba} library \cite{Lam2015} for computational speed optimization. The final rotation curve sums all component velocities according to each specified model.

\subsection{Disk}
\label{app:Disk}
The disk is derived differently from other Galactic components due to its thin cylindrical model. Hence, we must derive the general gravitational potential and circular velocity equation for an axisymmetrical cylindrical matter distribution, where the potential is evaluated at the $z=0$ plane.

\subsubsection{Newtonian potential}
The calculation of the Newtonian disk potential follows the derivation done by Ref. \cite{Mannheim2006}. The potential is given as  
\begin{equation}
	\label{pot Disk N}
	\Phi_{Disk,N}(r) = -\pi G \Sigma_0 r \left[I_0\left(\frac{r\alpha}{2} \right)K_1\left(\frac{r\alpha}{2}\right) -I_1\left(\frac{r\alpha}{2}\right)K_0\left(\frac{r\alpha}{2}\right) \right].
\end{equation}
By benefiting relations of derivatives of modified Bessel functions, we can evaluate the circular velocity, which is obtained as
\begin{equation}
	\label{app:almostfindisk}
	V_{Disk,N}^2 =  \pi G \Sigma_0 r^2\alpha \left[I_0\left(\frac{r\alpha}{2}\right)K_0\left(\frac{r\alpha}{2}\right) - I_1\left(\frac{r\alpha}{2}\right)K_1\left(\frac{r\alpha}{2}\right) \right].
\end{equation}
We can write this in terms of the disk mass, following Ref. \cite{Sofue2015}, as
\begin{equation}
	\label{app:diskmass}
	M_d = \int_0^\infty 2\pi r \Sigma(r) \odif{r} = 2\pi \Sigma_0 a_d^2,
\end{equation}
hence Eq. (\ref{app:almostfindisk}) can be written as
\begin{equation}
	V_{Disk,N}^2 = \frac{G M_d r^2}{2a_d^3} \left[I_0\left(\frac{r}{2a_d}\right)K_0\left(\frac{r}{2a_d}\right) - I_1\left(\frac{r}{2a_d}\right)K_1\left(\frac{r}{2a_d}\right) \right].
\end{equation}
This equation can then be computed with the help of the \texttt{special} functions from the \texttt{scipy} library, which contains the modified Bessel functions.
\subsubsection{Yukawa-correction potential}
We use instead the Green's function for the Helmholtz case from Ref. \cite{Conway2010}, evaluating in cylindrical coordinates for the $\varphi=0$ case, which causes the the $m$-summation to collapse. We substitute the disk density Eq. (\ref{bab2:rhodisk}) for observation points in the $z=0$ plane of the disk to obtain the modified gravity potential as
\begin{equation}
	\Phi_{Disk,Y}(r,z=0) = -2\pi G\beta \Sigma_0 \int_0^\infty e^{-\alpha r'} G_H^0(\lambda,r,r') r' \odif{r'},
\end{equation}
where the Fourier coefficients of the Helmholtz Green function becomes
\begin{equation}
	\label{app:G0Hnear}
	G_H^0(\lambda,r,r')= \frac{1}{\pi}\int_0^\pi\frac{\exp\left(-\frac{1}{\lambda}\sqrt{r^2+{r'}^2-2rr'\cos\psi}\right)}{\sqrt{r^2+{r'}^2-2rr'\cos\psi}}\odif{\psi}.
\end{equation}
We can also rewrite this in terms of the disk mass from Eq. (\ref{app:diskmass}) and its scale length as 
\begin{equation}
	\label{app:diskMGphi}
	\Phi_{Disk,Y}(r,z=0) = - \frac{G\beta M_d}{a_d^2} \int_0^\infty e^{- r'/a_d} G_H^0(\lambda,r,r') r' \odif{r'}.
\end{equation}
Because of the machine limit on floating-point numbers, the integral above may become divergent for large values of $r$. Thus, in such regions, we can use the following far-field expression for the Fourier coefficients when $m=0$, as seen in Ref. \cite{Conway2010},
\begin{equation}
	\label{app:G0Hfar}
	G_H^0(\gamma,\mathcal{K},r,r')= \frac{\mathcal{K}}{2\sqrt{rr'}}\exp{[\xi(1-\mathcal{K}^2/4)]}I_0(-\xi \mathcal{K} ^2/4),
\end{equation}
where 
\begin{equation}
	\mathcal{K} = \frac{\sqrt{4rr'}}{|r+r'|} ,
\end{equation}
and 
\begin{equation}
	\xi = -\frac{1}{\lambda}|r+r'|.
\end{equation}
The modified gravitational potential Eq. \ref{app:diskMGphi} is computed numerically using 20 logarithmically spaced points $r_{val}$ between data points $Rad_{\text{min}}$ and $Rad_{\text{max}}$. The $G^0_H$ function is selected based on the value of $|\xi|$, far-field expression from Eq. (\ref{app:G0Hfar}) for $|\xi| \ge 100$, and near-field expression Eq. (\ref{app:G0Hnear}) for $|\xi| < 100$. We introduce a small error term $\epsilon^{int}_{err} = 10^{-9}$ to stabilize integrals with radicands $r_{val}^2+{r'}^2-2r_{val}r\cos{\varphi} < 1\times 10^{-5}$. A custom \texttt{logintegral} function is used to evaluate the integral in Eq. (\ref{app:G0Hnear}) by replacing either integration bounds by $10^{\pm15}$ if it exceeds that value and applying the trapezoidal rule with \texttt{logsumexp} on 80 logarithmically-spaced points, $x_{val}$, between the adjusted bounds. Our calculations of $G^0_H$ have been tested to be consistent with the numerical results listed in Ref. \cite{Conway2010}.

We integrate Eq. (\ref{app:diskMGphi}) using the \texttt{scipy.quad} function, and thereafter transformed to logarithm-space $\odif{\ln{r_{val}}}$, and linearly interpolated and extrapolated to obtain the function $\Phi_{Disk,Y}(\ln{r})$. The velocity is numerically approximated using the five-point central difference method,
\begin{align}
	V_{Disk,Y}^2(\ln{r_{val}}) &= \odv{\Phi_{Disk,Y}}{\ln{r_{val}}}\\
	&\approx \frac{1}{12h}\left[ \Phi_{Disk,Y}(\ln{r_{val}}-2h)\right. \notag \\
	&\qquad\qquad \left. +8\Phi_{Disk,Y}(\ln{r_{val}}+h) \right.\notag  \\
    & \qquad \qquad \left. -\Phi_{Disk,Y}(\ln{r_{val}}+2h)\right].
\end{align}
To avoid oscillations for values of $V_{Disk,Y}^2 \approx 0$, we interpolate to obtain the function $V_{Disk,Y}^2(\ln{r})$, with cubic interpolation for the threshold $V^2_{Disk,Y}(\ln{r_{val}}) > 20$, and linear interpolation below it. The resulting function can be used to calculate $V^2_{Disk,Y}(\ln{Rad})$ with real data points $Rad$.


\subsection{Bulge}
\label{app:Bulge}
We model the bulge (and any additional galactic components aside from the disk) as spherically symmetric. Due to the non-elementary nature of the volume mass density profile of the bulge, subsequent calculations will be carried out numerically.
\subsubsection{Newtonian potential}
The Newtonian gravitational potential in spherical coordinates reduces to
\begin{align}
	\Phi_{N}(r) &= -4\pi G  \left[ \frac{1}{r}\int_0^r   \rho(r')  {r'}^{2} \odif{r'} + \int_r^\infty \rho(r') r' \odif{r'}  \right] \\
	&= -4\pi G  \left[ \frac{M(r)}{4\pi r} + \int_r^\infty \rho(r') r' \odif{r'}  \right],
\end{align}
Using the Leibniz integral rule, the second term vanishes as we take the derivative of the potential with respect to $r$, hence we obtain the circular velocity expression for a spherically symmetric mass distribution as
\begin{equation}
	\label{circ vel}
	V_{N}^2(r) = r\odv{\Phi_{N}(r)}{r} = \frac{G}{r}\int_0^r  \odif{M(r)} = \frac{4\pi G}{r} \int_0^r \rho(r') {r'}^2 \odif{r'}.
\end{equation}
The volume mass density of the bulge is the Abel transformation of the surface matter density \cite{Mannheim2006},
\begin{align}
	\label{rho bul}
	\rho_{Bul}(r') &=-\frac{1}{\pi}\int_{r'}^\infty\odv{I(x)}{x}\frac{\odif{x}}{\sqrt{x^2-{r'}^2}} \\
    &= \frac{b_n e^{b_n}I_e}{4\pi a_b^{1/4}}\int_{r'}^\infty\frac{e^{-b_n (x/a_b)^{1/4}}}{x^{3/4}\sqrt{x^2-{r'}^2}} \odif{x},
\end{align}
and thus upon substituting into Eq. (\ref{circ vel}), the velocity can be written as 
\begin{equation}
	\label{vBul N full}
	V_{Bul,N}^2(r) = \frac{b_n e^{b_n}I_e G}{a_b^{1/4}r} \int_0^r \int_{r'}^\infty\frac{e^{-b_n (x/a_b)^{1/4}}}{x^{3/4}\sqrt{x^2-{r'}^2}} {r'}^2 \odif{x,r'}.
\end{equation}
We can then rewrite this in terms of the total cylindrical or projected mass of the bulge predicted by the profile, as per Ref. \cite{Sofue2015}. Following the derivation from Ref. \cite{Vaccari2000}, we integrate Eq.(\ref{bab2:deVau}) as
\begin{align}
	M_b &= 2\pi \int_0^\infty I(r') r' \odif{r'}\\
	&= 2\pi e^{b_n} I_e \int_0^\infty \exp{-b_n\left(\frac{r'}{a_b}\right)^{1/n}} r' \odif{r'} \\
	&= \frac{8\pi e^{b_n} I_e a_b^2}{b_n^8} \Gamma(8)\\
	&= \eta I_e a_b^2,
\end{align}
\sloppy in which we used the dimensionless parameter $\eta \equiv 8\pi e^{b_n}\Gamma(8)/b_n^8 = 22.665$ defined in Ref. \cite{Sofue2015}. Accordingly, Eq. (\ref{vBul N full}) can be rewritten as 
\begin{equation}
	\label{app:Bulge_V_N}
	V_{Bul,N}^2(r) = \frac{b_n e^{b_n}M_b G}{\eta a_b^{9/4}r} \int_0^r \int_{r'}^\infty\frac{e^{-b_n (x/a_b)^{1/4}}}{x^{3/4}\sqrt{x^2-{r'}^2}} {r'}^2 \odif{x,r'}.
\end{equation}
To compute this equation, we follow the numerical method used for Yukawa-correction velocity in the disk component. 
\subsubsection{Yukawa-correction potential}
Using the radial Green's function while considering the symmetrical spherical nature, and substituting the $j_{0}(x) = \sin x/x$ and $h_{0}^{(1)}(x)= e^{ix}/ix$ forms of the spherical Bessel and Hankel functions, we acquire
\begin{align}
	\label{pot bulge int 2}
	\Phi_{Bul,Y}(r) 
	&= -\frac{4\pi G\beta\lambda}{r}\left[ e^{-\frac{r}{\lambda}}\int_0^r \rho(r')\sinh{\left(\frac{r'}{\lambda}\right)}r'\odif{r'}\right.\notag\\
	&\qquad \left. +\sinh{\left(\frac{r}{\lambda}\right)}\int_r^\infty \rho(r')e^{-\frac{r'}{\lambda}}r'\odif{r'}\right].
\end{align}
Substitute the bulge volume mass density Eq.(\ref{rho bul}) into Eq. (\ref{pot bulge int 2}) to obtain
\begin{equation}
	\Phi_{Bul,Y}(r) = -\frac{b_n e^{b_n} I_e G\beta\lambda}{a_b^{1/4}r}\left[ e^{-r/\lambda}\mathcal{I}_{1}^{Bul} +\sinh{\left(\frac{r}{\lambda}\right)}\mathcal{I}_{2}^{Bul}\right],
\end{equation}
where we have to find these integrals numerically
\begin{equation}
	\label{app:bulfirstint}
	\mathcal{I}_{Bul,1} = \int_0^r \int_{r'}^\infty\frac{e^{-b_n (x/ab)^{1/4}}}{x^{3/4}\sqrt{x^2-{r'}^2}}r'\sinh{\left(\frac{r'}{\lambda}\right)}\odif{x,r'},
\end{equation}
and
\begin{equation}
	\label{app:bulsecint}
	\mathcal{I}_{Bul,2} = \int_r^\infty \int_{r'}^\infty\frac{e^{-b_n (x/ab)^{1/4}}}{x^{3/4}\sqrt{x^2-{r'}^2}}r' e^{-r'/\lambda}\odif{x,r'}.
\end{equation}
The rotational velocity equation for the bulge can be derived straightforwardly as
\begin{align}
	\label{app:bulMGv}
	V_{Bul,Y}^{2}(r) &=  \frac{b_n e^{b_n} M_b G\beta\lambda }{\eta a_b^{9/4}r} \left[ e^{-r/\lambda}\left(1+\frac{r}{\lambda}\right)\mathcal{I}_{Bul,1} \right.\\
	&\qquad \left. + \left\{\sinh{\left(\frac{r}{\lambda}\right)} -\frac{r}{\lambda}\cosh{\left(\frac{r}{\lambda}\right)}\right\}\mathcal{I}_{Bul,2} \right],
\end{align}
We substitute $I_e=M_b/\eta a_b^2$ and compute its velocity numerically, similar to the Newtonian method. We utilize the logarithmic output of \texttt{logsumexp} to transform Eq. (\ref{app:bulMGv}) into 
\begin{align}
	V_{Bul,Y}^{2}(r_{val}) &=  \frac{b_n e^{b_n} M_b G\beta\lambda }{\eta a_b^{9/4}r_{val}} \left[  e^{-\frac{r_{val}}{\lambda}+ (1+\frac{r_{val}}{\lambda}) + \ln{\mathcal{I}_{Bul,1} } } \right.\notag \\
	&\qquad \left. +\frac12 \left\{e^{\frac{r_{val}}{\lambda} + \ln{\mathcal{I}_{Bul,2}}}  + e^{-\frac{r_{val}}{\lambda} + \ln{\mathcal{I}_{Bul,2}}} \right\} \right.\notag\\ 
	&\qquad  \left.  - \frac{r_{val}}{2\lambda}\left\{e^{\frac{r_{val}}{\lambda} + \ln{\mathcal{I}_{Bul,2}}}  + e^{-\frac{r_{val}}{\lambda} + \ln{\mathcal{I}_{Bul,2}}} \right\}  \right] ,
\end{align}
for easier calculations. 
\subsection{DM halo}
Since the DM Halo model is spherically symmetric, we can follow the same derivation as for the bulge component.
\label{app:DMHalo}
\subsubsection{Newtonian potential}
We can substitute Eq. (\ref{bab2:rhoDM}) to the circular velocity equation Eq. (\ref{circ vel}),
\begin{equation}
	V_{DM,N}^2 = \frac{4\pi G}{r} \int_0^r \frac{\rho_s r'^2}{\left(\frac{r'}{r_s}\right)\left(1+\frac{r'}{r_s}\right)^2}\odif{r'}. 
\end{equation}
and hence obtaining the velocity as
\begin{equation}
	\label{v DM Newt}
	V_{DM,N}^2  = \frac{4\pi G \rho_s r_s^3}{r} \left[\ln\left(1+\frac{r}{r_s}\right) - \frac{r/r_s}{1+\frac{r}{r_s}}\right].
\end{equation}

\subsubsection{Yukawa-correction Potential}
Using the NFW density Eq. (\ref{bab2:rhoDM}), we obtain the potential as
\begin{align}
	\label{DM main int}
	\mathcal{I}_{DM,Main} &= -\frac{\lambda^2}{r}\left[ e^{-\frac{r}{\lambda}} \int_0^r \frac{\sinh{\left( \frac{r'}{\lambda}\right)}}{(r'+r_s)^2} \odif{r'} \right. \notag \\
    & \qquad \qquad \left. + \sinh{\left( \frac{r}{\lambda}\right)} \int_r^\infty \frac{e^{-\frac{r'}{\lambda}} }{(r'+r_s)^2}\odif{r'} \right] \\
	&=  -\frac{\lambda}{2r}\left[e^{\frac{r_s+r}{\lambda}}\text{Ei}\left(-\frac{r+r_s}{\lambda}\right)-e^{\frac{r_s-r}{\lambda}}\text{Ei}\left(-\frac{r_s}{\lambda}\right)\right.\notag\\ 
	&\qquad \qquad \left. +e^{-\frac{r_s+r}{\lambda}}\left\{\text{Ei}\left(\frac{r+r_s}{\lambda}\right)-\text{Ei}\left(\frac{r_s}{\lambda}\right)\right\}\right]
\end{align}
so the Yukawa-correction gravitational potential results to
\begin{align}
	\Phi_{DM,Y}(r)&= \frac{2\pi G\beta \rho_s r_s^3}{r} \left[e^{\frac{r_s-r}{\lambda}}\text{Ei}\left(-\frac{r_s}{\lambda}\right) -e^{\frac{r+r_s}{\lambda}}\text{Ei}\left(-\frac{r+r_s}{\lambda}\right)\right.\notag\\ 
	&\qquad \left. +e^{-\frac{r+r_s}{\lambda}}\left\{\text{Ei}\left(\frac{r_s}{\lambda}\right)-\text{Ei}\left(\frac{r+r_s}{\lambda}\right)\right\}\right],
\end{align}
with a velocity of 
\begin{align}
	V_{DM,Y}^2 &= -\frac{2\pi G \beta\rho_{s}r_{s}^{3}}{r} \left[\frac{2r}{r+r_{s}}+e^{-\frac{r+r_{s}}{\lambda}}\left(1+\frac{r}{\lambda}\right) \left\{    \text{Ei}\left(\frac{r_{s}}{\lambda}\right) \right.\right. \notag \\
    &\qquad \left. \left. + e^{\frac{2r_s}{\lambda}} \text{Ei}\left(-\frac{r_{s}}{\lambda}\right) - \text{Ei}\left(\frac{r+r_{s}}{\lambda}\right) \right\}  \right.\notag \\
	&\qquad \left.   + e^{ \frac{r+r_{s}}{\lambda}}\left(\frac{r}{\lambda}-1\right)  \text{Ei}\left(-\frac{r+r_{s}}{\lambda}\right) \right],
\end{align}
matching the form in Ref. \cite{Almeida2018}. We numerically compute this using the exponential and Ei functions from the \texttt{mpmath} library \cite{mpmathdevelopmentteam2023}, converting results from \texttt{mpf} (real floats) back to \texttt{numpy.float64}, with \texttt{NaN} values set to 0.
\subsection{Black hole}
\label{app:BH}
The central SMBH can be modeled as a point mass in spherical coordinates, in which we modeled the mass $M_{BH}$ all contained in the center of the origin. We will consider $M_{BH}$ as a parameter during the fitting.
\subsubsection{Newtonian potential}
We can immediately substitute the mass of the black hole to the circular velocity equation derived in Eq.(\ref{circ vel}) as
\begin{align}
	V_{BH,N}^2 = \frac{GM_{BH}}{r}.
\end{align}
\subsubsection{Yukawa-correction potential}
Substituting the BH mass density in Eq. (\ref{bab2:rhoBH}), we acquire
\begin{align}
	\label{pot BH start}
	\Phi_{BH,Y}(r) &=  \frac{GM_{BH}\beta}{\lambda}\int_0^\infty j_0\left(\frac{ir_<}{\lambda}\right) h_0^{(1)}\left(\frac{ir_>}{\lambda}\right)\delta(r')\odif{r'} \\
    &=  \frac{GM_{BH}\beta}{\lambda} \mathcal{I}_{BH,Main}.
\end{align}
The integral can then be separated into two cases,
\begin{align}
	\mathcal{I}_{BH,Main} &= -\frac{\lambda^2}{r}\left[e^{-\frac{r}{\lambda}}\int_0^r \frac{\sinh(\frac{r'}{\lambda})}{r'} \delta(r')\odif{r'} \right. \notag\\ 
    &\qquad \qquad \left. +\sinh\left(\frac{r}{\lambda}\right)\int_r^\infty \frac{e^{-\frac{r'}{\lambda}}}{r'}\delta(r')\odif{r'} \right]\\
	&= -\frac{\lambda e^{-\frac{r}{\lambda}}}{r}.
\end{align}
Substitute back to Eq. (\ref{pot BH start}) to express the potential equation as
\begin{equation}
	\label{pot BH fin}
	\Phi_{BH,Y}(r) =  -\frac{GM_{BH}\beta e^{-\frac{r}{\lambda}}}{r},
\end{equation}
from which we can derive the velocity as
\begin{equation}
	V_{BH,Y}^2 = \frac{GM_{BH}\beta e^{-\frac{r}{\lambda}}(r+\lambda)}{r \lambda}.
\end{equation}
We can numerically compute both the Newtonian and Yukawa term velocities immediately. 

\subsubsection{Approach to handle non-physical solutions}
Before combining the velocities of each Galactic component to compute the total rotational velocity, as outlined in Chapter \ref{rotcurdecompose}, we circumvent potentially problematic values from each component. Here, problematic refers to values that could lead to unphysical rotation curves, such as imaginary velocities (or negative sign of squared velocities). To address this, if the velocity of any component yields negative or \texttt{NaN} values, we adjust so that the log-likelihood function returns a $-\infty$ likelihood. Furthermore, we verify whether the total velocity contains any negative or \texttt{NaN} values, and if so, the log-likelihood function will also return a $-\infty$ likelihood.

\section{Data selection and preprocess}
\subsection{GRAVITY}
\label{constrainingMBH}

The GRAVITY collaboration has measured the SMBH Sgr A* mass by observing stellar orbits within $9R_g$ (0.38 AU) from the center of the Milky Way using interferometry. To incorporate this data into our framework, we first calculate the circular velocity-equivalent of stellar orbits at $R=9R_g$ for the enclosed SMBH mass based on GRAVITY measurement (see Table 2 in Ref. \cite{Abuter2023}),
\begin{equation}
	V_{BH,enc}^2 = \frac{GM_{BH,enc}}{9R_g}.
\end{equation}
Uncertainties in $M_{BH}$ and $R_g$ propagate through $V_{BH}^2$, 
\begin{align}
	\sigma_{V_{BH,enc}^2}^2 &\approx \left. \sigma_{M_{BH,enc}}^2 \left( \pdv{V_{BH,enc}}{M_{BH,enc}} \right)^2 + \sigma_{R}^2 \left(\pdv{V_{BH,enc}}{R} \right)^2 \right|_{R=9R_g},\\ 
	&\approx  \sigma_{M_{BH,enc}}^2 \left( \frac{G}{9R_g} \right)^2 + \sigma_{9R_g}^2 \left(-\frac{GM_{BH,enc}}{(9R_g)^2} \right)^2.
\end{align}
Using the same method, we can find the standard deviation of the orbital velocity data as
\begin{equation}
	\sigma_{V_{BH,enc}} = \frac{\sigma_{V_{BH,enc}^2}}{2\sqrt{V_{BH,enc}^2}}.
\end{equation}
Finally, we use Eq. (\ref{loglikelihood}) to calculate the likelihood for the SMBH from the GRAVITY collaboration data to further constrain the $M_{BH}$ parameter in the inner regions. We treat $V_{BH}$ as a rotation curve data point at $R=9R_g$, from which we assume a Gaussian likelihood, 
\begin{equation}
\mathcal{L}(\Theta_{\text{BH}}) = \frac{1}{\sqrt{2\pi\sigma^2}} \exp{-\frac{(V_{\text{obs}} - V_{\text{pred}})^2}{2\sigma^2}}
\end{equation}
\sloppy where $\sigma = \sigma_{V_{\text{BH,enc}}}$, $V_{\text{obs}} = V_{\text{obs}}(9R_g)$, and $V_{\text{pred}} = V_{\text{pred}}(\Theta,\mathcal{M}_i(9R_g))$.

\subsection{Milky Way rotation curves}

One of the most comprehensive datasets of the rotation curve of the Milky Way galaxy has been derived by Y. Sofue \cite{Sofue2009,Sofue2015,Sofue2016,Sofue2020} (here, we refer to it as the Galactic RC). This dataset incorporates a running average derived from various sources, including disk objects such as interstellar gases or population I stars, observed in H$\alpha$, HI, and CO emission lines \cite{Sofue2016}. The methodology for deriving the Galactic RC is distinct from other spiral galaxies due to our
position within the Galaxy, and thus, different methods are used to derive the
rotational velocities in both the inner and outer regions. In the inner regions, where the tangential velocity is measurable, the rotational velocity is determined using the HI tangential velocity alongside the velocity measurements from CO emissions. Conversely, in the outer regions, the absence of a tangential point along the line of sight requires alternative methodologies, such as analyzing the geometry of the HI disk or utilizing known distances to galactic objects (e.g., HII regions). As the rotation curve is dependent on Galactic constants such as the Solar radius $R_0$ and its velocity $V_0$ \cite{Sofue2016}, the Galactic RC is continuously revised with updated measurements of these constants and with correction factors for non-coplanar objects. The latest iteration of the Galactic RC is the Sofue (2020) dataset, which exhibits more regional specificity than the Sofue (2017) dataset.

Meanwhile, the third data release of Gaia (Gaia DR3) in 2022 \cite{GaiaCollaboration2023} presents newer and more accurate astrometric information of Galactic stars, which can act as a basis for constructing the Galactic RC. The RC derivation from the Gaia DR3 set has been done in several studies, including those by Ref. \cite{Wang2022}, which uses statistical deconvolution of the parallax errors to the stars, and Ref. \cite{Zhou2023}, which determines precise distances to $\sim 54000$ thin-disk population stars selected from samples of luminous red giant branch stars. Although both results are limited to the 5 – 30 kpc range, we add these datasets to our study to examine their effects. 

The Gaia data set has a noticeably higher velocity than the derived Sofue data in the 5 - 30 kpc region. We applied a weighted running average across a 5-30 kpc region to reconcile this unresolved tension over all datasets. We define an asymmetric window $[R_i-\Delta_{left}, R_i+\Delta_{right}]$ covering every data point in the original Sofue dataset (2020), where $\Delta_{left} = \frac12(R_i-R_{i-1})$ (or $\Delta_{left,i} = \frac12(R_{i+1} - R_i)$ at the start point) and $\Delta_{right} = \frac12(R_{i+1} - R_i)$ (or $\Delta_{right,N_S} = \frac12(R_i-R_{i-1})$ at the end point).
The weighted average is taken from all data points that are covered by this window.
\begin{equation}
	V_{avg}(R_i) = \frac{w_SV_s+\sum_{k=1}^{N_{G}}w_GV_G(R_k)}{w_S+N_Gw_G},
\end{equation}
In which subscript $S$ stands for Sofue data and $G$ stands for Gaia data. We set $w_S=1$ and $w_G=1/N_G$, which ensures only 50\% importance for the Gaia dataset. We then calculate its accompanying weighted variance, $\sigma^2=\sum_{i}w_i(V_i-V_{avg})^2/\sum_{i}w_i$.

\section{Constraints of parameters}
The limits of parameters derived from posterior distributions are shown by Table \ref{tab:Posterior_MW} for the Milky Way and Table \ref{tab:Posterior_M31} for M31.

\begin{table*}[htbp]
	\centering
	\caption{Estimated $95\%$ CI of the marginalized posterior for all parameters of the Milky Way within the cases considered in this study, with the prior $\ln{(\lambda)} = [-4,10]$ for the "long range" case. In the Milky Way case, the bulge structure is assumed to have a de Voucaleurs profile with $n=4$. An asterisk ($^*$) indicates an unconstrained upper limit restricted by prior, while a dagger ($_\dag$) denotes the lower limit. If unconstrained from the upper or lower limit, the bound shows $95\%$ limit calculated from the opposite prior limit. For posteriors unconstrained from both upper and lower limits, the value represents the average of the distribution within the prior bounds.}
	\label{tab:Posterior_MW}
	\small
	\renewcommand{\arraystretch}{1.1}
	\setlength{\tabcolsep}{3pt}
	\begin{tabular}{c@{\hspace{6pt}}l@{\hspace{4pt}}l@{\hspace{4pt}}l@{\hspace{4pt}}l@{\hspace{4pt}}l@{\hspace{4pt}}l@{\hspace{4pt}}l}
		\hline\hline
		\multirow{3}{*}{Parameters} & \multicolumn{7}{c}{Milky Way} \\ \cline{2-8} 
		& \multirow{2}{*}{Newtonian} & \multicolumn{2}{c}{NTDMC} & \multicolumn{2}{c}{MG} & \multicolumn{2}{c}{No DM} \\ \cline{3-8}
		& & SR & LR & SR & LR & SR & LR \\[2pt] \hline
		$\beta$ & -- & $132^{+16.9\ast}_{-44.6}$ & $83.8^{+62.4}_{-74.0}$ & $9.85^{+3.71}_{-3.81}$ & $-0.583^{+0.0481}_{-0.0267}$ & $4.12^{+0.452}_{-0.753}$ & $1.27^{+0.475}_{-1.60}$ \\ \hline
		$\ln{(\lambda)}$ & -- & $-1.27^{+0.0377}_{-0.0345}$ & $8.45^{+2.40\ast}_{-3.49}$ & $-0.773^{+0.0556}_{-0.0527}$ & $2.69^{+0.492}_{-0.340}$ & $-0.591^{+0.00680}_{-0.0702}$ & $9.72^{+1.21\ast}_{-6.57}$ \\ \hline
		$M_b$ & $(4.04^{+0.239}_{-0.230})\times 10^{10}$ & $(7.36^{+45}_{-7.21\dag})\times 10^{8}$ & $(4.04^{+0.248}_{-0.231})\times 10^{10}$ & $(4.79^{+2.81}_{-1.32})\times 10^{10}$ & $(9.75^{+0.240\ast}_{-0.980})\times 10^{10}$ & $(9.83^{+0.166\ast}_{-0.619})\times 10^{10}$ & $(1.85^{+4.65}_{-0.305}) \times 10^{10}$ \\ \hline
		$a_b$ & $1.62^{+0.100}_{-0.0961}$ & $6.76^{+3.11\ast}_{-5.10}$ & $1.62^{+0.100}_{-0.0933}$ & $9.90^{+0.0988\ast}_{-0.341}$ & $1.61^{+0.0974}_{-0.0899}$ & $9.71^{+0.284\ast}_{-0.987}$ & $1.69^{+0.09}_{-0.0859}$ \\ \hline
		$M_d$ & $(2.05^{+0.374}_{-0.319})\times 10^{11}$ & $(8.38^{+0.881}_{-1.22})\times 10^{10}$ & $(2.03^{+0.358}_{-0.281})\times 10^{11}$ & $(8.69^{+1.61}_{-1.39})\times 10^{10}$ & $(2.33^{+0.554}_{-0.428})\times 10^{11}$ & $(2.51^{+0.240}_{-0.234})\times 10^{11}$ & $(1.48^{+2.31}_{-0.215})\times 10^{11}$ \\ \hline
		$a_d$ & $7.45^{+0.693}_{-0.622}$ & $3.85^{+0.278}_{-0.242}$ & $7.40^{+0.666}_{-0.586}$ & $5.52^{+0.538}_{-0.458}$ & $5.74^{+0.581}_{-0.526}$ & $9.07^{+0.451}_{-0.478}$ & $9.49^{+0.458}_{-0.977}$ \\ \hline
		$n$ & $4$ & $4$ & $4$ & $4$ & $4$ & $4$ & $4$ \\ \hline
		$\ln{(\rho_s)}$ & $11.1^{+1.51}_{-0.922}$ & $15.1^{+0.755}_{-0.278}$ & $6.77^{+2.14}_{-1.21}$ & $13.4^{+1.10}_{-1.28}$ & $11.4^{+1.32}_{-1.19}$ & -- & -- \\ \hline
		$r_s$ & $192^{+162}_{-117}$ & $26.8^{+4.78}_{-8.09}$ & $209^{+158}_{-125}$ & $60.2^{+70.3}_{-26.0}$ & $157^{+197}_{-86.3}$ & -- & -- \\ \hline
		$\ln{(M_{BH})}$ & $15.1^{+0.266}_{-0.310}$ & $15.2^{+0.244}_{-0.278}$ & $15.1^{+0.271}_{-0.310}$ & $12.9^{+0.520}_{-0.453}$ & $15.7^{+0.280}_{-0.323}$ & $13.6^{+0.291}_{-0.323}$ & $14.4^{+1.00}_{-0.436}$ \\ \hline\hline
	\end{tabular}
\end{table*}

\begin{table*}[htbp]
	\centering
	\caption{Similar to Table \ref{tab:Posterior_MW} for M31. Bulge and disk are assumed to follow a S\'ersic model with $n$, $a_b$, and $a_d$ priors derived from photometric observations.}
	\label{tab:Posterior_M31}
	\footnotesize
	\renewcommand{\arraystretch}{1.1}
	\setlength{\tabcolsep}{3pt}
	\begin{tabular}{c@{\hspace{4pt}}l@{\hspace{4pt}}l@{\hspace{4pt}}l@{\hspace{4pt}}l@{\hspace{4pt}}l@{\hspace{4pt}}l@{\hspace{4pt}}l}
		\hline\hline
		\multirow{3}{*}{Parameters} & \multicolumn{7}{c}{M31} \\ \cline{2-8} 
		& \multirow{2}{*}{Newtonian} & \multicolumn{2}{c}{NTDMC} & \multicolumn{2}{c}{MG} & \multicolumn{2}{c}{No DM} \\ \cline{3-8}
		& & SR & LR & SR & LR & SR & LR \\ \hline
		$\beta$ & -- & $34.5^{+100}_{-31.2}$ & $87.0^{+59.9\ast}_{-63.3}$ & $-0.541^{+2.59}_{-0.250}$ & $-0.0217^{+1.17}_{-0.687}$ & $-0.533^{+3.28}_{-0.229}$ & $-0.689^{+0.653}_{-0.0647}$ \\ \hline
		$\ln{(\lambda)}$ & -- & $5.03^{+1.33}_{-4.40}$ & $6.11^{+3.75\ast}_{-2.55}$ & $2.81^{+1.33}_{-4.40}$ & $4.27^{+5.17\ast}_{-1.71}$ & $2.18^{+4.51}_{-1.37}$ & $3.51^{+2.84}_{-1.06}$ \\ \hline
		$M_b$ & $(1.91^{+1.53}_{-0.233})\times 10^{10}$ & $(2.02^{+0.736}_{-0.461})\times 10^{10}$ & $(2.36^{+0.490}_{-0.674}) \times 10^{10}$ & $(5.35^{+4.44}_{-3.52}) \times 10^{10}$ & $(2.41^{+7.45\ast}_{-1.56}) \times 10^{10}$ & $(5.72^{+3.64}_{-5.08}) \times 10^{10}$ & $(8.27^{+1.71\ast}_{-6.38})\times 10^{10}$ \\ \hline
		$a_b$ & $0.712^{+0.568}_{-0.110}$ & $0.757^{+0.237}_{-0.216}$ & $0.868^{+0.152}_{-0.259}$ & $0.925^{+0.136}_{-0.205}$ & $0.822^{+0.487}_{-0.251}$ & $0.927^{+0.145}_{-0.351}$ & $0.889^{+0.208}_{-0.308}$ \\ \hline
		$M_d$ & $(1.63^{+0.408}_{-0.382})\times 10^{11}$ & $(1.91^{+0.398}_{-0.459})\times 10^{11}$ & $(1.77^{+0.330}_{-0.330}) \times 10^{11}$ & $(3.39^{+3.06}_{-1.73}) \times 10^{11}$ & $(2.02^{+3.85}_{-1.02}) \times 10^{11}$ & $(2.86^{+2.86}_{-2.25})\times 10^{11}$ & $(6.12^{+1.23}_{-3.77})\times 10^{11}$ \\ \hline
		$a_d$ & $5.17^{+0.840}_{-0.445}$ & $5.49^{+0.827}_{-0.788}$ & $5.38^{+0.858}_{-0.695}$ & $5.40^{+0.455}_{-0.581}$ & $5.54^{+0.925}_{-0.441}$ & $5.60^{+0.645}_{-0.892}$ & $5.56^{+0.793}_{-0.633}$ \\ \hline
		$n$ & $2.85^{+0.318}_{-0.363}$ & $2.72^{+0.435}_{-0.511}$ & $2.57^{+0.597}_{-0.469}$ & $2.52^{+0.411}_{-0.472}$ & $2.31^{+1.07}_{-0.424}$ & $2.66^{+0.535}_{-0.325}$ & $2.30^{+0.887}_{-0.316}$ \\ \hline
		$\ln{(\rho_s)}$ & $13.8^{+2.82}_{-3.36}$ & $8.42^{+3.45}_{-2.38}$ & $7.45^{+2.97}_{-1.83}$ & $10.0^{+3.94}_{-9.07}$ & $11.4^{+3.23}_{-7.08}$ & -- & -- \\ \hline
		$r_s$ & $50.9^{+287}_{-39.9\dag}$ & $193^{+186}_{-128}$ & $167^{+203\ast}_{-137\dag}$ & $309^{+71.8\ast}_{-272}$ & $163^{+211}_{-141\dag}$ & -- & -- \\ \hline
		$\ln{(M_{BH})}$ & $14.1^{+6.14\ast}_{-1.28}$ & $19.2^{+1.12\ast}_{-9.42}$ & $19.8^{+0.643\ast}_{-14.7}$ & $20.7^{+0.767}_{-6.59}$ & $20.1^{+1.51}_{-19.7\dag}$ & $20.4^{+0.684\ast}_{-18.5}$ & $21.3^{+0.540\ast}_{-17.3}$ \\ \hline\hline
	\end{tabular}
\end{table*}

\bibliographystyle{spphys}
\bibliography{paper}

\begin{thebibliography}{10}
\providecommand{\url}[1]{{#1}}
\providecommand{\urlprefix}{URL }
\expandafter\ifx\csname urlstyle\endcsname\relax
  \providecommand{\doi}[1]{DOI \discretionary{}{}{}#1}\else
  \providecommand{\doi}{DOI \discretionary{}{}{}\begingroup
  \urlstyle{rm}\Url}\fi

\bibitem{Bertone2018}
G.~Bertone, D.~Hooper, Reviews of Modern Physics \textbf{90}(4), 045002.
\newblock \doi{10.1103/revmodphys.90.045002}.
\newblock \urlprefix\url{http://dx.doi.org/10.1103/RevModPhys.90.045002}

\bibitem{Honma1997}
M.~Honma, Y.~Sofue, Publications of the Astronomical Society of Japan
  \textbf{49}(4), 453.
\newblock \doi{10.1093/pasj/49.4.453}.
\newblock
  \urlprefix\url{https://ui.adsabs.harvard.edu/abs/1997PASJ...49..453H}.
\newblock ADS Bibcode: 1997PASJ...49..453H

\bibitem{Cirelli2024}
M.~Cirelli, A.~Strumia, J.~Zupan,
  \urlprefix\url{https://arxiv.org/abs/2406.01705}

\bibitem{Will2018}
C.M. Will, \emph{{Theory and Experiment in Gravitational Physics}} (Cambridge
  University Press, 2018)

\bibitem{Bertone2005}
G.~Bertone, D.~Hooper, J.~Silk, Physics Reports \textbf{405}(5--6), 279.
\newblock \doi{10.1016/j.physrep.2004.08.031}.
\newblock \urlprefix\url{http://dx.doi.org/10.1016/j.physrep.2004.08.031}

\bibitem{Mannheim2006}
P.~Mannheim, Progress in Particle and Nuclear Physics \textbf{56}(2), 340.
\newblock \doi{10.1016/j.ppnp.2005.08.001}.
\newblock \urlprefix\url{http://dx.doi.org/10.1016/j.ppnp.2005.08.001}

\bibitem{Khelashvili2023}
M.~Khelashvili, A.~Rudakovskyi, S.~Hossenfelder, Monthly Notices of the Royal
  Astronomical Society \textbf{523}(3), 3393.
\newblock \doi{10.1093/mnras/stad1595}

\bibitem{Baker2015}
T.~Baker, D.~Psaltis, C.~Skordis, The Astrophysical Journal \textbf{802}(1),
  63.
\newblock \doi{10.1088/0004-637x/802/1/63}

\bibitem{Binney2008}
J.~Binney, S.~Tremaine, \emph{Galactic Dynamics: Second Edition} (Princeton
  University Press, 2008)

\bibitem{Almeida2018}
{\'A}.~de~Almeida, L.~Amendola, V.~Niro, Journal of Cosmology and Astroparticle
  Physics \textbf{2018}(08), 012 (2018).
\newblock \doi{10.1088/1475-7516/2018/08/012}.
\newblock \urlprefix\url{http://dx.doi.org/10.1088/1475-7516/2018/08/012}

\bibitem{Clifton2008}
T.~Clifton, Physical Review D \textbf{77}(2), 024041 (2008).
\newblock \doi{10.1103/physrevd.77.024041}

\bibitem{Stabile2011}
A.~Stabile, G.~Scelza, Physical Review D \textbf{84}(12), 124023 (2011).
\newblock \doi{10.1103/physrevd.84.124023}.
\newblock \urlprefix\url{http://dx.doi.org/10.1103/PhysRevD.84.124023}

\bibitem{Moffat2013}
J.W. Moffat, S.~Rahvar, Monthly Notices of the Royal Astronomical Society
  \textbf{436}(2), 1439 (2013).
\newblock \doi{10.1093/mnras/stt1670}.
\newblock \urlprefix\url{http://dx.doi.org/10.1093/mnras/stt1670}

\bibitem{Rahvar2014}
S.~Rahvar, B.~Mashhoon, Physical Review D \textbf{89}(10), 104011 (2014).
\newblock \doi{10.1103/PhysRevD.89.104011}.
\newblock \urlprefix\url{https://link.aps.org/doi/10.1103/PhysRevD.89.104011}

\bibitem{Henrichs2021}
J.~Henrichs, M.~Lembo, F.~Iocco, L.~Amendola, Physical Review D
  \textbf{104}(4), 043009 (2021).
\newblock \doi{10.1103/physrevd.104.043009}

\bibitem{Piazza2003}
F.~Piazza, C.~Marinoni, Physical Review Letters \textbf{91}(14), 141301 (2003).
\newblock \doi{10.1103/physrevlett.91.141301}.
\newblock \urlprefix\url{http://dx.doi.org/10.1103/PhysRevLett.91.141301}

\bibitem{Jusufi2023}
K.~Jusufi, G.~Leon, A.D. Millano, Physics of the Dark Universe \textbf{42},
  101318 (2023).
\newblock \doi{10.1016/j.dark.2023.101318}

\bibitem{Berezhiani2009}
Z.~Berezhiani, F.~Nesti, L.~Pilo, N.~Rossi, Journal of High Energy Physics
  \textbf{2009}(07), 083 (2009).
\newblock \doi{10.1088/1126-6708/2009/07/083}.
\newblock \urlprefix\url{http://dx.doi.org/10.1088/1126-6708/2009/07/083}

\bibitem{Murata2015}
J.~Murata, S.~Tanaka, Classical and Quantum Gravity \textbf{32}(3), 033001
  (2015).
\newblock \doi{10.1088/0264-9381/32/3/033001}.
\newblock \urlprefix\url{http://dx.doi.org/10.1088/0264-9381/32/3/033001}

\bibitem{Jovanovic2023}
P.~Jovanović, V.~Borka~Jovanović, D.~Borka, A.F. Zakharov, Journal of
  Cosmology and Astroparticle Physics \textbf{2023}(03), 056 (2023).
\newblock \doi{10.1088/1475-7516/2023/03/056}.
\newblock \urlprefix\url{http://dx.doi.org/10.1088/1475-7516/2023/03/056}

\bibitem{Hassan2024}
D.S. Hassan, M.D. Danarianto, A.~Sulaksono, in \emph{Proc. 10th Int. Seminar on
  Aerospace Science and Technology} (Springer, 2025), pp. 213--222.
\newblock \doi{10.1007/978-981-96-1344-1_23}

\bibitem{Sofue2020}
Y.~Sofue,   (2020).
\newblock \urlprefix\url{https://arxiv.org/abs/2004.11688}

\bibitem{deRham2017}
C.~de~Rham, J.T. Deskins, A.J. Tolley, S.Y. Zhou, Reviews of Modern Physics
  \textbf{89}(2), 025004 (2017).
\newblock \doi{10.1103/revmodphys.89.025004}.
\newblock \urlprefix\url{http://dx.doi.org/10.1103/RevModPhys.89.025004}

\bibitem{Fischbach1986}
E.~Fischbach, D.~Sudarsky, A.~Szafer, C.~Talmadge, S.H. Aronson, Physical
  Review Letters \textbf{56}(1), 3 (1986).
\newblock \doi{10.1103/PhysRevLett.56.3}.
\newblock \urlprefix\url{https://link.aps.org/doi/10.1103/PhysRevLett.56.3}

\bibitem{Gibbons1981}
G.W. Gibbons, B.F. Whiting, Nature \textbf{291}(5817), 636 (1981).
\newblock \doi{10.1038/291636a0}.
\newblock \urlprefix\url{https://doi.org/10.1038/291636a0}

\bibitem{Fujii1971}
Y.~Fujii, Nature \textbf{234}, 5 (1971)

\bibitem{Murata2014}
J.~Murata, S.~Tanaka, Classical and Quantum Gravity \textbf{32}(3), 033001
  (2015).
\newblock \doi{10.1088/0264-9381/32/3/033001}

\bibitem{Lee2020}
J.G. Lee, E.G. Adelberger, T.S. Cook, S.M. Fleischer, B.R. Heckel, Physical
  Review Letters \textbf{124}(10), 101101 (2020).
\newblock \doi{10.1103/PhysRevLett.124.101101}

\bibitem{Schmidt2007}
H.J. Schmidt, International Journal of Geometric Methods in Modern Physics
  \textbf{04}(02), 209 (2007).
\newblock \doi{10.1142/s0219887807001977}.
\newblock \urlprefix\url{http://dx.doi.org/10.1142/S0219887807001977}

\bibitem{Sotiriou2006}
T.P. Sotiriou, Classical and Quantum Gravity \textbf{23}, 5117 (2006).
\newblock \doi{10.1088/0264-9381/23/17/003}

\bibitem{Sanders1986}
R.H. Sanders, Astronomy and Astrophysics \textbf{154}, 135 (1986).
\newblock \urlprefix\url{https://ui.adsabs.harvard.edu/abs/1986A&A...154..135S}

\bibitem{Damour2002}
T.~Damour, F.~Piazza, G.~Veneziano, Physical Review D \textbf{66}(4), 046007
  (2002).
\newblock \doi{10.1103/physrevd.66.046007}

\bibitem{Freeman1970}
K.C. Freeman, The Astrophysical Journal \textbf{160}, 811 (1970).
\newblock \doi{10.1086/150474}

\bibitem{Sofue2015}
Y.~Sofue, Publications of the Astronomical Society of Japan \textbf{67}(4), 75
  (2015).
\newblock \doi{10.1093/pasj/psv042}.
\newblock \urlprefix\url{http://dx.doi.org/10.1093/pasj/psv042}

\bibitem{Rix2013}
H.W. Rix, J.~Bovy, The Astronomy and Astrophysics Review \textbf{21}(1), 61
  (2013).
\newblock \doi{10.1007/s00159-013-0061-8}.
\newblock \urlprefix\url{http://dx.doi.org/10.1007/s00159-013-0061-8}

\bibitem{Adibekyan2011}
V.Z. Adibekyan, N.C. Santos, S.G. Sousa, G.~Israelian, Astronomy and
  Astrophysics \textbf{535}, L11 (2011).
\newblock \doi{10.1051/0004-6361/201118240}

\bibitem{Bovy2012}
J.~Bovy, H.W. Rix, D.W. Hogg, The Astrophysical Journal \textbf{751}(2), 131
  (2012).
\newblock \doi{10.1088/0004-637x/751/2/131}.
\newblock \urlprefix\url{http://dx.doi.org/10.1088/0004-637X/751/2/131}

\bibitem{Sofue2009}
Y.~Sofue, M.~Honma, T.~Omodaka, Publications of the Astronomical Society of
  Japan \textbf{61}(2), 227 (2009).
\newblock \doi{10.1093/pasj/61.2.227}.
\newblock \urlprefix\url{http://dx.doi.org/10.1093/pasj/61.2.227}

\bibitem{deVaucouleurs1948}
G.~de~Vaucouleurs, Annales d'Astrophysique \textbf{11}, 247 (1948)

\bibitem{Sersic1963}
J.L. Sérsic, Boletín de la Asociación Argentina de Astronomía La Plata
  Argentina \textbf{6}, 41 (1963)

\bibitem{Capaccioli1989}
M.~Capaccioli, in \emph{World of Galaxies (Le Monde des Galaxies)}, ed. by
  J.~Corwin, Harold~G., L.~Bottinelli (1989), pp. 208--227

\bibitem{Baes2010}
M.~Baes, G.~Gentile, Astronomy and Astrophysics \textbf{525}, A136 (2010).
\newblock \doi{10.1051/0004-6361/201015716}.
\newblock \urlprefix\url{http://dx.doi.org/10.1051/0004-6361/201015716}

\bibitem{Baes2011}
M.~Baes, E.~Van~Hese, Astronomy and Astrophysics \textbf{534}, A69 (2011).
\newblock \doi{10.1051/0004-6361/201117708}.
\newblock \urlprefix\url{http://dx.doi.org/10.1051/0004-6361/201117708}

\bibitem{Benson2010}
A.J. Benson, Physics Reports \textbf{495}(2--3), 33 (2010).
\newblock \doi{10.1016/j.physrep.2010.06.001}.
\newblock \urlprefix\url{http://dx.doi.org/10.1016/j.physrep.2010.06.001}

\bibitem{Navarro1996}
J.F. Navarro, C.S. Frenk, S.D.M. White, The Astrophysical Journal \textbf{462},
  563 (1996).
\newblock \doi{10.1086/177173}

\bibitem{Ghez1998}
A.M. Ghez, B.L. Klein, M.~Morris, E.E. Becklin, The Astrophysical Journal
  \textbf{509}(2), 678 (1998).
\newblock \doi{10.1086/306528}.
\newblock \urlprefix\url{http://dx.doi.org/10.1086/306528}

\bibitem{GaiaCollaboration2023}
{Gaia Collaboration}, A.~Vallenari, A.G.A. Brown, T.~Prusti, J.H.J. de~Bruijne,
  F.~Arenou, et~al., Astronomy and Astrophysics \textbf{674}, A1 (2023).
\newblock \doi{10.1051/0004-6361/202243940}.
\newblock \urlprefix\url{http://dx.doi.org/10.1051/0004-6361/202243940}

\bibitem{Speagle2020}
J.S. Speagle, Monthly Notices of the Royal Astronomical Society
  \textbf{493}(3), 3132 (2020).
\newblock \doi{10.1093/mnras/staa278}.
\newblock \urlprefix\url{http://dx.doi.org/10.1093/mnras/staa278}

\bibitem{Kass1995}
R.E. Kass, A.E. Raftery, Journal of the American Statistical Association
  \textbf{90}(430), 773 (1995).
\newblock \doi{10.1080/01621459.1995.10476572}

\bibitem{Skilling2006}
J.~Skilling, Bayesian Analysis \textbf{1}(4), 833 (2006).
\newblock \doi{10.1214/06-BA127}.
\newblock \urlprefix\url{https://doi.org/10.1214/06-BA127}

\bibitem{Higson2018}
E.~Higson, W.~Handley, M.~Hobson, A.~Lasenby, Statistics and Computing
  \textbf{29}(5), 891 (2018).
\newblock \doi{10.1007/s11222-018-9844-0}.
\newblock \urlprefix\url{http://dx.doi.org/10.1007/s11222-018-9844-0}

\bibitem{Sofue2016}
Y.~Sofue, Publications of the Astronomical Society of Japan \textbf{69}(1), 1
  (2016).
\newblock \doi{10.1093/pasj/psw103}

\bibitem{Burton1978}
W.B. Burton, M.A. Gordon, Astronomy and Astrophysics \textbf{63}, 7 (1978).
\newblock \urlprefix\url{https://api.semanticscholar.org/CorpusID:116387020}

\bibitem{Fich1989}
M.~Fich, L.~Blitz, A.~Stark, The Astrophysical Journal \textbf{342}, 272
  (1989).
\newblock \doi{10.1086/167591}

\bibitem{Clemens1985}
D.P. Clemens, The Astrophysical Journal \textbf{295}, 422 (1985).
\newblock \doi{10.1086/163386}

\bibitem{Petrovskaia1986}
I.V. Petrovskaia, P.~Teerikorpi, Astronomy and Astrophysics \textbf{163}(1--2),
  39 (1986)

\bibitem{Merrifield1992}
M.R. Merrifield, The Astronomical Journal \textbf{103}, 1552 (1992).
\newblock \doi{10.1086/116168}

\bibitem{Brand1993}
J.~Brand, L.~Blitz, Astronomy and Astrophysics \textbf{275}, 67 (1993)

\bibitem{Wang2022}
H.F. Wang, {\v{Z}}.~Chrob{\'a}kov{\'a}, M.~L{\'o}pez-Corredoira,
  F.~Sylos~Labini, The Astrophysical Journal \textbf{942}(1), 12 (2022).
\newblock \doi{10.3847/1538-4357/aca27c}.
\newblock \urlprefix\url{http://dx.doi.org/10.3847/1538-4357/aca27c}

\bibitem{Zhou2023}
Y.~Zhou, X.~Li, Y.~Huang, H.~Zhang, The Astrophysical Journal \textbf{946}(2),
  73 (2023).
\newblock \doi{10.3847/1538-4357/acadd9}.
\newblock \urlprefix\url{http://dx.doi.org/10.3847/1538-4357/acadd9}

\bibitem{Abuter2023}
R.~Abuter, N.~Aimar, P.~Amaro~Seoane, {GRAVITY Collaboration}, Astronomy and
  Astrophysics \textbf{677}, L10 (2023).
\newblock \doi{10.1051/0004-6361/202347416}.
\newblock \urlprefix\url{http://dx.doi.org/10.1051/0004-6361/202347416}

\bibitem{Courteau2011}
S.~Courteau, L.M. Widrow, M.~McDonald, P.~Guhathakurta, K.M. Gilbert, Y.~Zhu,
  R.L. Beaton, S.R. Majewski, The Astrophysical Journal \textbf{739}(1), 20
  (2011).
\newblock \doi{10.1088/0004-637x/739/1/20}

\bibitem{Naik2018}
A.P. Naik, E.~Puchwein, A.C. Davis, C.~Arnold, Monthly Notices of the Royal
  Astronomical Society \textbf{480}(4), 5211 (2018).
\newblock \doi{10.1093/mnras/sty2199}.
\newblock \urlprefix\url{https://doi.org/10.1093/mnras/sty2199}

\bibitem{Tan2024}
Y.~Tan, Y.~Lu, Phys. Rev. D \textbf{109}, 044047 (2024).
\newblock \doi{10.1103/PhysRevD.109.044047}.
\newblock \urlprefix\url{https://link.aps.org/doi/10.1103/PhysRevD.109.044047}

\bibitem{AlBaidhany2020}
I.A. Al-Baidhany, S.S. Chiad, W.A. Jabbar, A.K. Al-kadumi, N.F. Habubi, H.L.
  Mansour, in \emph{American Institute of Physics Conference Series},
  \emph{American Institute of Physics Conference Series}, vol. 2290 (AIP,
  2020), \emph{American Institute of Physics Conference Series}, vol. 2290, p.
  050050.
\newblock \doi{10.1063/5.0027838}.
\newblock \urlprefix\url{https://ui.adsabs.harvard.edu/abs/2020AIPC.2290e0050A}

\bibitem{Sanders1984}
R.H. Sanders, Astronomy and Astrophysics \textbf{136}(2), L21 (1984).
\newblock \urlprefix\url{https://ui.adsabs.harvard.edu/abs/1984A&A...136L..21S}

\bibitem{GRAVITYCollaboration2025}
{The GRAVITY Collaboration}, et~al., arXiv e-prints arXiv:2504.02908 (2025).
\newblock \doi{10.48550/arXiv.2504.02908}.
\newblock \urlprefix\url{https://ui.adsabs.harvard.edu/abs/2025arXiv250402908T}

\bibitem{Khoury2003}
J.~Khoury, A.~Weltman, Physical Review Letters \textbf{93}(17), 171104 (2003).
\newblock \doi{10.1103/physrevlett.93.171104}

\bibitem{Lelli2016}
F.~Lelli, S.S. McGaugh, J.M. Schombert, The Astronomical Journal
  \textbf{152}(6), 157 (2016).
\newblock \doi{10.3847/0004-6256/152/6/157}

\bibitem{Haubner2024}
K.~Haubner, F.~Lelli, E.~Di~Teodoro, F.~Duey, S.~McGaugh, J.~Schombert, K.M.
  Hess, {the Apertif Team},   (2024).
\newblock \doi{10.48550/arXiv.2411.13329}

\bibitem{Harris2020}
C.R. Harris, et~al., Nature \textbf{585}(7825), 357 (2020).
\newblock \doi{10.1038/s41586-020-2649-2}.
\newblock \urlprefix\url{https://doi.org/10.1038/s41586-020-2649-2}

\bibitem{Virtanen2020}
P.~Virtanen, et~al., Nature Methods \textbf{17}(3), 261 (2020).
\newblock \doi{10.1038/s41592-019-0686-2}.
\newblock \urlprefix\url{http://dx.doi.org/10.1038/s41592-019-0686-2}

\bibitem{mpmathdevelopmentteam2023}
{The mpmath development team}, \emph{mpmath: a Python library for
  arbitrary-precision arithmetic} (2023).
\newblock \url{http://mpmath.org/}

\bibitem{Lam2015}
S.K. Lam, A.~Pitrou, S.~Seibert, in \emph{Proc. 2nd Workshop on LLVM Compiler
  Infrastructure in HPC} (2015), pp. 7:1--7:6.
\newblock \doi{10.1145/2833157.2833162}

\bibitem{Conway2010}
J.T. Conway, H.S. Cohl, Zeitschrift für Angewandte Mathematik und Physik
  \textbf{61}(3), 425 (2010).
\newblock \doi{10.1007/s00033-009-0039-6}

\bibitem{Vaccari2000}
M.~Vaccari, Gaia galaxy survey: Simulated galaxy observations.
\newblock Master's thesis, Università degli Studi di Padova (2000).
\newblock
  \urlprefix\url{https://www.mattiavaccari.net/research/masterthesis/masterthesis/thesis.html}

\end{thebibliography}

\end{document}